\PassOptionsToPackage{unicode}{hyperref}
\PassOptionsToPackage{hyphens}{url}
\PassOptionsToPackage{dvipsnames,svgnames,x11names}{xcolor}
\documentclass[
  12pt]{article}

\usepackage{amsmath,amssymb}
\usepackage{amsthm}
\usepackage{bm}
\usepackage{mathtools}

\usepackage{psfrag}
\usepackage{enumerate}
\usepackage{algorithm}
\usepackage{algpseudocode}
\usepackage{tabularx}

\allowdisplaybreaks
\usepackage{iftex}
\ifPDFTeX
  \usepackage[T1]{fontenc}
  \usepackage[utf8]{inputenc}
  \usepackage{textcomp} 
\else 
  \usepackage{unicode-math}
  \defaultfontfeatures{Scale=MatchLowercase}
  \defaultfontfeatures[\rmfamily]{Ligatures=TeX,Scale=1}
\fi
\usepackage{lmodern}
\ifPDFTeX\else  
\fi
\IfFileExists{upquote.sty}{\usepackage{upquote}}{}
\IfFileExists{microtype.sty}{
  \usepackage[]{microtype}
  \UseMicrotypeSet[protrusion]{basicmath} 
}{}
\makeatletter
\@ifundefined{KOMAClassName}{
  \IfFileExists{parskip.sty}{%
    \usepackage{parskip}
  }{
    \setlength{\parindent}{0pt}
    \setlength{\parskip}{6pt plus 2pt minus 1pt}}
}{
  \KOMAoptions{parskip=half}}
\makeatother
\usepackage{xcolor}
\makeatletter
\ifx\paragraph\undefined\else
  \let\oldparagraph\paragraph
  \renewcommand{\paragraph}{
    \@ifstar
      \xxxParagraphStar
      \xxxParagraphNoStar
  }
  \newcommand{\xxxParagraphStar}[1]{\oldparagraph*{#1}\mbox{}}
  \newcommand{\xxxParagraphNoStar}[1]{\oldparagraph{#1}\mbox{}}
\fi
\ifx\subparagraph\undefined\else
  \let\oldsubparagraph\subparagraph
  \renewcommand{\subparagraph}{
    \@ifstar
      \xxxSubParagraphStar
      \xxxSubParagraphNoStar
  }
  \newcommand{\xxxSubParagraphStar}[1]{\oldsubparagraph*{#1}\mbox{}}
  \newcommand{\xxxSubParagraphNoStar}[1]{\oldsubparagraph{#1}\mbox{}}
\fi
\makeatother

\usepackage{longtable,booktabs,array}
\usepackage{calc} 
\usepackage{etoolbox}
\makeatletter
\patchcmd\longtable{\par}{\if@noskipsec\mbox{}\fi\par}{}{}
\makeatother
\IfFileExists{footnotehyper.sty}{\usepackage{footnotehyper}}{\usepackage{footnote}}
\makesavenoteenv{longtable}
\usepackage{graphicx}
\makeatletter
\def\maxwidth{\ifdim\Gin@nat@width>\linewidth\linewidth\else\Gin@nat@width\fi}
\def\maxheight{\ifdim\Gin@nat@height>\textheight\textheight\else\Gin@nat@height\fi}
\makeatother
\setkeys{Gin}{width=\maxwidth,height=\maxheight,keepaspectratio}
\makeatletter
\def\fps@figure{htbp}
\makeatother

\makeatletter
\@ifpackageloaded{caption}{}{\usepackage{caption}}
\AtBeginDocument{%
\ifdefined\contentsname
  \renewcommand*\contentsname{Table of contents}
\else
  \newcommand\contentsname{Table of contents}
\fi
\ifdefined\listfigurename
  \renewcommand*\listfigurename{List of Figures}
\else
  \newcommand\listfigurename{List of Figures}
\fi
\ifdefined\listtablename
  \renewcommand*\listtablename{List of Tables}
\else
  \newcommand\listtablename{List of Tables}
\fi
\ifdefined\figurename
  \renewcommand*\figurename{Figure}
\else
  \newcommand\figurename{Figure}
\fi
\ifdefined\tablename
  \renewcommand*\tablename{Table}
\else
  \newcommand\tablename{Table}
\fi
}
\@ifpackageloaded{float}{}{\usepackage{float}}
\floatstyle{ruled}
\@ifundefined{c@chapter}{\newfloat{codelisting}{h}{lop}}{\newfloat{codelisting}{h}{lop}[chapter]}
\floatname{codelisting}{Listing}

\makeatother
\makeatletter
\@ifpackageloaded{caption}{}{\usepackage{caption}}
\@ifpackageloaded{subcaption}{}{\usepackage{subcaption}}
\makeatother

\ifLuaTeX
  \usepackage{selnolig}  
\fi
\usepackage[]{natbib}
\IfFileExists{xurl.sty}{\usepackage{xurl}}{} 
\newcommand{\anon}{1}

\newtheorem{theorem}{Theorem}
\newtheorem{lemma}[theorem]{Lemma}
\newtheorem{proposition}[theorem]{Proposition}
\newtheorem{corollary}[theorem]{Corollary}

\theoremstyle{definition}
\newtheorem{assumption}{Assumption}
\newtheorem{definition}{Definition}

\theoremstyle{remark}
\newtheorem{remark}{Remark}

\usepackage{hyperref}
\usepackage{bookmark}
\usepackage[nameinlink,capitalise,noabbrev]{cleveref}

\hypersetup{
    pdftitle={Learning Sequential Decisions from Multiple Sources via Group-Robust Markov Decision Processes},
    pdfauthor={Mingyuan Xu; Zongqi Xia; Tianxi Cai; Doudou Zhou; Nian Si},
    pdfkeywords={multi-source reinforcement learning; offline reinforcement learning; d-rectangular uncertainty set; pessimistic value iteration},
  colorlinks=true,
  linkcolor={Blue},
  filecolor={Blue},
  citecolor={Blue},
  urlcolor={Blue},
  pdfcreator={LaTeX via pandoc}}

\usepackage{bbm}
\usepackage{mathrsfs}
\usepackage{multirow}

\begin{document}

\def\spacingset#1{\renewcommand{\baselinestretch}%
{#1}\small\normalsize} \spacingset{1}

\if1\anon
{
  \title{\bf Learning Sequential Decisions from Multiple Sources via Group-Robust Markov Decision Processes}
  \author{\normalsize
    Mingyuan Xu$^{1}$,
    Zongqi Xia$^{2}$,
    Tianxi Cai$^{3,4}$,
    Doudou Zhou$^{1}$\thanks{Corresponding author. Email: ddzhou@nus.edu.sg},
    Nian Si$^{5}$\thanks{Corresponding author. Email: niansi@ust.hk}
    \\[0.6em]
    \parbox{0.95\linewidth}{\centering\normalsize
      $^{1}$Department of Statistics and Data Science, National University of Singapore\\
      $^{2}$Department of Neurology, University of Pittsburgh\\
      $^{3}$Department of Biostatistics, Harvard T.H. Chan School of Public Health\\
      $^{4}$Department of Biomedical Informatics, Harvard Medical School\\
      $^{5}$Department of Industrial Engineering and Decision Analytics, Hong Kong University of Science and Technology
    }
  }
  \date{}
  \maketitle
} \fi

\if0\anon
{
  \bigskip
  \bigskip
  \bigskip
  \begin{center}
    {\LARGE\bf Title}
\end{center}
  \medskip
} \fi

\bigskip
\begin{abstract}
Learning a common policy from multiple source environments requires accounting for cross-source heterogeneity, and the offline setting introduces an additional challenge of limited data coverage. We formulate this problem as a distributionally robust Markov decision process under a shared linear MDP representation with source-specific transition and reward components. The uncertainty set is constructed through feature-wise mixtures of these components, yielding a $d$-rectangular structure that captures cross-source heterogeneity and admits a robust Bellman recursion. Building on this formulation, we develop Group-Robust Pessimistic Value Iteration (GR-PEVI), an offline policy-learning algorithm that combines feature-wise worst-case aggregation with a data-dependent pessimism penalty. Under a robust partial coverage condition, we establish finite-sample suboptimality guarantees for GR-PEVI and extend the analysis to clustered sources, characterizing the trade-off between the statistical gain from pooling similar sources and the approximation error induced by within-cluster heterogeneity. Simulation studies demonstrate the effectiveness of GR-PEVI, with empirical behavior consistent with the theoretical analysis. We apply GR-PEVI to a multi-source electronic health record study of multiple sclerosis treatment decisions to illustrate its use in a clinical setting.
\end{abstract}

\noindent%
\vfill

\newpage
\spacingset{1.8} 

\section{Introduction}
\label{sec:intro}
Sequential decision data are increasingly assembled across multiple sources, particularly in healthcare, where longitudinal records accumulate across hospitals or health systems~\citep{pollard2018eicu,forrest2021pcornet}. Combining data from multiple sources not only increases sample size but also provides evidence on a broader range of patient populations, treatment patterns, and care pathways than is typically available from a single institution \citep{hripcsak2016characterizing}. These data can inform a common treatment policy across healthcare systems that maps patient states to treatment decisions to improve long-term outcomes. 

Learning this policy presents two main challenges. First, online experimentation in healthcare can be costly, risky, or infeasible, so policy learning often relies on fixed observational trajectories collected under existing source-specific behavior policies~\citep{zhou2024federated,schweighofer2022dataset}. These policies may vary across hospitals, regions, and physician groups~\citep{corallo2014systematic} and cover only a limited portion of the state--action space relevant to a target policy, giving rise to the distribution-shift challenge of offline reinforcement learning (RL)~\citep{levine2020offline}. Second, sources may also differ in their transition and reward mechanisms, so the data need not arise from a single common Markov decision process (MDP)~\citep{shi2023provably}. Such differences mean that the same treatment may have different effects for patients in the same state, beyond differences in treatment choices across behavior policies.

A policy that performs well on average across sources may still perform poorly in some individual sources, because gains in one source can compensate for losses in another. Group distributionally robust optimization (DRO) formalizes this concern in non-sequential learning by optimizing worst-group performance~\citep{sagawa2019distributionally}. For the multi-source problem considered here, each group corresponds to a source MDP, and a policy's performance on that group is measured by its expected return under the corresponding MDP. A group-robust MDP formulation captures this principle by considering an uncertainty set that contains the source MDPs and optimizing the worst-case return over this set. Because each source MDP belongs to the uncertainty set, the resulting worst-case return is a lower bound on the policy's return in every source.

To formulate this group-robust learning problem, we consider a distributionally robust Markov decision process (DRMDP) under a shared linear MDP representation~\citep{jin2020provably}, in which the sources share a common feature map while their transition and reward components may vary. We refer to this formulation as a group-linear DRMDP. Allowing the source-mixture weights to vary across feature coordinates, with common weights for the transition and reward components within each coordinate, yields a $d$-rectangular uncertainty set that contains every source model and admits a robust Bellman recursion within the shared linear function class. This set also includes combinations of components that need not coincide with any individual source, so the robust objective protects against a broader class of environments induced by cross-source heterogeneity. To our knowledge, this is the first work to study common-policy learning across heterogeneous source MDPs in offline RL under a group-DRO formulation.

In the offline setting, the source-specific quantities in the Bellman recursion must be estimated from fixed trajectories. Limited coverage can make these estimates unreliable for actions rarely taken in the data, even when cross-source heterogeneity is accounted for in the objective. We develop Group-Robust Pessimistic Value Iteration (GR-PEVI) to implement the robust recursion from offline data. GR-PEVI estimates the source-specific coefficients of the linear Bellman backup and combines them through feature-wise worst-case aggregation. A data-dependent pessimism penalty then adjusts the estimated values for finite-sample uncertainty under limited coverage.

Under a robust partial coverage condition, we establish finite-sample robust suboptimality guarantees with explicit dependence on the feature dimension, horizon, and source-specific sample sizes. Because estimation uncertainty must be controlled across all sources under the feature-wise worst-case aggregation, the error bound is determined by the smallest source sample size, with the number of sources entering only through a logarithmic factor. To improve learning when individual sources have limited data, we develop a cluster-level extension that pools data within prespecified groups of similar sources. We derive an error bound in which pooled sample sizes replace individual source sample sizes in the statistical error term, while an additional approximation term accounts for within-cluster heterogeneity.

Simulation studies examine empirical performance, finite-sample scaling, and the effect of pessimism. We also apply GR-PEVI to a multi-source electronic health record (EHR) study of treatment decisions in multiple sclerosis (MS), where held-out off-policy evaluation suggests improved worst-source performance relative to the baseline methods.

\subsection{Related Work}
Multi-source RL methods differ in which policy they aim to learn. In the offline setting, some methods use cross-source information to improve source-specific or task-specific policies~\citep{zhou2024federated,bai2024pessimistic}, while others use multiple source MDPs or tasks to learn a policy for a designated target task~\citep{shi2023provably,bose2025offline}. Outside the offline setting, \citet{zhang2025pessimism} study zero-shot transfer RL, using multiple source environments to learn a policy for a target environment. In contrast, we use data from all sources to learn a single policy intended to perform across all source MDPs.

Common-policy learning has been studied in federated offline RL when the sources share one underlying MDP but collect data under different behavior policies~\citep{woo2024federated}. When the source MDPs differ, existing work has considered both aggregate and worst-case criteria for evaluating a common policy across sources. Aggregate objectives include the weighted value across a finite set of models in multi-model MDPs~\citep{steimle2021multimodel}, the average performance across heterogeneous environments in online federated RL~\citep{wang2024momentum}, and an average-case objective for zero-shot generalization RL~\citep{wang2025zeroshot}. Worst-case objectives have been studied in online multi-task RL~\citep{xu2023group} and online federated RL~\citep{hwang2025federated}. With fixed offline data, \citet{carranza2025robust} minimize worst-case regret over mixtures of the source distributions, but for single-stage decisions; in an MDP, the policy also changes the state distributions at which later rewards are received.

A central element of a robust worst-case objective is the uncertainty set, which specifies the environments over which performance is evaluated. In the robust extension of \citet{shi2023provably}, policy performance is evaluated over a KL-divergence neighborhood around a center MDP representing the population mean of the source MDPs. \citet{hwang2025federated} build the uncertainty set around the average source transition model and enlarge it to contain all source transition models as well as plausible perturbations. Both constructions are centered at an average model. We instead form the uncertainty set from mixtures of the source-specific transition and reward components, analogous to the use of source-distribution mixtures by \citet{carranza2025robust} in the single-stage setting. Our set therefore contains every source model without requiring a center model or a neighborhood radius.

How the uncertainty set is structured also determines whether robust dynamic programming remains tractable~\citep{iyengar2005robust,nilim2005robust}. The $r$-rectangular structure of \citet{goyal2023robust}, termed $d$-rectangularity in linear MDPs with $d$ feature coordinates, allows coupling across states while retaining tractability. Robust offline RL under $d$-rectangular uncertainty has been studied for linear MDPs~\citep{ma2022distributionally,liu2024minimax,wang2024sample}, where uncertainty is constructed through divergence- or distance-based neighborhoods around nominal per-feature transition components. Our uncertainty set is also $d$-rectangular, because the mixture weights are chosen separately across feature coordinates, but it is built from source-specific components rather than from neighborhoods of a nominal model.

The uncertainty set addresses cross-source heterogeneity, but offline learning must also contend with limited coverage of the fixed datasets. Pessimistic methods address this by learning conservatively where the data provide weak support~\citep{jin2021pessimism,uehara2021pessimistic}. DRO-based offline RL provides another approach, using uncertainty sets to construct conservative value estimates, with guarantees for the learned policy's performance in the underlying MDP~\citep{wang2024achieving,panaganti2025bridging}. When performance is itself defined over an uncertainty set, \citet{blanchet2023double} establish guarantees under a robust partial coverage condition using a double-pessimism framework. In GR-PEVI, pessimism and robustness play separate roles: the pessimism penalty controls estimation error in the robust Bellman backup, while the source-mixture uncertainty set specifies the environments over which the policy is evaluated. Our coverage condition extends that of \citet{blanchet2023double} to multiple sources (Section~\ref{sec:upper_bound}).

The remainder of the paper is organized as follows. Section~\ref{sec:model_method} introduces the group-linear DRMDP formulation and GR-PEVI. Section~\ref{sec:theory} develops the finite-sample theory for the proposed method, and Section~\ref{sec:cross-source} presents the cluster-level extension and its theoretical analysis. Sections~\ref{sec:simulation} and~\ref{sec:real-data} report simulation studies and a multi-source EHR study, respectively. Section~\ref{sec:discussion} concludes with a discussion. All proofs and additional implementation details are provided in the Supplementary Material.

\section{Model and Method}
\label{sec:model_method}
\subsection{Multi-Source Linear MDP and Offline Data}
\label{sec:linear-mdp}
We use $\mathbf I_d$ for the $d\times d$ identity matrix, $(\mathbf M)_{ii}$ for the $i$th diagonal entry of $\mathbf M$, $\mathbf 0_d$ for the zero vector in $\mathbb R^d$, and $\mathbf e_i$ for the $i$th standard basis vector, with dimensions understood from context. For an integer $n$, let $[n]:=\{1,\ldots,n\}$ and $\Delta^{n-1}:=\{\mathbf v\in\mathbb R^n:\mathbf v\ge \mathbf 0_n,\ \|\mathbf v\|_1=1\}$. For symmetric matrices $\mathbf A$ and $\mathbf B$, write $\mathbf A\succeq\mathbf B$ ($\mathbf A\succ\mathbf B$) if $\mathbf A-\mathbf B$ is positive semidefinite (positive definite). For a positive definite matrix $\mathbf M$, define $\|\mathbf v\|_{\mathbf M}:=\sqrt{\mathbf v^\top\mathbf M\mathbf v}$. For $x,y\in\mathbb R$, let $\min\{x,y\}^+:=\max\{\min\{x,y\},0\}$. For a probability measure $\mu$ and a bounded measurable function $V$, write $\langle\mu,V\rangle:=\int_{\mathcal S}V(s')\,\mu(\mathrm ds')$. 

Suppose there are $K$ source environments, each of which may represent a hospital. Source $k\in[K]$ is modeled as an episodic MDP $(\mathcal S,\mathcal A,H, P^k,r^k)$, where $\mathcal S$ and $\mathcal A$ are the state and action spaces, respectively, $H$ is the horizon, $P^k=\{P_h^k\}_{h=1}^H$ are the transition kernels, and $r^k=\{r_h^k\}_{h=1}^H$ are the expected reward functions with $r_h^k:\mathcal S\times\mathcal A\to[0,1]$. We assume that $\mathcal S$ is a measurable space and $\mathcal A$ is a nonempty compact metric space. For each $h\in[H]$ and $(s,a)\in\mathcal S\times\mathcal A$, $P_h^k(\cdot\mid s,a)$ is a probability measure on $\mathcal S$. Rewards are bounded in $[0,1]$, with conditional mean $r_h^k(s,a)$ given $s_h=s$ and $a_h=a$. For example, in a healthcare application, the MDP can model a patient's longitudinal care process. The state may encode the patient's current clinical condition, the action may represent a treatment or clinical decision, the reward may quantify the resulting health outcome, and the transition kernel characterizes how the patient's condition evolves in response to the chosen treatment.

We assume that the source-specific MDPs share a known measurable feature map $\boldsymbol{\phi}:\mathcal S\times\mathcal A\to\mathbb R^d$. This assumption is natural in many healthcare applications. Different sources may correspond to different hospitals, but patients are usually described using a common set of clinically meaningful variables, such as age, vital signs, laboratory measurements, comorbidities, and treatment indicators. These variables provide a common representation of state--action pairs across sources, even though the underlying patient populations, treatment effects, and disease dynamics may differ substantially. 

Following the simplex-valued linear MDP formulation of \citet[Example~2.2]{jin2020provably}, we require $\boldsymbol{\phi}(s,a)\in\Delta^{d-1}$ for all $(s,a)$. We further assume that $a\mapsto\boldsymbol{\phi}(s,a)$ is continuous for every $s\in\mathcal S$. For every $h\in[H]$ and $k\in[K]$, there exist $\boldsymbol{\mu}_h^{k}=(\mu_{h,1}^{k},\ldots,\mu_{h,d}^{k})^\top$, where each $\mu_{h,i}^{k}$ is a probability measure on $\mathcal S$, and $\boldsymbol{\theta}_h^{k}\in[0,1]^d$ such that
\begin{equation}
P_h^{k}(\cdot\mid s,a)
=
\sum_{i=1}^d
\phi_i(s,a)\mu_{h,i}^{k}(\cdot),
\qquad
r_h^{k}(s,a)
=
\boldsymbol{\phi}(s,a)^\top\boldsymbol{\theta}_{h}^{k}.
\label{source_wise_linear}
\end{equation}
Thus, $\|\boldsymbol{\theta}_h^k\|_2\le\sqrt d$. The shared feature map and source-specific transition--reward components are treated as part of the model specification, relative to which the robust uncertainty set below is defined.

We observe offline trajectories $\mathcal D=\{\mathcal D_k\}_{k=1}^K$, where source $k$ provides $N_k$ trajectories:
\[
\mathcal D_k
=
\Bigl\{
\bigl\{(s_h^{(k,\tau)},a_h^{(k,\tau)},r_h^{(k,\tau)})\bigr\}_{h=1}^{H}
\Bigr\}_{\tau=1}^{N_k}.
\]
These trajectories are generated by the source-specific MDP in \eqref{source_wise_linear} under behavior policies that may be unknown and may differ across sources. We assume that the datasets $\{\mathcal D_k\}_{k=1}^K$ are independent across sources. Within each source, the data collection process is required to satisfy the following compliance condition. For each $h\in[H]$, $k\in[K]$, and $\tau\in[N_k]$, let $\mathcal F_{h,k}^{\tau-1}:=\sigma\left(\left\{(s_h^{(k,j)},a_h^{(k,j)})\right\}_{j=1}^{\tau},\left\{(r_h^{(k,j)},s_{h+1}^{(k,j)})\right\}_{j=1}^{\tau-1}
\right)$. We require $\mathbb E\left[
r_h^{(k,\tau)}\mid \mathcal F_{h,k}^{\tau-1}
\right]=r_h^k\left(s_h^{(k,\tau)},a_h^{(k,\tau)}\right)$ and, for every measurable set $B\subseteq\mathcal S$,
$\mathbb P\left(s_{h+1}^{(k,\tau)}\in B\mid \mathcal F_{h,k}^{\tau-1}\right)=P_h^k\left(B\mid s_h^{(k,\tau)},a_h^{(k,\tau)}\right)$, almost surely.

\subsection{Robust Formulation and Learning Objective}
Under the specification \eqref{source_wise_linear}, cross-source heterogeneity at stage $h$ and feature coordinate $i$ is represented by $\{(\mu_{h,i}^k,\theta_{h,i}^k)\}_{k=1}^K$. To efficiently leverage information from these heterogeneous sources, we must specify how the source-specific components may be combined. However, a policy optimized for a fixed aggregation of source models may perform poorly in individual sources. This motivates a robust formulation in which the uncertainty set consists of models obtained by varying the source-mixture weights, and policies are evaluated by their worst-case performance over this set. Specifically, for each stage $h$ and coordinate $i$, we associate a source-mixture vector $\boldsymbol{\alpha}_{h,i}\in\Delta^{K-1}$ that is shared across state--action pairs and applied to both the transition and reward components of coordinate $i$, while the weights may vary across coordinates. For each $h\in[H]$, define
\begin{align}
\mathcal U_h
:=\Bigg\{\,(P_h,r_h):\ &
\exists\,
(\boldsymbol{\alpha}_{h,1},\ldots,\boldsymbol{\alpha}_{h,d})
\in(\Delta^{K-1})^d
\ \text{such that, for every }(s,a)\in\mathcal S\times\mathcal A,
\nonumber\\
& r_h(s,a)
=
\sum_{i=1}^d
\phi_i(s,a)
\sum_{k=1}^K
\alpha_{h,i}^{k}\theta_{h,i}^{k},
\nonumber\\
& P_h(\cdot\mid s,a)
=
\sum_{i=1}^d
\phi_i(s,a)
\sum_{k=1}^K
\alpha_{h,i}^{k}\mu_{h,i}^{k}(\cdot)
\,\Bigg\}.
\label{eq:robust_set}
\end{align}
This construction contains every source MDP as a special case while also allowing feature-wise combinations of source mechanisms. The resulting finite-horizon robust MDP $\left(\mathcal S,\mathcal A,H,\{\mathcal U_h\}_{h=1}^H\right)$ has a $d$-rectangular uncertainty structure~\citep{goyal2023robust}, with the same source-mixture weights used for the transition and reward components within each coordinate. Across stages, we take the uncertainty set to be $\prod_{h=1}^H\mathcal U_h$, so model choices are not coupled across stages. This stage-wise product structure, together with feature-wise rectangularity, yields the robust Bellman recursion below.

Given a Markov policy $\pi=\{\pi_h\}_{h=1}^H$, the robust value function $V_h^\pi$ and $Q$-function $Q_h^\pi$ at stage $h$ are defined as
\begin{align}
V_h^\pi(s)
&=
\inf_{\{(P_t,r_t)\}_{t=h}^H\in\prod_{t=h}^H\mathcal U_t}
\mathbb E_{\{P_t\}_{t=h}^H,\pi}
\left[\sum_{t=h}^H r_t(s_t,a_t)\mid s_h=s\right],
\label{eq:value-function}\\
Q_h^\pi(s,a)
&=
\inf_{\{(P_t,r_t)\}_{t=h}^H\in\prod_{t=h}^H\mathcal U_t}
\mathbb E_{\{P_t\}_{t=h}^H,\pi}
\left[\sum_{t=h}^H r_t(s_t,a_t)\mid s_h=s,\ a_h=a\right].
\label{eq:Q-function}
\end{align}
Here the adversary selects an admissible stage-specific model sequence $\{(P_t,r_t)\}_{t=h}^H$, independently of the realized trajectory history, to minimize the expected return. 

\begin{proposition}[Robust Bellman Equations]\label{prop:robust-bellman}
For the stage-wise product uncertainty set $\prod_{h=1}^H\mathcal U_h$, with each $\mathcal U_h$ defined in~\eqref{eq:robust_set}, the value and $Q$-functions of any Markov policy $\pi$ satisfy
\begin{align}
Q_h^\pi(s,a)
&=
\inf_{(P_h,r_h)\in\mathcal U_h}
\left\{
r_h(s,a)
+
\mathbb E_{s'\sim P_h(\cdot\mid s,a)}
\left[V_{h+1}^\pi(s')\right]
\right\},
\label{robust-bellman-equation-1}\\
V_h^\pi(s)
&=
\mathbb E_{a\sim\pi_h(\cdot\mid s)}
\left[Q_h^\pi(s,a)\right].
\label{robust-bellman-equation-2}
\end{align}
For the finite-horizon setting considered here, the terminal condition is $V_{H+1}^\pi\equiv 0$.
\end{proposition}

After substituting \eqref{eq:robust_set} into \eqref{robust-bellman-equation-1}, the current-stage infimum decomposes coordinate-wise because the weights range over $(\Delta^{K-1})^d$ and $\phi_i(s,a)\ge0$. Since each resulting objective is linear over $\Delta^{K-1}$, we obtain
\begin{align}
Q_h^\pi(s,a)
&=
\sum_{i=1}^d \phi_i(s,a)
\inf_{\boldsymbol{\alpha}_{h,i}\in\Delta^{K-1}}
\sum_{k=1}^K \alpha_{h,i}^k
\left[
\theta_{h,i}^k+
\left\langle\mu_{h,i}^k,V_{h+1}^\pi\right\rangle
\right]
\nonumber\\
&=
\sum_{i=1}^d \phi_i(s,a)
\min_{k\in[K]}
\left[
\theta_{h,i}^k+
\left\langle\mu_{h,i}^k,V_{h+1}^\pi\right\rangle
\right].
\label{eq:q_in_our_setting}
\end{align}
Hence, the robust Bellman operator can be written as
\begin{equation}
(\mathbb B_hV)(s,a)
=
\boldsymbol{\phi}(s,a)^\top\mathbf w_h(V),
\qquad
\left(\mathbf{w}_h(V)\right)_i
=
\min_{k\in[K]}
\left[
\theta_{h,i}^k+\langle\mu_{h,i}^k,V\rangle
\right].
\label{eq:Bellman-Operator}
\end{equation}
Thus, for any fixed bounded measurable $V$, the robust Bellman backup remains linear in the shared feature map $\boldsymbol{\phi}$.

We now state the learning objective. Let $\Pi:=\{\pi=\{\pi_h\}_{h=1}^H: \pi_h(\cdot\mid s)\in\mathcal P(\mathcal A) \text{ for all } s\in\mathcal S,\ h\in[H]\}$ denote the class of Markov policies, where $\mathcal P(\mathcal A)$ is the set of Borel probability measures on $\mathcal A$ and each $\pi_h$ is required to be a measurable probability kernel. Define the robust optimal value function by $V_h^\star(s):=\sup_{\pi\in\Pi}V_h^\pi(s)$. Let $\pi^\star$ denote a robust optimal policy satisfying $V_h^{\pi^\star}(s)=V_h^\star(s)$ for all $h\in[H]$ and $s\in\mathcal S$. A deterministic Markov policy with this property exists; see Supplementary Section~\ref{proof:existence}. Given a policy $\widehat{\pi}$ learned from the offline data, define its robust suboptimality by
\begin{equation}
\label{eq:subopt-def}
\operatorname{SubOpt}(\widehat{\pi};s)
:=V_1^{\pi^\star}(s)-V_1^{\widehat{\pi}}(s),
\qquad s\in\mathcal S.
\end{equation}
Our goal is to learn a policy with provably controlled robust suboptimality using only the pre-collected datasets.

\subsection{Algorithm}
\label{sec:algorithm}

Implementing the robust Bellman operator in \eqref{eq:Bellman-Operator} from offline data requires estimating, for each source, the coefficients of the shared feature map in the Bellman backup. These coefficients may differ across sources because their transition and reward mechanisms can differ. Their estimates may also have different levels of uncertainty because sample sizes and coverage can vary across sources. GR-PEVI retains the source-specific estimates for feature-wise robust aggregation and accounts for differences in estimation uncertainty through a data-dependent pessimism penalty.

For a bounded measurable function $V$, define the source-specific Bellman coefficient vector $\boldsymbol{\nu}_h^k(V):=
\bigl(\theta_{h,i}^k+\langle\mu_{h,i}^k,V\rangle\bigr)_{i=1}^d$. Under \eqref{source_wise_linear}, the corresponding source-$k$ Bellman backup is $\boldsymbol{\phi}(s,a)^\top\boldsymbol{\nu}_h^k(V)$. Given $\widehat V_{h+1}$, we estimate $\boldsymbol{\nu}_h^k(\widehat V_{h+1})$ by ridge regression using the source-$k$ Bellman targets
\[
y_h^{(k,\tau)}
:=
\begin{cases}
r_h^{(k,\tau)}
+\widehat V_{h+1}\bigl(s_{h+1}^{(k,\tau)}\bigr),
& h<H,\\
r_H^{(k,\tau)},
& h=H.
\end{cases}
\]
For a regularization parameter $\lambda>0$, the ridge estimator is
\begin{equation}
\begin{aligned}
\widehat{\boldsymbol{\nu}}_h^k
&=
\arg\min_{\boldsymbol{\nu}\in\mathbb R^d}
\left\{
\sum_{\tau=1}^{N_k}
\left[
y_h^{(k,\tau)}
-
\boldsymbol{\phi}\bigl(
s_h^{(k,\tau)},a_h^{(k,\tau)}
\bigr)^\top\boldsymbol{\nu}
\right]^2
+
\lambda\|\boldsymbol{\nu}\|_2^2
\right\}\\
&=
\bigl(\boldsymbol{\Lambda}_h^k\bigr)^{-1}
\sum_{\tau=1}^{N_k}
\boldsymbol{\phi}\bigl(
s_h^{(k,\tau)},a_h^{(k,\tau)}
\bigr)y_h^{(k,\tau)},
\label{eq:estimate-nu}
\end{aligned}
\end{equation}
where
\begin{equation}
\boldsymbol{\Lambda}_h^k
=
\sum_{\tau=1}^{N_k}
\boldsymbol{\phi}\bigl(
s_h^{(k,\tau)},a_h^{(k,\tau)}
\bigr)
\boldsymbol{\phi}\bigl(
s_h^{(k,\tau)},a_h^{(k,\tau)}
\bigr)^\top
+
\lambda\mathbf I_d.
\label{eq:Lambda}
\end{equation}

The feature-wise structure of the robust operator determines how these estimates are combined. Following the structure of \eqref{eq:Bellman-Operator}, we define $\left(\widehat{\mathbf{w}}_h\right)_i:=\min _{k \in[K]}\left(\widehat{\boldsymbol{\nu}}_h^k\right)_i$ for $i \in[d]$, and construct the empirical robust backup
\begin{equation}
(\widehat{\mathbb B}_h\widehat V_{h+1})(s,a)
:=
\boldsymbol{\phi}(s,a)^\top\widehat{\mathbf w}_h.
\label{eq:estimate-bellman-backup}
\end{equation}
The minimizing source may differ across feature coordinates, so the resulting backup need not coincide with the estimated Bellman backup of any individual source.

Coordinate-wise minimization implements the robust aggregation, but does not by itself control estimation error in the resulting backup. To calibrate the pessimistic correction, we use the following notion of uncertainty quantification, where $\mathbb P_{\mathcal D}$ denotes the probability law of the offline dataset $\mathcal D$ and $\xi\in(0,1)$ is a confidence parameter.

\begin{definition}[$\xi$-uncertainty quantifier \citep{jin2021pessimism}]
\label{def:uncertainty}
A possibly data-dependent collection $\{\Gamma_h\}_{h=1}^H$, with $\Gamma_h:\mathcal S\times\mathcal A\to[0,\infty)$, is a $\xi$-uncertainty quantifier with respect to $\mathbb P_{\mathcal D}$ if the event
\begin{equation}
\mathcal E=\Big\{\ \big|(\widehat{\mathbb B}_h \widehat V_{h+1})(s,a)-(\mathbb B_h \widehat V_{h+1})(s,a)\big|\ \le\ \Gamma_h(s,a),\ \forall (s,a)\in\mathcal S\times\mathcal A,\ \forall h\in[H]\ \Big\}
\label{uncertainty-quantifier-1}
\end{equation}
satisfies $\mathbb P_{\mathcal D}(\mathcal E)\ge 1-\xi$.
\end{definition}

The uncertainty quantifier must account for the effect of estimation errors on the coordinate-wise minimum across sources. Because the empirical and population minima may be attained at different sources, we must control the coefficient estimation errors simultaneously across sources. For each coordinate $i$, the inequality
\[
\left|
\min_{k\in[K]}(\widehat{\boldsymbol{\nu}}_h^k)_i
-
\min_{k\in[K]}
\bigl(\boldsymbol{\nu}_h^k(\widehat V_{h+1})\bigr)_i
\right|\leq
\max_{k\in[K]}
\left|
(\widehat{\boldsymbol{\nu}}_h^k)_i
-
\bigl(\boldsymbol{\nu}_h^k(\widehat V_{h+1})\bigr)_i
\right|
\]
bounds the estimation error of the coordinate-wise minimum by the largest source-specific coefficient estimation error. 

We use the following data-dependent uncertainty quantifier: for a scaling parameter $\beta>0$, define
\begin{equation}
\Gamma_h(s,a):=
\beta\sum_{i=1}^d
\phi_i(s,a)
\max_{k\in[K]}
\|\mathbf{e}_i\|_{(\boldsymbol{\Lambda}_h^k)^{-1}}=
\beta\sum_{i=1}^d
\max_{k\in[K]}
\|\phi_i(s,a)\mathbf{e}_i\|_{(\boldsymbol{\Lambda}_h^k)^{-1}}.
\label{eq:pessimism-penalty}
\end{equation}
The matrices $\boldsymbol{\Lambda}_h^k$ capture both the source-specific sample sizes and the information available along different feature directions. Thus, $\Gamma_h$ captures the largest source-specific estimation uncertainty within each feature coordinate. We show in Section~\ref{sec:theory} that, for an appropriate choice of $\beta$, the resulting collection $\{\Gamma_h\}_{h=1}^H$ is a valid $\xi$-uncertainty quantifier.

GR-PEVI aggregates the source-specific Bellman estimates according to \eqref{eq:estimate-bellman-backup} and applies $\Gamma_h$ as a data-dependent pessimism penalty before selecting the greedy policy. Specifically, starting from $\widehat V_{H+1}\equiv 0$, the algorithm performs the following pessimistic update for $h=H,\ldots,1$:
\begin{align*}
\overline Q_h(s,a)
&=
(\widehat{\mathbb B}_h\widehat V_{h+1})(s,a)-\Gamma_h(s,a),
\qquad
\widehat Q_h(s,a)
=
\min\{\overline Q_h(s,a),H-h+1\}^{+},\\
\widehat{\pi}_h(\cdot\mid s)
&\in
\underset{\nu\in\mathcal P(\mathcal A)}{\arg\max}\,
\mathbb E_{a\sim\nu}
\left[\widehat Q_h(s,a)\right],
\qquad
\widehat V_h(s)
=
\mathbb E_{a\sim\widehat\pi_h(\cdot\mid s)}
\left[\widehat Q_h(s,a)\right].
\end{align*}
At each stage, $\widehat\pi_h$ is chosen to be a measurable deterministic maximizer. The analysis assumes that the greedy maximization is solved exactly. The resulting policy is $\widehat\pi=\{\widehat\pi_h\}_{h=1}^H$, and the complete procedure is summarized in Algorithm~\ref{alg:pess-source-ridge}.

\begin{algorithm}[htbp]
\caption{Group-Robust Pessimistic Value Iteration (GR-PEVI)}
\label{alg:pess-source-ridge}
\begin{algorithmic}[1]
\Require feature map $\boldsymbol{\phi}$, horizon $H$, regularizer $\lambda>0$, pessimism parameter $\beta>0$, and datasets $\mathcal D=\{\mathcal D_k\}_{k=1}^K$.
\State $\widehat V_{H+1}(\cdot)\gets 0$
\For{$h=H,H-1,\ldots,1$}
  \State \textbf{(A) Source-wise ridge estimation and uncertainty calculation}
  \For{$k=1$ \textbf{to} $K$}
     \State $y_h^{(k,\tau)} \gets r_h^{(k,\tau)}+\widehat V_{h+1}(s_{h+1}^{(k,\tau)})$ for $h<H$, and $y_H^{(k,\tau)} \gets r_H^{(k,\tau)}$, for all $\tau\in[N_k]$
     \State $\boldsymbol{\Lambda}_h^k \gets \sum_{\tau=1}^{N_k}\boldsymbol{\phi}\big(s_h^{(k,\tau)},a_h^{(k,\tau)}\big)\,
            \boldsymbol{\phi}\big(s_h^{(k,\tau)},a_h^{(k,\tau)}\big)^{ \top}+\lambda\mathbf I_d$
     \State $\widehat{\boldsymbol{\nu}}_h^k \gets (\boldsymbol{\Lambda}_h^k)^{-1}
             \left(\sum_{\tau=1}^{N_k}\boldsymbol{\phi}\big(s_h^{(k,\tau)},a_h^{(k,\tau)}\big)\,
            y_h^{(k,\tau)}\right)$
    \State $\mathbf u_h^k\gets\sqrt{\operatorname{diag}\big((\boldsymbol{\Lambda}_h^k)^{-1}\big)}$
    \Comment{$[\mathbf u_h^k]_i=\|\mathbf e_i\|_{(\boldsymbol{\Lambda}_h^k)^{-1}}$}
  \EndFor
  \State \textbf{(B) Coordinate-wise min/max across sources}
  \For{$i=1$ \textbf{to} $d$} \State $(\widehat{\mathbf{w}}_h)_i \gets \min_{k\in[K]} (\widehat{\boldsymbol{\nu}}_h^k)_i$ \State $(\mathbf{m}_h)_i
\gets
\max_{k\in[K]}(\mathbf u_h^k)_i$
  \EndFor
  \State \textbf{(C) Compute $Q/\pi/V$ with data-dependent pessimism}
\Statex \hspace{\algorithmicindent} All updates below are defined pointwise.
\State $\overline Q_h(s,a)\gets
\boldsymbol{\phi}(s,a)^\top\widehat{\mathbf w}_h
-\beta\boldsymbol{\phi}(s,a)^\top\mathbf m_h$
\State $\widehat Q_h(s,a)\gets
\min\{\overline Q_h(s,a),H-h+1\}^{+}$
\State $\widehat\pi_h(\cdot\mid s)\in
\underset{\nu\in\mathcal P(\mathcal A)}{\arg\max}\,
\mathbb E_{a\sim\nu}[\widehat Q_h(s,a)]$
\hfill {\small$\triangleright$ greedy}
\State $\widehat V_h(s)\gets
\mathbb E_{a\sim\widehat\pi_h(\cdot\mid s)}
[\widehat Q_h(s,a)]$
\EndFor
\State \textbf{Output:} $\widehat\pi=\{\widehat\pi_h\}_{h=1}^H$
\end{algorithmic}
\end{algorithm}

\section{Theory}
\label{sec:theory}
This section establishes finite-sample guarantees for GR-PEVI. We first derive a robust suboptimality decomposition in terms of model evaluation error. We then show that the feature-wise pessimism penalty controls the model evaluation errors with high probability and obtain an explicit rate under a robust partial coverage condition. 

\subsection{Decomposition of Suboptimality for Robust MDP}
\label{sec:decomposition}
We derive a three-term upper bound for the robust suboptimality in \eqref{eq:subopt-def}. Following the terminology of \citet{jin2021pessimism}, we refer to these terms as \emph{spurious correlation}, \emph{intrinsic uncertainty}, and \emph{optimization error}.

Algorithm~\ref{alg:pess-source-ridge} produces an estimated $Q$-function $\widehat Q_h:\mathcal S\times\mathcal A\to\mathbb R$ and value function $\widehat V_h:\mathcal S\to\mathbb R$ from $\mathcal D$. We define the pointwise model evaluation error at stage $h\in[H]$ by
\begin{equation}
\label{eq:iota-def}
\iota_h(s,a)\ :=\ \big(\mathbb{B}_h \widehat{V}_{h+1}\big)(s,a)\ -\ \widehat{Q}_h(s,a)\,,
\end{equation}
which measures the difference between the robust Bellman backup of $\widehat V_{h+1}$ and the estimated $Q$-value at $(s,a)$. Since both $\widehat V_{h+1}$ and $\widehat Q_h$ depend on $\mathcal D$, $\iota_h$ is data-dependent. 

\begin{lemma}[Decomposition of Suboptimality for Robust MDP]
\label{lemma:decomposition}
Let $\pi^\star\in\Pi$ be a robust optimal policy, and let $\widehat{\pi}$ be the policy returned by Algorithm~\ref{alg:pess-source-ridge}. For any initial state $s\in\mathcal S$, the suboptimality satisfies
\begin{align*} 
        \operatorname{SubOpt}(\widehat{\pi} ; s) 
        &= V_1^{\pi^\star}(s)-V_1^{\widehat{\pi}}(s) \\ 
        &\leq \underbrace{-\sum_{h=1}^H \mathbb{E}^{\widehat{\pi}, P^{\widehat{\pi}, \natural}}\left[\iota_h\left(s_h, a_h\right) \mid s_1=s\right]}_{\text{(i): Spurious Correlation}}\\
         &\quad + \underbrace{\sum_{h=1}^H \mathbb{E}^{\pi^\star, P^{\pi^\star, \dagger}}\left[\iota_h\left(s_h, a_h\right) \mid s_1=s\right]}_{\text{(ii): Intrinsic Uncertainty}}\\
        &\quad +\underbrace{\sum_{h=1}^H \mathbb{E}^{\pi^\star, P^{\pi^\star, \dagger}}\left[\mathbb E_{a\sim\pi_h^\star(\cdot\mid s_h)}
        \left[\widehat Q_h(s_h,a)\right]-\mathbb E_{a\sim\widehat\pi_h(\cdot\mid s_h)}\left[\widehat Q_h(s_h,a)\right] \mid s_1=s\right]}_{\text{(iii): Optimization Error}}.
\end{align*}
Here $\mathbb E^{\pi,P^{\pi,\diamond}}[\cdot\mid s_1=s]$ denotes expectation under policy $\pi$ and a possibly data-dependent admissible transition-kernel sequence $P^{\pi,\diamond}$, where $\diamond\in\{\natural,\dagger\}$ distinguishes the two sequences appearing in the decomposition. Their explicit constructions are given in Supplementary Section~\ref{app:proofs-decomposition}.
\end{lemma}

One challenge in the analysis of this robust formulation is that the stage-wise differences involve two worst-case backups whose minimizing transition models need not coincide. We therefore control each such difference under a common admissible transition model and apply this argument stage by stage, yielding the sequences $P^{\widehat\pi,\natural}$ and $P^{\pi^\star,\dagger}$. These sequences determine the transition dynamics in the expectations appearing in terms~(i)--(iii). Specifically, term~(i) involves spurious correlation because both the learned policy $\widehat\pi$ and the model evaluation error $\iota_h$ are data-dependent. Term~(ii) uses the robust optimal policy $\pi^\star$ and thus avoids dependence through the learned policy. Term~(iii) is nonpositive because $\widehat\pi_h$ is greedy with respect to $\widehat Q_h$.

\subsection{Suboptimality Upper Bound}
\label{sec:upper_bound}
We now apply the decomposition lemma (Lemma~\ref{lemma:decomposition}) to GR-PEVI. With high probability, the pessimism penalty ensures that the model evaluation error is nonnegative and uniformly bounded, so that term~(i) is nonpositive and term~(ii) is controlled.

\begin{theorem}[Suboptimality]
\label{thm:suboptimality}
Let $\widehat\pi$ be the policy returned by Algorithm~\ref{alg:pess-source-ridge} with
\[
\lambda=1,\qquad
\beta=c\,dH\sqrt{\zeta},
\qquad
\zeta=\log(2dKHN_{\max}/\xi),
\]
where $c>0$ is an absolute constant, $\xi\in(0,1)$, and $N_{\max}=\max_{k\in[K]}N_k$. Then, with probability at least $1-\xi$, for any $s\in\mathcal S$,
\[
\operatorname{SubOpt}(\widehat\pi;s)
\leq
2\beta
\sum_{h=1}^H\sum_{i=1}^d
\mathbb E^{\pi^\star,P^{\pi^\star,\dagger}}
\left[
\max_{k\in[K]}
\left\|
\phi_i(s_h,a_h)\mathbf e_i
\right\|_{(\boldsymbol{\Lambda}_h^k)^{-1}}
\mid s_1=s
\right],
\]
where $\mathbb E^{\pi^\star,P^{\pi^\star,\dagger}}$ is defined in Lemma~\ref{lemma:decomposition}.
\end{theorem}
Theorem~\ref{thm:suboptimality} shows that the robust suboptimality of GR-PEVI is controlled by the accumulated feature-wise estimation uncertainty. This form differs from the standard (non-robust) linear-MDP bound, where the Bellman estimation error can typically be reduced to a least-squares estimation error and controlled by an elliptical uncertainty term; see, e.g., \citet{jin2021pessimism}. In our multi-source robust setting, each coordinate of the robust Bellman coefficient is obtained by minimizing over the candidate sources, and the source attaining the minimum may vary across feature coordinates and between the population and empirical quantities. Consequently, the overall Bellman estimation error cannot be associated with a single source-specific design matrix and must instead be controlled coordinate-wise. This leads to the feature-wise uncertainty term in Theorem~\ref{thm:suboptimality}, where each feature coordinate is governed by the largest source-specific estimation uncertainty.

To translate this uncertainty bound into an explicit convergence rate in terms of the sample sizes, we impose the following robust partial coverage condition for each feature coordinate.

\begin{assumption}[Robust partial coverage]
\label{partial-coverage}
Fix an initial state $s\in\mathcal S$. There exists a constant $c^\dagger>0$ such that, for every $k\in[K]$, $i\in[d]$, $h\in[H]$, and $P=(P_1,\ldots,P_H)\in\prod_{t=1}^H\mathcal P_t$,
\begin{equation}
\frac{1}{N_k}\boldsymbol{\Lambda}_h^k
\succeq
\frac{\lambda}{N_k}\mathbf I_d
+
c^\dagger
\mathbb E_{(s_h,a_h)\sim d_{P,h}^{\pi^\star,s}}
\left[
\bigl(\phi_i(s_h,a_h)\mathbf e_i\bigr)
\bigl(\phi_i(s_h,a_h)\mathbf e_i\bigr)^\top
\right],
\end{equation}
where $\boldsymbol{\Lambda}_h^k$ is defined in \eqref{eq:Lambda}, $\mathcal P_t:=\{P_t:\exists\,r_t\text{ such that }(P_t,r_t)\in\mathcal U_t\}$, and $d_{P,h}^{\pi^\star,s}$ is the stage-$h$ state--action visitation distribution under $\pi^\star$ and $P$, conditional on the initial state $s_1=s$.
\end{assumption}

Assumption~\ref{partial-coverage} extends the robust partial coverage condition of \citet[Assumption~6.2]{blanchet2023double} to the multi-source setting. For the fixed initial state $s$, each source-specific empirical covariance must provide sufficient coverage along every feature coordinate. The coverage requirement is partial because it is imposed only on the state--action distributions induced by the robust optimal policy $\pi^\star$, and robust because it must hold uniformly over all admissible transition-kernel sequences. The following corollary combines Theorem~\ref{thm:suboptimality}
with this coverage condition for the realized data to obtain an explicit sample-size rate. 

\begin{corollary}[Suboptimality with partial coverage]
\label{corollary-suboptimality}
Fix an initial state $s\in\mathcal S$. Let $\widehat\pi$ be the policy returned by Algorithm~\ref{alg:pess-source-ridge} with $\lambda$ and $\beta$ as in Theorem~\ref{thm:suboptimality}, and let $N_{\min}:=\min_{k\in[K]}N_k$. Then, with probability at least $1-\xi$, the following implication holds: if the realized offline dataset satisfies the coverage condition in Assumption~\ref{partial-coverage} for the fixed initial state $s$, then
\[
\operatorname{SubOpt}(\widehat\pi;s)
\leq
c^{\ddagger}d^2H^2
\sqrt{
\frac{\log(2dKHN_{\max}/\xi)}
{N_{\min}}
},
\]
where $c^{\ddagger}>0$ depends only on $c^\dagger$ and the absolute constant $c$.
\end{corollary}

The dependence on $N_{\min}$ arises because the estimation error must be controlled uniformly across sources, so the rate is governed by the smallest source-specific sample size.

\section{Cluster-Level Extension}
\label{sec:cross-source}
In practice, some sources may have small sample sizes but be highly similar to other sources. In such cases, the bound in Corollary~\ref{corollary-suboptimality} can be driven by the smallest source-specific sample size, motivating the pooling of sufficiently similar sources. Such pooling can increase the effective sample size at the cost of approximation error due to within-cluster heterogeneity. We therefore consider a cluster-level extension in which the sources are partitioned into $M$ prespecified clusters $\{\mathcal C_m\}_{m=1}^M$ and each cluster is treated as a ``super-source'' for estimation. The robust objective remains defined over the original $K$ sources, and we do not assume that the sources within a cluster share a common underlying MDP.

To quantify within-cluster heterogeneity, we consider the discrepancy among sources within each cluster. Let $\mathcal B_{h+1}:= \left\{ V:\mathcal S\to[0,H-h] \colon V \text{ is measurable} \right\}$, and define
\[
\Delta_{h,m}
:=
\sup_{V\in\mathcal B_{h+1}}
\max_{k,k'\in\mathcal C_m}
\left\|
\boldsymbol{\nu}_h^k(V)
-
\boldsymbol{\nu}_h^{k'}(V)
\right\|_2,
\qquad
\Delta_h:=\max_{m\in[M]}\Delta_{h,m}.
\]
The quantity $\Delta_{h,m}$ measures the worst-case pairwise discrepancy among sources within cluster $\mathcal C_m$. Since the clipped value estimates from the cluster-level procedure belong to $\mathcal B_{h+1}$, $\Delta_{h,m}$ uniformly controls the within-cluster discrepancies encountered in the analysis.

For each cluster $m\in[M]$, define the pooled dataset $\mathcal D_{\mathcal C_m}:=\bigcup_{k\in\mathcal C_m}\mathcal D_k$ and $N_m:=\sum_{k\in\mathcal C_m}N_k$. The cluster-level procedure applies GR-PEVI to $\{\mathcal D_{\mathcal C_m}\}_{m=1}^M$, with the source index replaced by the cluster index. Define
\[
\boldsymbol{\Lambda}_h^{\mathcal C_m}
:=
\sum_{k\in\mathcal C_m}
\sum_{\tau=1}^{N_k}
\boldsymbol{\phi}(s_h^{(k,\tau)},a_h^{(k,\tau)})
\boldsymbol{\phi}(s_h^{(k,\tau)},a_h^{(k,\tau)})^\top
+\lambda \mathbf I_d.
\]
The complete cluster-level procedure is presented in Supplementary Section~\ref{app:cluster-algorithm}.

\begin{assumption}[Cluster-level robust partial coverage] \label{partial-coverage-cluster} 
Fix an initial state $s\in\mathcal S$. There exists a constant $c'>0$ such that, for every $m\in[M]$, $i\in[d]$, $h\in[H]$, and $P=(P_1,\ldots,P_H)\in\prod_{t=1}^H\mathcal P_t$, the empirical covariance matrix of cluster $m$ satisfies
\[
\frac{1}{N_m}\boldsymbol{\Lambda}_h^{\mathcal C_m}
\succeq
\frac{\lambda}{N_m}\mathbf I_d
+
c'
\mathbb E_{(s_h,a_h)\sim d_{P,h}^{\pi^\star,s}}
\left[
\bigl(\phi_i(s_h,a_h)\mathbf e_i\bigr)
\bigl(\phi_i(s_h,a_h)\mathbf e_i\bigr)^\top
\right],
\]
with $\mathcal P_t$ and $d_{P,h}^{\pi^\star,s}$ as in Assumption~\ref{partial-coverage}.
\end{assumption}
\begin{remark}
For the same fixed initial state $s$, Assumption~\ref{partial-coverage-cluster} is implied by Assumption~\ref{partial-coverage}. Indeed, for any fixed $i\in[d]$, $h\in[H]$, and admissible transition-kernel sequence $P$, Assumption~\ref{partial-coverage} gives, for every $k\in\mathcal C_m$,
\[
\sum_{\tau=1}^{N_k}
\boldsymbol{\phi}\left(s_h^{(k,\tau)},a_h^{(k,\tau)}\right)
\boldsymbol{\phi}\left(s_h^{(k,\tau)},a_h^{(k,\tau)}\right)^\top
\succeq
N_kc^\dagger
\mathbb E_{(s_h,a_h)\sim d_{P,h}^{\pi^\star,s}}
\left[
\bigl(\phi_i(s_h,a_h)\mathbf e_i\bigr)
\bigl(\phi_i(s_h,a_h)\mathbf e_i\bigr)^\top
\right].
\]
Summing over $k\in\mathcal C_m$ yields $\boldsymbol{\Lambda}_h^{\mathcal C_m}
\succeq
\lambda\mathbf I_d
+
N_mc^\dagger
\mathbb E_{(s_h,a_h)\sim d_{P,h}^{\pi^\star,s}}
\left[
\bigl(\phi_i(s_h,a_h)\mathbf e_i\bigr)
\bigl(\phi_i(s_h,a_h)\mathbf e_i\bigr)^\top
\right]$. Hence Assumption~\ref{partial-coverage-cluster} holds with $c'=c^\dagger$.
\end{remark}

\begin{corollary}[Cluster-level suboptimality]
\label{cor:cluster-level-subopt}
Fix an initial state $s\in\mathcal S$. Let $\widehat\pi^{\mathrm{pool}}$ be the policy returned by GR-PEVI applied to $\mathcal D^{\mathrm{pool}}
=
\{\mathcal D_{\mathcal C_m}\}_{m=1}^M$, with 
\[
\lambda=1,\qquad
\beta^{\mathrm{pool}}
=
c\,dH\sqrt{\zeta^{\mathrm{pool}}},
\qquad
\zeta^{\mathrm{pool}}
=
\log\left(2dMHN_{\max}^{\mathrm{pool}}/\xi\right),
\]
where $c>0$ is an absolute constant, $\xi\in(0,1)$ is the confidence
parameter, $N_{\max}^{\mathrm{pool}}:=\max_{m\in[M]}N_m$, and $N_{\min}^{\mathrm{pool}}:=\min_{m\in[M]}N_m$. Then, with probability at least $1-\xi$, the following implication holds: if the realized offline dataset satisfies Assumption~\ref{partial-coverage-cluster} for the fixed initial state $s$, then
\[
\operatorname{SubOpt}(\widehat\pi^{\mathrm{pool}};s)
\leq
c^\natural d^2H^2
\sqrt{
\frac{
\log\left(2dMHN_{\max}^{\mathrm{pool}}/\xi\right)
}{
N_{\min}^{\mathrm{pool}}
}
}
+
C^\natural\sum_{h=1}^H\Delta_h,
\]
where $c^\natural>0$ depends only on $c$ and $c'$, and
\[
C^\natural
=
2\left(
1+
\sup_{h\in[H]}
\sup_{(\tilde s,a)\in\mathcal S\times\mathcal A}
\sum_{i=1}^d
\phi_i(\tilde s,a)
\sqrt{
\frac{1}{
1/N_{\max}^{\mathrm{pool}}+c'\sigma_{h,i}^2
}
}
\right),
\]
with $\sigma_{h,i}^2
:=
\sup_{P\in\prod_{t=1}^H\mathcal P_t}
\mathbb E_{(s_h,a_h)\sim d_{P,h}^{\pi^\star,s}}
\left[
\phi_i(s_h,a_h)^2
\right]$, where $\mathcal P_t$ is as in Assumption~\ref{partial-coverage-cluster}. 
\end{corollary}

\begin{remark}
The coefficient $C^\natural$ in Corollary~\ref{cor:cluster-level-subopt} quantifies how within-cluster heterogeneity is amplified through the feature representation and coverage. Feature coordinates with larger $\sigma_{h,i}^2$ incur less discrepancy amplification, whereas coordinates that are weakly represented under every admissible state--action visitation distribution induced by $\pi^\star$ can lead to a larger coefficient. Since $C^\natural$ may depend on $N_{\max}^{\mathrm{pool}}$, the heterogeneity term is not generally a sample-size-independent approximation floor.
\end{remark}


Corollary~\ref{cor:cluster-level-subopt} replaces the source-wise sample-size factor $N_{\min}$ in Corollary~\ref{corollary-suboptimality} by the pooled quantity $N_{\min}^{\mathrm{pool}}\geq N_{\min}$, while introducing the approximation term $C^\natural\sum_{h=1}^H\Delta_h$ due to within-cluster heterogeneity. Thus, up to constants and logarithmic factors, the result characterizes a finite-sample trade-off: pooling can improve the bound when the reduction in statistical error from the larger effective sample size dominates this additional heterogeneity term.

\section{Simulation}
\label{sec:simulation}
We conduct two main simulation studies. Section~\ref{sec:benchmark-comparison} compares GR-PEVI with alternative multi-source learning strategies under source heterogeneity. Section~\ref{subsec:worst-case-scaling} examines how the robust suboptimality of GR-PEVI scales with the minimum source sample size and how this scaling depends on the pessimism penalty. An additional study in Supplementary Section~\ref{app:cluster-simulation} examines the cluster-level extension and illustrates the trade-off between the benefit of pooling similar sources and the cost of within-cluster discrepancy.

\subsection{Robust Performance under Source Heterogeneity}
\label{sec:benchmark-comparison}
We compare GR-PEVI with four baselines to examine how different strategies for combining information across heterogeneous sources affect robust policy performance. Pooled PEVI aggregates all source data and applies standard pessimistic value iteration (PEVI)~\citep{jin2021pessimism}. Avg-PEVI, motivated by \citet{shi2023provably}, runs PEVI separately within each source and uniformly selects among the source-specific policy recommendations at each state--stage pair. Majority-PEVI uses the same source-specific PEVI fits but selects the most frequently recommended action. Operator-Avg PEVI, motivated by the averaged Bellman-operator aggregation in \citet{zhang2025pessimism}, computes source-specific pessimistic Bellman backups using a shared continuation value and averages them before taking the greedy action. Complete implementation details are provided in Supplementary Section~\ref{app:baseline-details}.

We generate episodic linear MDPs with $K=3$ sources, $\mathcal S=[0,1]^3$, $|\mathcal A|=10$, and $H=7$. The feature map $\boldsymbol\phi(s,a)\in\mathbb R^{30}$ is sparse and action-specific: the nonnegative state vector is normalized to unit $\ell_1$ norm and placed in the block corresponding to action $a$, with all other blocks set to zero. Rewards and transitions follow the linear MDP model in Section~\ref{sec:linear-mdp}, with source- and stage-specific reward and transition components inducing cross-source heterogeneity. Offline trajectories are collected under a uniform behavior policy with initial states sampled uniformly from $\mathcal S$. Additional details are provided in Supplementary Section~\ref{sec:supp-main-simulation}.

All methods use $\lambda=1$, $\xi=0.05$, and $c=5\times10^{-4}$, with sample sizes $(N_1,N_2,N_3)=(3000,2000,5000)$. Performance is measured by robust suboptimality in \eqref{eq:subopt-def}. We compute $V_1^{\pi^{\star}}$ and $V_1^{\widehat{\pi}}$ by backward recursion under the true robust Bellman operators, approximating transition expectations with $50,000$ Monte Carlo draws. We run $50$ replications under fixed heterogeneous source MDPs, regenerating the offline dataset in each replication, and estimate expected suboptimality over $200$ initial states sampled uniformly from $\mathcal S$.

\begin{figure}[tb]
  \centering
  \includegraphics[width=0.9\linewidth]{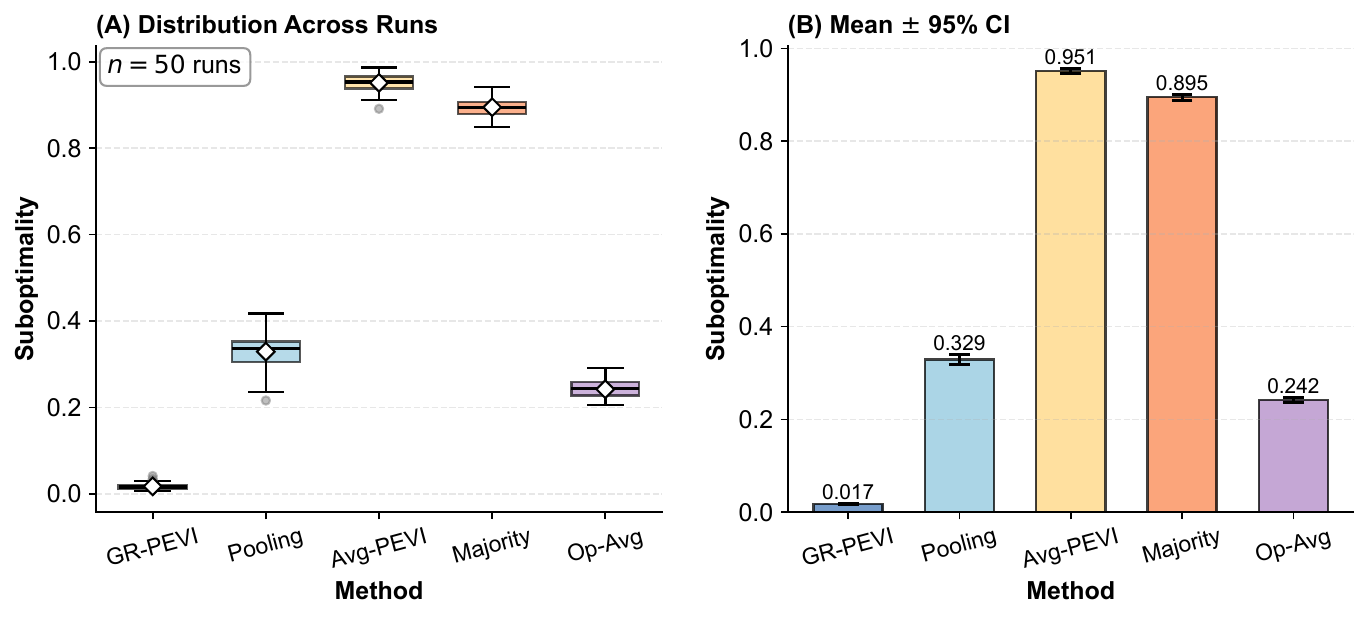}
  \caption{Comparison of robust suboptimality across methods:
(A) Distribution of suboptimality over 50 replications.
(B) Mean suboptimality with 95\% confidence intervals.}
\label{fig:comparison}
\end{figure}

Figure~\ref{fig:comparison} shows that GR-PEVI achieves the lowest mean robust suboptimality, with Operator-Avg PEVI as the strongest baseline.
Pooled PEVI does not explicitly account for source heterogeneity. Avg-PEVI and Majority-PEVI combine source-specific recommendations without guarding against cross-source variation. Operator-Avg PEVI retains source-specific Bellman information but averages the backups rather than aggregating them robustly. The improvement of GR-PEVI is consistent with its objective of optimizing policy performance under worst-case environments in the uncertainty set. Under the milder transition heterogeneity in Supplementary Section~\ref{sec:supp-mild-heterogeneity}, the ordering is unchanged, but the gaps narrow.

\subsection{Worst-Case Scaling under Imbalanced Coverage}
\label{subsec:worst-case-scaling}
We next examine how the robust suboptimality of GR-PEVI scales with $N_{\min}$ and how this scaling depends on the penalty constant $c$. Corollary~\ref{corollary-suboptimality} gives, up to problem-dependent constants and logarithmic factors, an $N_{\min}^{-1/2}$ upper bound on the robust suboptimality under Assumption~\ref{partial-coverage}. We examine this scaling empirically on a family of multi-source linear MDPs indexed by a gap parameter $\delta$, with a common coverage constant $c^\dagger$. At each sample size, we report the largest average robust suboptimality over a fixed grid of $\delta$ values.

We consider $K=4$ sources sharing the state space $\mathcal S=\{s_0,s_1,s_2\}$, the action space $\mathcal A=\{b_1,b_2,b_3\}$, and the horizon $H=7$, where $s_1$ and $s_2$ are absorbing states. Each trajectory starts from $s_0$ at stage $h=1$, and the transition probabilities from $s_0$ in source $k\in[K]$ are
\[
P_1^k(s_1\mid s_0,b_j)=p_j^k,
\qquad
P_1^k(s_2\mid s_0,b_j)=1-p_j^k,
\qquad j=1,2,3.
\]
Rewards depend only on the current state, with $r(s_0)=r(s_2)=0$ and $r(s_1)=1$. The feature map $\boldsymbol{\phi}(s,a)\in\mathbb R^5$ is defined by
\[
\boldsymbol{\phi}(s_0,b_j)=\mathbf{e}_j,\qquad
\boldsymbol{\phi}(s_1,a)=\mathbf{e}_4,\qquad
\boldsymbol{\phi}(s_2,a)=\mathbf{e}_5,
\]
for $j\in[3]$ and $a\in\mathcal A$. Cross-source heterogeneity is introduced through the transition kernels at $h=1$. Let $(\eta_1,\eta_2,\eta_3,\eta_4)=(0,0.03,0.06,0.09)$. We set, for $k\in[K]$,
\[
p_1^k=\tfrac12+\delta+\eta_k,
\qquad
p_2^k=\tfrac12-\delta+\eta_{5-k},
\qquad
p_3^k=
\begin{cases}
0.45, & k=3,\\
0.85, & k\neq 3,
\end{cases}
\]
where $\delta\in(0,0.25]$ is a gap parameter indexing the problem instance. Since $s_1$ and $s_2$ are absorbing with rewards $1$ and $0$, respectively, it follows from \eqref{eq:q_in_our_setting} that the robust $Q$-value of $b_j$ at $s_0$ is $(H-1)\min_{k\in[K]}p_j^k$. Because $\min_k p_1^k=0.5+\delta$, $\min_k p_2^k=0.5-\delta$, and $\min_k p_3^k=0.45$, action $b_1$ is robustly optimal for every $\delta>0$. For a deterministic learned policy $\widehat\pi$, the robust suboptimality is therefore
\[
\operatorname{SubOpt}(\widehat\pi;s_0)
=
\begin{cases}
0, & \widehat\pi_1(s_0)=b_1,\\
2(H-1)\delta, & \widehat\pi_1(s_0)=b_2,\\
(H-1)(\delta+0.05), & \widehat\pi_1(s_0)=b_3.
\end{cases}
\]

The first-stage action counts are specified in advance and randomly permuted across trajectories. In source $k$, actions $b_1$ and $b_2$ each receive $\omega_k n$ observations, with $(\omega_1,\omega_2,\omega_3,\omega_4)=(1,1.5,2,3)$ and $n\in\{50,150,500,1500,5000,15000\}$, whereas action $b_3$ receives $10$ observations in every source. Thus $N_k=2\omega_k n+10$, and $N_{\min}=2n+10$ ranges from $110$ to $30{,}010$. Actions at later stages are drawn uniformly at random and do not affect the features. 

This design covers the state--action pairs visited by the robust optimal policy while leaving $b_3$ sparsely sampled. At stage $h=1$, the robust optimal policy selects $b_1$, which accounts for at least $5/11$ of the first-stage observations in every source by construction. At later stages, both absorbing states have visitation probabilities bounded away from zero uniformly over $(n,\delta)$. The fixed count of $b_3$ does not conflict with Assumption~\ref{partial-coverage}, since the robust optimal policy never selects $b_3$ at $s_0$. Thus, with high probability, Assumption~\ref{partial-coverage} holds with a common coverage constant $c^\dagger$ independent of $n$ and $\delta$.

We take $\delta$ on a grid $\mathcal G$ of $36$ geometrically spaced values in $[5\times10^{-4},0.25]$, which is fixed across all sample sizes. For each $n$, we use $R=500$ independent replications, with common random numbers across the values of $\delta$ within each replication. We run GR-PEVI with $\lambda=1$, $\xi=0.05$, and
$c\in\{0,0.01,0.02,0.03\}$, where $c=0$ removes the pessimism
penalty. All values of $c$ are applied to the same datasets. At each sample size, we report
\[
\widehat{\mathcal E}(c,N_{\min})
:=\max_{\delta\in\mathcal G}\frac1R\sum_{r=1}^R
\operatorname{SubOpt}\!\left(\widehat\pi_{c,n,\delta}^{(r)};s_0\right),
\]
where $\widehat\pi_{c,n,\delta}^{(r)}$ denotes the policy learned in the $r$th replication under $(c,n,\delta)$. We construct $95\%$ percentile confidence intervals for $\widehat{\mathcal E}(c,N_{\min})$ and its log--log slope from $2{,}000$ bootstrap resamples of the replication indices, using the same resampled indices across all $c$, $n$, and $\delta$. In each bootstrap resample, the maximizing $\delta$ is reselected, and the log--log slope of $\widehat{\mathcal E}(c,N_{\min})$ against $N_{\min}$ is refitted by least squares over all six sample sizes.

\begin{figure}[tb]
  \centering
  \includegraphics[width=0.9\linewidth]{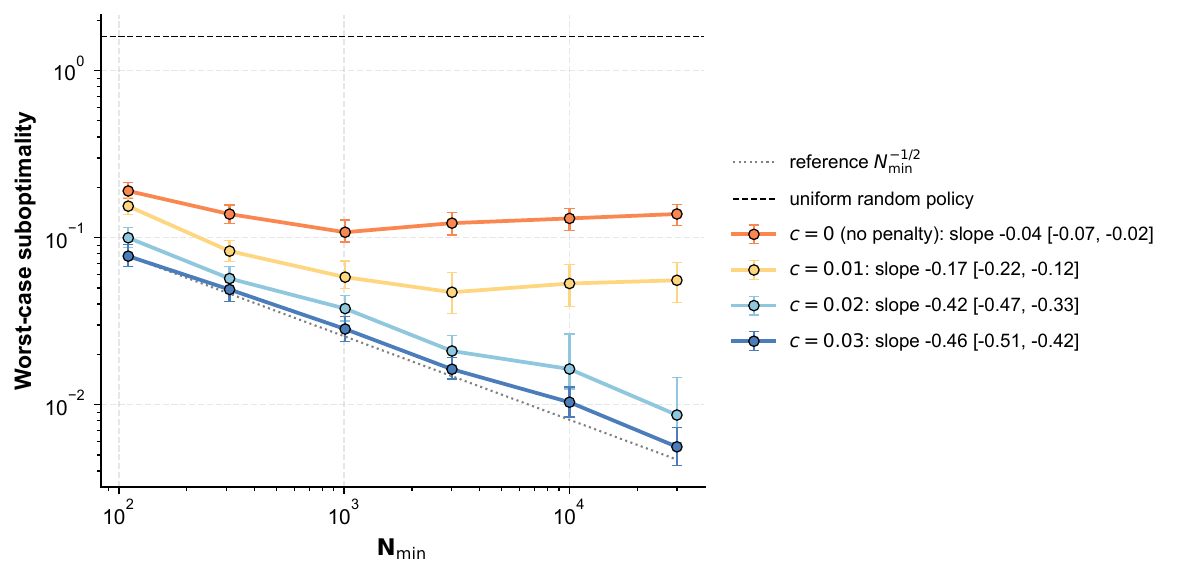}
   \caption{Worst-case robust suboptimality $\widehat{\mathcal E}(c,N_{\min})$ versus $N_{\min}$ for different values of $c$, based on 500 replications. Error bars and slope intervals are 95\% bootstrap confidence intervals. The dotted line shows a reference $N_{\min}^{-1/2}$ rate, and the dashed line shows the worst-case robust suboptimality of the uniform random policy.}
  \label{fig:worst-case-scaling}
\end{figure}

Figure~\ref{fig:worst-case-scaling} plots $\widehat{\mathcal E}(c,N_{\min})$ against $N_{\min}$ on a log--log scale for each value of $c$. Without the pessimism penalty ($c=0$), $\widehat{\mathcal E}(c,N_{\min})$ decreases only at small sample sizes and then remains at a persistent plateau. This plateau arises because $b_3$ receives only $10$ observations per source, so its estimate remains noisy as $N_{\min}$ increases. Since $b_3$ has a high transition probability to $s_1$ in the other three sources, overestimation in source~3 can make it appear optimal even after taking the coordinate-wise minimum over sources. As $c$ increases, the decrease of $\widehat{\mathcal E}(c,N_{\min})$ becomes more pronounced: with $c=0.01$, it levels off at a lower value, whereas with $c=0.02$ and $c=0.03$, the observed decrease is close to the $N_{\min}^{-1/2}$ reference rate over the sample sizes considered.

\section{Application to MS Treatment}
\label{sec:real-data}
\subsection{Data Sources and Trajectory Construction}
We considered sequential disease-modifying therapy (DMT) decisions for patients with multiple sclerosis (MS), a chronic neurological disease associated with progressive disability~\citep{stephen2020treatment}. DMTs differ in efficacy and safety~\citep{mcginley2021diagnosis}, and patient populations and treatment patterns may vary across healthcare systems and subgroups. Combining data from these heterogeneous sources supports learning a single treatment policy, but strong average performance may mask poor performance in a particular source. We therefore applied GR-PEVI to learn a common sequential DMT policy under a group-robust objective that accounts for cross-source heterogeneity. 

We analyzed longitudinal EHR data for patients with MS from Mass General Brigham (MGB) and the University of Pittsburgh Medical Center (UPMC). The data included structured EHR features, diagnosis and medication information, clinical concepts extracted from unstructured notes using natural language processing (NLP), demographic variables, and Patient-Determined Disease Steps (PDDS) scores. PDDS ranges from 0 to 8, with higher values indicating greater disability~\citep{Lo2005treatment}. To define multiple analysis sources while maintaining adequate source-specific sample sizes, we divided the MGB cohort into MGB-young and MGB-old at its median age of 42.04 years, computed among all MGB patients before applying the trajectory eligibility criteria described below, while retaining UPMC as a separate source.

We identified DMT histories from RxNorm-coded prescription records. DMTs were classified as standard- or high-efficacy according to Supplementary Table~\ref{tab:ms_dmt}. Trajectories began at the first month with a mapped DMT. Patients without such a record or with fewer than 24 consecutive monthly records thereafter were excluded. For each eligible patient, the 24-month follow-up was divided into $H=4$ non-overlapping six-month decision stages. Each stage was summarized by the final monthly record of its interval, from which the state, DMT action, and PDDS-based reward were constructed. Table~\ref{tab:cohort-characteristics} summarizes the resulting analytic cohort. Within each source, patients were randomly split into training and test sets in an approximately $80:20$ ratio.

\begin{table}[t]
\centering
\caption{Patient characteristics by source. Continuous variables are reported as mean (SD), and categorical variables as $n$ (\%).}
\label{tab:cohort-characteristics}
\footnotesize
\begin{tabular}{lrrrr}
\toprule
Characteristic & Overall & MGB-young & MGB-old & UPMC \\
\midrule
Patients, $n$
& 5,700 & 2,370 & 1,925 & 1,405 \\

Female sex, $n$ (\%)
& 4,244 (74.5\%) & 1,750 (73.8\%) & 1,398 (72.6\%) & 1,096 (78.0\%) \\

Age, mean (SD)
& 40.9 (11.4) & 32.3 (6.5) & 51.3 (6.7) & 41.2 (11.0) \\

White race, $n$ (\%)
& 4,943 (86.9\%) & 1,941 (81.9\%) & 1,748 (90.8\%) & 1,254 (90.0\%) \\

Standard-efficacy DMT, $n$ (\%)
& 4,357 (76.4\%) & 1,851 (78.1\%) & 1,459 (75.8\%) & 1,047 (74.5\%) \\

High-efficacy DMT, $n$ (\%)
& 1,343 (23.6\%) & 519 (21.9\%) & 466 (24.2\%) & 358 (25.5\%) \\
\bottomrule
\end{tabular}

\vspace{2pt}
\scriptsize
\textit{Note:} DMT categories are defined at trajectory initiation. Race percentages are calculated among patients with observed race.
\end{table}

\subsection{Construction of RL Elements}
For the state representation, we retained variables shared by MGB and UPMC, comprising 262 structured EHR features and 672 NLP-derived features. After reserving 41 RxNorm medication indicators for action encoding, the remaining 893 clinical variables were projected onto 24 principal components. Within each train--test split, PCA was fitted using training data only, with at most $50,000$ patient-month records per source to balance source contributions, and then applied to both training and test data. The principal components were augmented with age, sex, and recorded race. Age was standardized using the training data, sex was encoded as a male indicator, and race was encoded as 1 for White, 0 for non-White, and 0.5 if missing, yielding a 27-dimensional state vector.

The action space was $\mathcal{A}=\{0,1,2\}$, corresponding to no DMT, standard-efficacy DMT, and high-efficacy DMT, respectively. Monthly actions were assigned from RxNorm-coded medication records, with high-efficacy DMT taking precedence when both categories were present. If a month contained recorded medications but none mapped to either DMT category, action 0 was assigned. If no coded medication was recorded in a month, the most recently assigned action was carried forward. The stage-wise reward was defined as $r_h=1-\mathrm{PDDS}_h/8$, where $\mathrm{PDDS}_h$ is PDDS in the final month of each six-month interval. Thus, $r_h\in[0,1]$ with larger values indicating lower disability burden. 

To obtain a nonnegative representation while allowing simple nonlinear effects, we constructed an action-specific simplex feature map. Let $\boldsymbol{x}\in\mathbb R^{25}$ contain the 24 principal components and standardized age, and define 
$\boldsymbol{g}(s)
=
\bigl(
1,\boldsymbol{x}^+,\boldsymbol{x}^-,
\boldsymbol{x}^{\circ 2},
\mathrm{sex},\mathrm{race}
\bigr)$, where $x_j^+=\max(x_j,0)$, $x_j^-=\max(-x_j,0)$, and $\boldsymbol{x}^{\circ 2}$ denotes the elementwise square. We normalized $\widetilde{\boldsymbol g}(s)=\boldsymbol g(s)/\|\boldsymbol g(s)\|_1$ and defined $\boldsymbol{\phi}(s,a)$ by placing $\widetilde{\boldsymbol g}(s)$ in the block for action $a$ and zeros elsewhere. The feature map is nonnegative, has unit $\ell_1$ norm, and dimension $d=234$.

\subsection{Policy Learning and Off-Policy Evaluation}
We compared GR-PEVI with the four baselines from Section~\ref{sec:benchmark-comparison}. All methods used the same patient-level splits and RL setup, with $\lambda=1$, $\xi=0.05$, and $c=10^{-5}$. Each learned policy was evaluated on the held-out trajectories using a stabilized, self-normalized variant of the augmented inverse-probability-weighted estimator, denoted WAIPW~\citep{jiang2016doubly,thomas2016data}. The outcome regression was estimated source-wise by linear fitted $Q$-evaluation (FQE) using backward ridge regressions on the training data. The residual correction used cumulative importance weights based on the estimated behavior propensities. Estimated propensities were floored at $b_{\min}=0.1$, and cumulative importance weights were truncated at $\rho_{\max}=100$. These stabilization steps can introduce bias, so WAIPW values are interpreted as stabilized empirical off-policy evaluation criteria. Further details are provided in Supplementary Section~\ref{appendix:ope}.

As an empirical measure of robustness across the observed sources, we used the minimum held-out WAIPW value across MGB-young, MGB-old, and UPMC as the evaluation criterion. We repeated the analysis over 50 random splits. As shown in Table~\ref{tab:worst-source}, GR-PEVI attained the highest worst-source WAIPW value in 30 of the 50 splits, and its mean worst-source WAIPW exceeded that of the strongest baseline, Operator-Avg PEVI, by 0.0507. 

\begin{table}[t]
\centering
\caption{Worst-source WAIPW over 50 train--test splits. SD is computed across splits; win count is the number of splits with the highest value; GR-PEVI win rate is the proportion of paired splits in which GR-PEVI outperformed each baseline.}
\label{tab:worst-source}
\small
\setlength{\tabcolsep}{6pt}
\renewcommand{\arraystretch}{1.08}
\begin{tabular}{@{}lccccc@{}}
\toprule
Method & Mean & SD & Median & Win count & GR-PEVI win rate \\
\midrule
GR-PEVI
& \textbf{3.2042}
& 0.0760
& \textbf{3.2107}
& \textbf{30/50}
& -- \\

Operator-Avg PEVI
& 3.1535
& 0.0803
& 3.1594
& 7/50
& 72\% \\

Avg-PEVI
& 3.1463
& 0.0814
& 3.1364
& 6/50
& 82\% \\

Majority-PEVI
& 3.1319
& 0.1050
& 3.1287
& 7/50
& 76\% \\

Pooled PEVI
& 3.0151
& 0.0803
& 3.0276
& 0/50
& 98\% \\
\bottomrule
\end{tabular}
\end{table}

To examine the source-specific patterns underlying the worst-source results, Table~\ref{tab:source-wise-waipw} reports the relative improvements of each learned policy's WAIPW estimate over the empirical cumulative reward under the observed behavior policy, expressed as a percentage. For each source and split, the empirical cumulative reward was calculated as the average of $\sum_{h=1}^H r_h$ over the held-out trajectories. GR-PEVI achieved positive mean relative improvements for all three sources, with the largest improvements among the methods in MGB-young and UPMC. It outperformed Avg-PEVI, Majority-PEVI, and Pooled PEVI in terms of mean relative improvement for every source. Operator-Avg PEVI attained the largest mean improvement in MGB-old but had a negative mean improvement in MGB-young.

\begin{table}[t]
\centering
\caption{Source-specific relative improvements (\%) over the observed behavior policy across 50 train--test splits. Entries report the mean (SD) across splits; bold entries indicate the largest mean in each source.}
\label{tab:source-wise-waipw}
\small
\setlength{\tabcolsep}{7pt}
\renewcommand{\arraystretch}{1.08}
\begin{tabular}{@{}lccc@{}}
\toprule
Method & MGB-young & MGB-old & UPMC \\
\midrule
GR-PEVI            & $\boldsymbol{+3.04}$ (1.85) & $+1.99$ (2.51)          & $\boldsymbol{+3.83}$ (4.34) \\
Operator-Avg PEVI  & $-1.07$ (2.08)          & $\boldsymbol{+2.72}$ (2.72) & $+3.14$ (4.76) \\
Avg-PEVI           & $+0.37$ (2.10)          & $+0.80$ (2.98)          & $+3.11$ (5.58) \\
Majority-PEVI      & $-0.07$ (2.57)          & $+0.26$ (3.69)          & $+3.13$ (6.11) \\
Pooled PEVI        & $-3.08$ (2.18)          & $-3.32$ (2.25)          & $-3.44$ (3.30) \\
\bottomrule
\end{tabular}
\end{table}

\section{Discussion}
\label{sec:discussion}
In this paper, we study the problem of learning a common policy from heterogeneous offline data sources under a worst-case performance objective. We formulate this problem as a group-linear DRMDP using a shared linear MDP representation and a feature-wise uncertainty set constructed from source-specific transition and reward components. This construction captures cross-source heterogeneity and admits a robust Bellman recursion. Building on this formulation, GR-PEVI combines feature-wise robust aggregation with data-dependent pessimism, and our finite-sample analysis characterizes its suboptimality under a robust partial coverage condition. The cluster-level extension further quantifies the trade-off between the statistical benefit of pooling similar sources and the approximation error induced by within-cluster heterogeneity. A natural direction for future work is to develop richer uncertainty structures that incorporate dependence across feature components while preserving tractability, as well as data-driven procedures for identifying source clusters.

We apply GR-PEVI to multi-source EHR data on MS treatment decisions and, in Supplementary Section~\ref{appendix:sepsis}, to a six-source sepsis cohort. In both settings, GR-PEVI showed favorable worst-source performance relative to the comparison methods. These applications rely on feature engineering to construct a shared linear representation across sources. Feature engineering guided by domain knowledge can provide an interpretable linear approximation, and data-adaptive representation learning may yield features that better capture the underlying transition and reward structure. Extending the framework beyond linear function approximation is another important direction for future work.

\section*{Acknowledgements}
The work described in this paper was partially supported by a grant from the ANR/RGC Joint Research Scheme sponsored by the Research Grants Council of the Hong Kong Special Administrative Region, China and French National Research Agency (Project No. A-HKUST603/25).

{\footnotesize 
\bibliography{Bibliography-MM-MC}
}

\clearpage

\setcounter{section}{0}
\setcounter{equation}{0}
\setcounter{theorem}{0}
\setcounter{table}{0}
\setcounter{figure}{0}
\setcounter{algorithm}{0}

\renewcommand{\thesection}{S\arabic{section}}
\renewcommand{\theequation}{S\arabic{equation}}
\renewcommand{\thetheorem}{S\arabic{theorem}}
\renewcommand{\thetable}{S\arabic{table}}
\renewcommand{\thefigure}{S\arabic{figure}}
\renewcommand{\thealgorithm}{S\arabic{algorithm}}

\def\theHsection{supp.\arabic{section}}
\def\theHequation{supp.\arabic{equation}}
\def\theHtheorem{supp.\arabic{theorem}}
\def\theHtable{supp.\arabic{table}}
\def\theHfigure{supp.\arabic{figure}}
\def\theHalgorithm{supp.\arabic{algorithm}}

\begin{center}
{\LARGE\bf Supplementary Material for\\``Learning Sequential Decisions from Multiple Sources via Group-Robust Markov Decision Processes''}
\end{center}
  \medskip

\spacingset{1.8} 

This supplementary material provides additional technical, numerical, and implementation details for the main paper. Section~\ref{app:proofs} contains the proofs of all theoretical results. Section~\ref{app:cluster-algorithm} presents the complete cluster-level GR-PEVI algorithm. Section~\ref{sec:supp-simulations} provides supplementary details for the simulation studies, including additional specifications for the main multi-source experiment, as well as a sensitivity analysis under milder transition heterogeneity and a numerical study of the cluster-level extension. Section~\ref{app:real-data-details} provides additional details for the MS analysis, including the DMT classification, PDDS imputation, and the implementation of FQE and WAIPW for off-policy evaluation. Section~\ref{appendix:sepsis} presents an additional real-data analysis of fluid-management decisions in a multi-source sepsis cohort.

\section{Proofs}
\label{app:proofs}
\subsection{Proof of Proposition~\ref{prop:robust-bellman}}
\label{app:proofs-robust-bellman}
\begin{proof}[Proof of Proposition~\ref{prop:robust-bellman}]
We prove the following stronger statement: There exists a set of worst-case parameters
$\left\{\boldsymbol{\alpha}^{\pi,\mathrm{wc}}_h\right\}_{h=1}^H$ (the superscript ``wc'' stands for worst case), with $\boldsymbol{\alpha}^{\pi,\mathrm{wc}}_h=\left\{\boldsymbol{\alpha}^{\pi,\mathrm{wc}}_{h,i}\right\}_{i=1}^d$ and $\boldsymbol{\alpha}^{\pi,\mathrm{wc}}_{h,i}\in\Delta^{K-1}$, such that

1. The robust Bellman equations hold:
$$
V_h^\pi(s)=\mathbb{E}_{a \sim \pi_h(\cdot \mid s)}\left[Q_h^\pi(s, a)\right],\qquad\forall s\in\mathcal{S}
$$
$$
Q_h^\pi(s, a)=\inf _{\left(P_h, r_h\right) \in \mathcal{U}_h}\left\{r_h(s, a)+\mathbb{E}_{s^{\prime} \sim P_h(\cdot \mid s, a)}\left[V_{h+1}^\pi\left(s^{\prime}\right)\right]\right\},\qquad\forall(s,a)\in\mathcal{S}\times\mathcal{A}
$$
2. The following expressions for robust value functions hold:
$$
\begin{aligned}
V_h^\pi(s) & =V_h^\pi\left(s ;\left\{\boldsymbol{\alpha}_t^{\pi,\mathrm{wc}}\right\}_{t=h}^H\right)=\mathbb{E}_{\left\{P_t^{\pi,\mathrm{wc}}\right\}_{t=h}^H, \pi}\left[\sum_{t=h}^H r_t^{\pi,\mathrm{wc}}\left(s_t, a_t\right) \mid s_h=s\right] \\
Q_h^\pi(s, a) & =Q_h^\pi\left(s,a ;\left\{\boldsymbol{\alpha}_t^{\pi,\mathrm{wc}}\right\}_{t=h}^H\right)=\mathbb{E}_{\left\{P_t^{\pi,\mathrm{wc}}\right\}_{t=h}^H, \pi}\left[\sum_{t=h}^H r_t^{\pi,\mathrm{wc}}\left(s_t, a_t\right) \mid s_h=s, a_h=a\right]
\end{aligned}
$$
where for all $h\in[H]$,
$$
P_h^{\pi,\mathrm{wc}}(\cdot \mid s, a)=\sum_{i=1}^d \phi_i(s, a) \sum_{k=1}^K \alpha^{k,\pi,\mathrm{wc}}_{h, i} \mu_{h, i}^k(\cdot)
$$
$$
r_h^{\pi,\mathrm{wc}}(s, a)=\sum_{i=1}^d \phi_i(s, a) \sum_{k=1}^K \alpha^{k,\pi,\mathrm{wc}}_{h, i} \theta_{h, i}^k
$$
The base case $h=H$ follows from the argument below with $V_{H+1}^\pi\equiv 0$; in this case, the continuation term vanishes and the coordinate-wise selector is obtained by minimizing the immediate reward factors $\theta_{H,i}^k$ over $k\in[K]$. Now suppose that conclusions 1 and 2 hold for step $h+1$, so that there exist $\left\{\boldsymbol{\alpha}_t^{\pi,\mathrm{wc}}\right\}_{t=h+1}^H$ such that the following condition holds for all $s\in\mathcal S$:
\begin{equation}
    V_{h+1}^\pi(s)  =V_{h+1}^\pi\left(s ;\left\{\boldsymbol{\alpha}_t^{\pi,\mathrm{wc}}\right\}_{t=h+1}^H\right)=\mathbb{E}_{\left\{P_t^{\pi,\mathrm{wc}}\right\}_{t=h+1}^H, \pi}\left[\sum_{t=h+1}^H r_t^{\pi,\mathrm{wc}}\left(s_t, a_t\right) \mid s_{h+1}=s\right]
    \label{h+1}
\end{equation}
Then by the definition of the $Q$ function, we have for all $(s,a)\in\mathcal{S}\times\mathcal{A}$,
{\footnotesize
\begin{align}
    Q_h^\pi(s, a)&=\inf _{\left\{\left(P_t, r_t\right)\right\}_{t=h}^H \in \prod_{t=h}^H \mathcal{U}_t} \mathbb{E}_{\left\{P_t\right\}_{t=h}^H, \pi}\left[\sum_{t=h}^H r_t\left(s_t, a_t\right) \mid s_h=s, a_h=a\right]\nonumber\\
    &=\inf _{\left\{\left(P_t, r_t\right)\right\}_{t=h}^H \in \prod_{t=h}^H \mathcal{U}_t}\left\{ r_h(s,a)+\mathbb{E}_{s'\sim P_h(\cdot\mid s,a)}\left[\mathbb{E}_{\left\{P_t\right\}_{t=h+1}^H, \pi}\left[\sum_{t=h+1}^H r_t\left(s_t, a_t\right) \mid s_{h+1}=s'\right]\right]\right\}\nonumber\\
    &\leq\inf_{\left(P_h, r_h\right) \in \mathcal{U}_h}\left\{ r_h(s,a)+\mathbb{E}_{s'\sim P_h(\cdot\mid s,a)}\left[\mathbb{E}_{\left\{P_t^{\pi,\mathrm{wc}}\right\}_{t=h+1}^H, \pi}\left[\sum_{t=h+1}^H r_t^{\pi,\mathrm{wc}}\left(s_t, a_t\right) \mid s_{h+1}=s'\right]\right] \right\}\label{eq:second}\\
    &=\inf_{\left(P_h, r_h\right) \in \mathcal{U}_h}\left\{ r_h(s,a)+\mathbb{E}_{s'\sim P_h(\cdot\mid s,a)}\left[V_{h+1}^\pi\left(s' ;\left\{\boldsymbol{\alpha}_t^{\pi,\mathrm{wc}}\right\}_{t=h+1}^H\right)\right] \right\}\nonumber\\
    &=\inf_{\left(P_h, r_h\right) \in \mathcal{U}_h}\left\{ r_h(s,a)+\mathbb{E}_{s'\sim P_h(\cdot\mid s,a)}\left[V_{h+1}^\pi\left(s'\right)\right]\right\}\nonumber\\
    &=\inf_{\left(P_h, r_h\right) \in \mathcal{U}_h}\left\{ r_h(s,a)+\mathbb{E}_{s'\sim P_h(\cdot\mid s,a)}\left[\inf _{\left\{\left(P_t, r_t\right)\right\}_{t=h+1}^H \in \prod_{t=h+1}^H \mathcal{U}_t} \mathbb{E}_{\left\{P_t\right\}_{t=h+1}^H, \pi}\left[\sum_{t=h+1}^H r_t\left(s_t, a_t\right) \mid s_{h+1}=s'\right]\right]\right\}\nonumber\\
    &\leq \inf _{\left\{\left(P_t, r_t\right)\right\}_{t=h}^H \in \prod_{t=h}^H \mathcal{U}_t}\left\{ r_h(s,a)+\mathbb{E}_{s'\sim P_h(\cdot\mid s,a)}\left[\mathbb{E}_{\left\{P_t\right\}_{t=h+1}^H, \pi}\left[\sum_{t=h+1}^H r_t\left(s_t, a_t\right) \mid s_{h+1}=s'\right]\right]\right\}\nonumber\\
    &=Q_h^\pi(s, a)\nonumber
\end{align}
}
Therefore,
\begin{align*}
    Q_h^\pi(s, a)&=\inf _{\left(P_h, r_h\right) \in \mathcal{U}_h}\left\{r_h(s, a)+\mathbb{E}_{s^{\prime} \sim P_h(\cdot \mid s, a)}\left[V_{h+1}^\pi\left(s^{\prime}\right)\right]\right\}
\end{align*}
This proves the $Q$-function part of conclusion 1 for step $h$. 

We now construct the current-step worst-case selector explicitly. For each
$i\in[d]$ and $k\in[K]$, define
\[
c_{h,i}^{k,\pi}
:=
\theta_{h,i}^k
+
\int_{\mathcal S} V_{h+1}^\pi(s')\,\mu_{h,i}^k(ds').
\]
The quantity $c_{h,i}^{k,\pi}$ depends on $h,i,k$ and the continuation value
$V_{h+1}^\pi$, but it does not depend on the current state-action pair
$(s,a)$. For each $i\in[d]$, choose
\[
k_{h,i}^{\pi,\mathrm{wc}}
\in
\arg\min_{k\in[K]} c_{h,i}^{k,\pi},
\]
with ties broken arbitrarily, and set
\[
\alpha_{h,i}^{k,\pi,\mathrm{wc}}
:=
\mathbf 1\{k=k_{h,i}^{\pi,\mathrm{wc}}\},
\qquad k\in[K].
\]
Then $\boldsymbol{\alpha}_{h,i}^{\pi,\mathrm{wc}}\in\Delta^{K-1}$ for every
$i\in[d]$.

For any $(P_h,r_h)\in\mathcal U_h$ represented by
$\{\boldsymbol{\alpha}_{h,i}\}_{i=1}^d$, we have, for every
$(s,a)\in\mathcal S\times\mathcal A$,
\begin{align*}
& r_h(s,a)
+
\mathbb E_{s'\sim P_h(\cdot\mid s,a)}
\left[V_{h+1}^\pi(s')\right] \\
&\quad =
\sum_{i=1}^d
\phi_i(s,a)
\sum_{k=1}^K
\alpha_{h,i}^k
\left[
\theta_{h,i}^k
+
\int_{\mathcal S}V_{h+1}^\pi(s')\,\mu_{h,i}^k(ds')
\right] \\
&\quad =
\sum_{i=1}^d
\phi_i(s,a)
\sum_{k=1}^K
\alpha_{h,i}^k c_{h,i}^{k,\pi} \\
&\quad \ge
\sum_{i=1}^d
\phi_i(s,a)
\min_{k\in[K]} c_{h,i}^{k,\pi},
\end{align*}
where the inequality uses $\phi_i(s,a)\ge 0$ and
$\boldsymbol{\alpha}_{h,i}\in\Delta^{K-1}$. The lower bound is attained by
the selector $\boldsymbol{\alpha}_{h}^{\pi,\mathrm{wc}}$ defined above. Therefore,
for every $(s,a)$,
\begin{align*}
&\inf_{\{\boldsymbol{\alpha}_{h,i}\in\Delta^{K-1}\}_{i=1}^d}
\left\{
r_h(s,a)
+
\mathbb E_{s'\sim P_h(\cdot\mid s,a)}
\left[V_{h+1}^\pi(s')\right]
\right\} \\
&\quad =
\sum_{i=1}^d
\phi_i(s,a)
\min_{k\in[K]}
\left[
\theta_{h,i}^k
+
\int_{\mathcal S} V_{h+1}^\pi(s')\,\mu_{h,i}^k(ds')
\right] \\
&\quad =
\sum_{i=1}^d
\phi_i(s,a)
\left[
\theta_{h,i}^{k_{h,i}^{\pi,\mathrm{wc}}}
+
\int_{\mathcal S} V_{h+1}^\pi(s')\,
\mu_{h,i}^{k_{h,i}^{\pi,\mathrm{wc}}}(ds')
\right].
\end{align*}
Thus the same selector $\boldsymbol{\alpha}_h^{\pi,\mathrm{wc}}$ attains the
one-step infimum simultaneously for all state-action pairs $(s,a)$.

Define
\[
P_h^{\pi,\mathrm{wc}}(\cdot\mid s,a)
=
\sum_{i=1}^d
\phi_i(s,a)\,
\mu_{h,i}^{k_{h,i}^{\pi,\mathrm{wc}}}(\cdot),
\]
and
\[
r_h^{\pi,\mathrm{wc}}(s,a)
=
\sum_{i=1}^d
\phi_i(s,a)\,
\theta_{h,i}^{k_{h,i}^{\pi,\mathrm{wc}}}.
\]
Equivalently, these are induced by the mixture weights
$\{\boldsymbol{\alpha}_{h,i}^{\pi,\mathrm{wc}}\}_{i=1}^d$. Hence $(P_h^{\pi,\mathrm{wc}},r_h^{\pi,\mathrm{wc}})\in\mathcal U_h$.

The construction of $\boldsymbol{\alpha}_h^{\pi,\mathrm{wc}}$ further shows that this one-step infimum is attained by the model $(P_h^{\pi,\mathrm{wc}},r_h^{\pi,\mathrm{wc}})$ defined above. Together with the induction hypothesis that the future selectors
$\{(P_t^{\pi,\mathrm{wc}},r_t^{\pi,\mathrm{wc}})\}_{t=h+1}^H$ attain $V_{h+1}^\pi$, we obtain
\begin{equation}
    Q_h^\pi(s, a)=\mathbb{E}_{\left\{P_t^{\pi,\mathrm{wc}}\right\}_{t=h}^H, \pi}\left[\sum_{t=h}^H r_t^{\pi,\mathrm{wc}}\left(s_t, a_t\right) \mid s_h=s, a_h=a\right]=Q_h^\pi\left(s,a ;\left\{\boldsymbol{\alpha}_t^{\pi,\mathrm{wc}}\right\}_{t=h}^H\right)\label{eq:3}
\end{equation}
This proves the $Q$-function part of conclusion 2 for step $h$. Now, we consider the value function:
\begin{align}
    V_h^\pi(s)&=\inf _{\left\{\left(P_t, r_t\right)\right\}_{t=h}^H \in \prod_{t=h}^H \mathcal{U}_t} \mathbb{E}_{\left\{P_t\right\}_{t=h}^H, \pi}\left[\sum_{t=h}^H r_t\left(s_t, a_t\right) \mid s_h=s\right]\nonumber\\
    &=\inf_{\left\{(P_t,r_t)\right\}_{t=h}^H
\in \prod_{t=h}^H\mathcal U_t}
\mathbb E_{a\sim\pi_h(\cdot\mid s)}
\left\{
\mathbb E_{\left\{P_t\right\}_{t=h}^H,\pi}
\left[
\sum_{t=h}^H r_t(s_t,a_t)
\mid s_h=s,\ a_h=a
\right]
\right\}\nonumber\\
    &\leq \mathbb E_{a\sim\pi_h(\cdot\mid s)}
\left\{ \mathbb{E}_{\left\{P_t^{\pi,\mathrm{wc}}\right\}_{t=h}^H, \pi}\left[\sum_{t=h}^H r_t^{\pi,\mathrm{wc}}\left(s_t, a_t\right) \mid s_h=s, a_h=a\right]\right\}\nonumber\\
    &=\mathbb E_{a\sim\pi_h(\cdot\mid s)}
\left[Q_h^\pi(s,a)\right]\nonumber\\
    &=\mathbb E_{a\sim\pi_h(\cdot\mid s)}
\left\{\inf _{\left\{\left(P_t, r_t\right)\right\}_{t=h}^H \in \prod_{t=h}^H \mathcal{U}_t} \mathbb{E}_{\left\{P_t\right\}_{t=h}^H, \pi}\left[\sum_{t=h}^H r_t\left(s_t, a_t\right) \mid s_h=s, a_h=a\right]\right\}\nonumber\\
    &\leq \inf _{\left\{\left(P_t, r_t\right)\right\}_{t=h}^H \in \prod_{t=h}^H \mathcal{U}_t} \mathbb E_{a\sim\pi_h(\cdot\mid s)}
\left\{\mathbb{E}_{\left\{P_t\right\}_{t=h}^H, \pi}\left[\sum_{t=h}^H r_t\left(s_t, a_t\right) \mid s_h=s, a_h=a\right]\right\}\nonumber\\
    &=\inf _{\left\{\left(P_t, r_t\right)\right\}_{t=h}^H \in \prod_{t=h}^H \mathcal{U}_t}\mathbb{E}_{\left\{P_t\right\}_{t=h}^H, \pi}\left[\sum_{t=h}^H r_t\left(s_t, a_t\right) \mid s_h=s\right]=V_h^\pi(s)\nonumber
\end{align}
The key point is that the selector
$\boldsymbol{\alpha}_h^{\pi,\mathrm{wc}}$ constructed above is independent of the realized action $a$. Hence the same current-step model
$(P_h^{\pi,\mathrm{wc}},r_h^{\pi,\mathrm{wc}})$ attains the one-step infimum for every $a\in\mathcal A$. Together with the induction hypothesis for the future selectors $\{(P_t^{\pi,\mathrm{wc}},r_t^{\pi,\mathrm{wc}})\}_{t=h+1}^H$, the model
sequence $\{(P_t^{\pi,\mathrm{wc}},r_t^{\pi,\mathrm{wc}})\}_{t=h}^H$ is admissible and attains the robust value after averaging over $a\sim\pi_h(\cdot\mid s)$. Therefore all inequalities in the display above
collapse to equalities, and we conclude that
\begin{equation}
V_h^\pi(s)=\mathbb E_{a\sim\pi_h(\cdot\mid s)}\left[Q_h^\pi(s,a)\right].
\label{eq:4}
\end{equation}
This proves the value function part of conclusion 1 for step $h$. By combining \eqref{eq:3} and \eqref{eq:4}, we have
$$
 V_h^\pi(s)=\mathbb{E}_{\left\{P_t^{\pi,\mathrm{wc}}\right\}_{t=h}^H, \pi}\left[\sum_{t=h}^H r_t^{\pi,\mathrm{wc}}\left(s_t, a_t\right) \mid s_h=s\right]
$$
This proves the value function part of conclusion 2 for step $h$. By induction, conclusions 1 and 2 hold for all $h\in[H]$. According to conclusion 1, we get
\begin{align*}
    V_h^\pi(s)&=\mathbb{E}_{a \sim \pi_h(\cdot \mid s)}\left[Q_h^\pi(s, a)\right]\\
    &=\mathbb{E}_{a \sim \pi_h(\cdot \mid s)}\left[\inf _{\left(P_h, r_h\right) \in \mathcal{U}_h}\left\{r_h(s, a)+\mathbb{E}_{s^{\prime} \sim P_h(\cdot \mid s, a)}\left[V_{h+1}^\pi\left(s^{\prime}\right)\right]\right\}\right]
\end{align*}
By the attainment property of $(P_h^{\pi,\mathrm{wc}},r_h^{\pi,\mathrm{wc}})$ established above, we have
\begin{align*}
   V_h^\pi(s)&=\mathbb{E}_{a \sim \pi_h(\cdot \mid s)}\left[\inf _{\left(P_h, r_h\right) \in \mathcal{U}_h}\left\{r_h(s, a)+\mathbb{E}_{s^{\prime} \sim P_h(\cdot \mid s, a)}\left[V_{h+1}^\pi\left(s^{\prime}\right)\right]\right\}\right]\\
    &=\mathbb{E}_{a \sim \pi_h(\cdot \mid s)}\left[r^{\pi,\mathrm{wc}}_h(s, a)+\mathbb{E}_{s^{\prime} \sim P_h^{\pi,\mathrm{wc}}(\cdot \mid s, a)}\left[V_{h+1}^\pi\left(s^{\prime}\right)\right]\right]\\
    &=\inf _{\left(P_h, r_h\right) \in \mathcal{U}_h}\mathbb{E}_{a \sim \pi_h(\cdot \mid s)}\left[r_h(s, a)+\mathbb{E}_{s^{\prime} \sim P_h(\cdot \mid s, a)}\left[V_{h+1}^\pi\left(s^{\prime}\right)\right]\right]
\end{align*}
The third equality is because
\begin{align*}
   &\mathbb{E}_{a \sim \pi_h(\cdot \mid s)}\left[\inf _{\left(P_h, r_h\right) \in \mathcal{U}_h}\left\{r_h(s, a)+\mathbb{E}_{s^{\prime} \sim P_h(\cdot \mid s, a)}\left[V_{h+1}^\pi\left(s^{\prime}\right)\right]\right\}\right]\\
   =&\mathbb{E}_{a \sim \pi_h(\cdot \mid s)}\left[r^{\pi,\mathrm{wc}}_h(s, a)+\mathbb{E}_{s^{\prime} \sim P_h^{\pi,\mathrm{wc}}(\cdot \mid s, a)}\left[V_{h+1}^\pi\left(s^{\prime}\right)\right]\right]\\
    \leq& \inf _{\left(P_h, r_h\right) \in \mathcal{U}_h}\mathbb{E}_{a \sim \pi_h(\cdot \mid s)}\left[r_h(s, a)+\mathbb{E}_{s^{\prime} \sim P_h(\cdot \mid s, a)}\left[V_{h+1}^\pi\left(s^{\prime}\right)\right]\right]
\end{align*}
and
\begin{equation*}
    \begin{split}
        &\inf_{(P_h,r_h)\in\mathcal U_h}
\mathbb{E}_{a\sim\pi_h(\cdot\mid s)}
\left[
   r_h(s,a)
   + \mathbb{E}_{s'\sim P_h(\cdot\mid s,a)} \left[ V_{h+1}^\pi(s') \right]
\right]\\
&\qquad\le
\mathbb{E}_{a\sim\pi_h(\cdot\mid s)}
\left[
   r_h^{\pi,\mathrm{wc}}(s,a)
   + \mathbb{E}_{s'\sim P_h^{\pi,\mathrm{wc}}(\cdot\mid s,a)} \left[V_{h+1}^\pi(s')\right]
\right]
    \end{split}
\end{equation*}

This completes the proof of the robust Bellman equations under the $d$-rectangular uncertainty set:
$$
\begin{aligned}
Q_h^\pi(s, a) & =\inf _{\left(P_h, r_h\right) \in \mathcal{U}_h}\left\{r_h(s, a)+\mathbb{E}_{s^{\prime} \sim P_h(\cdot \mid s, a)}\left[V_{h+1}^\pi\left(s^{\prime}\right)\right]\right\}, \\
V_h^\pi(s) & =\mathbb{E}_{a \sim \pi_h(\cdot \mid s)}\left[Q_h^\pi(s, a)\right].
\end{aligned}
$$
\end{proof}

\subsection{Proof of Existence of the Robust Optimal Policy}
\label{proof:existence}
\begin{proof}[Proof of Existence]
We first construct a candidate optimal value function by backward induction. Set $W_{H+1}\equiv 0$. For $h=H,H-1,\ldots,1$, suppose that $W_{h+1}$ is measurable and takes values in $[0,H-h]$. For each $s\in\mathcal S$, define
\[
W_h(s)
:=
\max_{a\in\mathcal A}
\inf_{(P_h,r_h)\in\mathcal U_h}
\left\{
r_h(s,a)
+
\mathbb E_{s'\sim P_h(\cdot\mid s,a)}
\left[W_{h+1}(s')\right]
\right\}.
\]
By the feature-wise representation of the robust Bellman operator, the expression being maximized equals
\[
\sum_{i=1}^d \phi_i(s,a)
\min_{k\in[K]}
\left\{
\theta_{h,i}^k
+
\int_{\mathcal S}W_{h+1}(s')\,\mu_{h,i}^k(ds')
\right\}.
\]
The coefficients do not depend on $(s,a)$. Since
$\boldsymbol{\phi}$ is measurable in $(s,a)$ and continuous in $a$ for every fixed $s$, this expression has the same measurability and continuity properties. Since $\mathcal A$ is a nonempty compact metric space, the maximum is attained. The measurable maximum theorem~\citep[Theorem~18.19]{aliprantis2006infinite} further implies that $W_h$ is measurable and that there exists a measurable map $a_h^\dagger:\mathcal S\to\mathcal A$ satisfying
\[
a_h^\dagger(s)
\in
\arg\max_{a\in\mathcal A}
\inf_{(P_h,r_h)\in\mathcal U_h}
\left\{
r_h(s,a)
+
\mathbb E_{s'\sim P_h(\cdot\mid s,a)}
\left[W_{h+1}(s')\right]
\right\},
\qquad s\in\mathcal S.
\]
Moreover, since the rewards lie in $[0,1]$ and $W_{h+1}$ takes values in $[0,H-h]$, $W_h$ takes values in $[0,H-h+1]$. This completes the backward construction.

Define a deterministic Markov policy $\pi^\dagger=\{\pi_h^\dagger\}_{h=1}^H$ by letting $\pi_h^\dagger(\cdot\mid s)$ be the point mass at $a_h^\dagger(s)$. The measurability of each $a_h^\dagger$ ensures that $\pi^\dagger\in\Pi$. For any fixed policy $\pi\in\Pi$, Proposition~\ref{prop:robust-bellman} gives the robust Bellman recursion for $\{V_h^\pi\}_{h=1}^H$, which we use below.

Next, we show that for any $\pi\in\Pi$ and any $(h,s)\in[H]\times\mathcal S$, 
\[
V_h^\pi(s)\le W_h(s).
\]
The claim holds trivially for $h=H+1$, since
$V_{H+1}^\pi\equiv W_{H+1}\equiv 0$.
Suppose that
\[
V_{h+1}^\pi(s')\le W_{h+1}(s'),
\qquad \forall s'\in\mathcal S.
\]
By Proposition~\ref{prop:robust-bellman}, for any $s\in\mathcal S$,
\begin{align*}
V_h^\pi(s)
&=
\mathbb E_{a\sim\pi_h(\cdot\mid s)}
\Bigg[
\inf_{(P_h,r_h)\in\mathcal U_h}
\left\{
r_h(s,a)
+
\mathbb E_{s'\sim P_h(\cdot\mid s,a)}
\left[V_{h+1}^\pi(s')\right]
\right\}
\Bigg]\\
&\le
\mathbb E_{a\sim\pi_h(\cdot\mid s)}
\Bigg[
\inf_{(P_h,r_h)\in\mathcal U_h}
\left\{
r_h(s,a)
+
\mathbb E_{s'\sim P_h(\cdot\mid s,a)}
\left[W_{h+1}(s')\right]
\right\}
\Bigg]\\
&\le
\max_{a\in\mathcal A}
\inf_{(P_h,r_h)\in\mathcal U_h}
\left\{
r_h(s,a)
+
\mathbb E_{s'\sim P_h(\cdot\mid s,a)}
\left[W_{h+1}(s')\right]
\right\}\\
&=
W_h(s).
\end{align*}
Thus, by backward induction,
\[
V_h^\pi(s)\le W_h(s)
\]
for every $\pi\in\Pi$, $h\in[H]$, and $s\in\mathcal S$.

It remains to show that the constructed policy $\pi^\dagger$ attains the candidate value function $W_h$, that is,
\[
V_h^{\pi^\dagger}(s)=W_h(s),
\qquad \forall (h,s)\in[H]\times\mathcal S.
\]
We prove this by backward induction on $h$. The claim holds for $h=H+1$ since $V_{H+1}^{\pi^\dagger}\equiv W_{H+1}\equiv 0$.
Assume that
\[
V_{h+1}^{\pi^\dagger}(s')=W_{h+1}(s'),
\qquad \forall s'\in\mathcal S.
\]
Since $\pi_h^\dagger(\cdot\mid s)$ is the point mass at $a_h^\dagger(s)$, Proposition~\ref{prop:robust-bellman} gives, for any $s\in\mathcal S$,
\begin{align*}
V_h^{\pi^\dagger}(s)
&=
\inf_{(P_h,r_h)\in\mathcal U_h}
\left\{
r_h\bigl(s,a_h^\dagger(s)\bigr)
+
\mathbb E_{s'\sim P_h(\cdot\mid s,a_h^\dagger(s))}
\left[V_{h+1}^{\pi^\dagger}(s')\right]
\right\}\\
&=
\inf_{(P_h,r_h)\in\mathcal U_h}
\left\{
r_h\bigl(s,a_h^\dagger(s)\bigr)
+
\mathbb E_{s'\sim P_h(\cdot\mid s,a_h^\dagger(s))}
\left[W_{h+1}(s')\right]
\right\}\\
&=
\max_{a\in\mathcal A}
\inf_{(P_h,r_h)\in\mathcal U_h}
\left\{
r_h(s,a)
+
\mathbb E_{s'\sim P_h(\cdot\mid s,a)}
\left[W_{h+1}(s')\right]
\right\}\\
&=
W_h(s),
\end{align*}
where the third equality follows from the definition of
$a_h^\dagger(s)$. Thus, by backward induction,
\[
V_h^{\pi^\dagger}(s)=W_h(s),
\qquad \forall h\in[H],\ s\in\mathcal S.
\]

Combining this with the inequality
$V_h^\pi(s)\le W_h(s)$ for every $\pi\in\Pi$, we obtain
\[
W_h(s)
=
V_h^{\pi^\dagger}(s)
\le
\sup_{\pi\in\Pi}V_h^\pi(s)
\le
W_h(s).
\]
Therefore,
\[
W_h(s)
=
\sup_{\pi\in\Pi}V_h^\pi(s)
=
V_h^\star(s),
\qquad \forall h\in[H],\ s\in\mathcal S.
\]
Hence $\pi^\dagger$ attains the robust optimal value function and is an
optimal deterministic Markov policy. Taking $\pi^\star=\pi^\dagger$
completes the proof.

\end{proof}

\subsection{Proof of Lemma~\ref{lemma:decomposition}}
\label{app:proofs-decomposition}
\begin{proof}[Proof of Lemma~\ref{lemma:decomposition}]
Fix a realization of the offline dataset $\mathcal D$, and hence of the data-dependent quantities $\{\widehat V_h,\widehat Q_h,\widehat\pi_h\}_{h=1}^H$. For notational convenience, define the policy evaluation operators $\mathbb J_h$ and $\widehat{\mathbb J}_h$ for any bounded measurable function $f:\mathcal S\times\mathcal A\to\mathbb R$ by
\[
(\mathbb J_h f)(s)
:=
\mathbb E_{a\sim\pi_h^\star(\cdot\mid s)}[f(s,a)],
\qquad
(\widehat{\mathbb J}_h f)(s)
:=
\mathbb E_{a\sim\widehat\pi_h(\cdot\mid s)}[f(s,a)].
\]
By the definition of suboptimality,
\[
\operatorname{SubOpt}(\widehat\pi;s)
=
V_1^{\pi^\star}(s)-V_1^{\widehat\pi}(s)
=
\bigl[V_1^{\pi^\star}(s)-\widehat V_1(s)\bigr]
+
\bigl[\widehat V_1(s)-V_1^{\widehat\pi}(s)\bigr].
\]

We first bound $V_1^{\pi^\star}(s)-\widehat V_1(s)$. For any stage $h$, we have
\begin{align*}
V_h^{\pi^\star}(s)-\widehat V_h(s)
&=
(\mathbb J_h Q_h^{\pi^\star})(s)
-
(\widehat{\mathbb J}_h\widehat Q_h)(s)\\
&=
(\mathbb J_h Q_h^{\pi^\star})(s)
-
(\mathbb J_h\widehat Q_h)(s)
+
(\mathbb J_h\widehat Q_h)(s)
-
(\widehat{\mathbb J}_h\widehat Q_h)(s)\\
&=
\mathbb J_h\left(Q_h^{\pi^\star}-\widehat Q_h\right)(s)
+
\underbrace{
\mathbb E_{a\sim\pi_h^\star(\cdot\mid s)}
[\widehat Q_h(s,a)]
-
\mathbb E_{a\sim\widehat\pi_h(\cdot\mid s)}
[\widehat Q_h(s,a)]
}_{\omega_h(s)}.
\end{align*}
Recall the definition of the model evaluation error $\iota_h$. We have
\[
\widehat Q_h(s,a)
=
(\mathbb B_h\widehat V_{h+1})(s,a)-\iota_h(s,a).
\]
Using Proposition~\ref{prop:robust-bellman}, we obtain
\[
Q_h^{\pi^\star}(s,a)-\widehat Q_h(s,a)
=
(\mathbb B_hV_{h+1}^{\pi^\star})(s,a)
-
(\mathbb B_h\widehat V_{h+1})(s,a)
+
\iota_h(s,a).
\]
Therefore,
\[
V_h^{\pi^\star}(s)-\widehat V_h(s)
=
\mathbb J_h
\left(
\mathbb B_hV_{h+1}^{\pi^\star}
-
\mathbb B_h\widehat V_{h+1}
\right)(s)
+
(\mathbb J_h\iota_h)(s)
+
\omega_h(s).
\]
By the definition of the robust Bellman operator,
\[
(\mathbb B_h V)(s,a)
=
\inf_{(P_h,r_h)\in\mathcal U_h}
\left\{
r_h(s,a)
+
\mathbb E_{s'\sim P_h(\cdot\mid s,a)}[V(s')]
\right\}.
\]
Using $\inf_x f(x)-\inf_x g(x)\leq\sup_x\{f(x)-g(x)\}$,
we obtain
\begin{align*}
&(\mathbb B_hV_{h+1}^{\pi^\star})(s,a)
-
(\mathbb B_h\widehat V_{h+1})(s,a)\\
&\qquad\leq
\sup_{(P_h,r_h)\in\mathcal U_h}
\mathbb E_{s'\sim P_h(\cdot\mid s,a)}
\left[
V_{h+1}^{\pi^\star}(s')
-
\widehat V_{h+1}(s')
\right].
\end{align*}
By the same coordinate-wise selection argument as in the proof of Proposition~\ref{prop:robust-bellman}, the feature-wise rectangular structure of $\mathcal U_h$ yields an admissible pair $(P_h^\dagger,r_h^\dagger)\in\mathcal U_h$ whose transition component $P_h^\dagger$ simultaneously attains this supremum for all $(s,a)$. Define
\[
(\mathbb E_h^\dagger f)(s,a)
:=
\mathbb E_{s'\sim P_h^\dagger(\cdot\mid s,a)}[f(s')].
\]
Hence,
\[
(\mathbb B_hV_{h+1}^{\pi^\star})(s,a)
-
(\mathbb B_h\widehat V_{h+1})(s,a)
\leq
\bigl(\mathbb E_h^\dagger
[V_{h+1}^{\pi^\star}-\widehat V_{h+1}]\bigr)(s,a).
\]
Now, we get
\[
V_h^{\pi^\star}(s)-\widehat V_h(s)
\le
\bigl(
\mathbb J_h\mathbb E_h^\dagger
[V_{h+1}^{\pi^\star}-\widehat V_{h+1}]
\bigr)(s)
+
(\mathbb J_h\iota_h)(s)
+
\omega_h(s).
\]
Recursively expanding this inequality over $h\in[H]$ and using the monotonicity of $\mathbb J_h$ and $\mathbb E_h^\dagger$, we obtain
\begin{align*}
V_1^{\pi^\star}(s)-\widehat V_1(s)
&\le
\left(
\prod_{h=1}^H
\mathbb J_h\mathbb E_h^\dagger
\right)
\left(
V_{H+1}^{\pi^\star}-\widehat V_{H+1}
\right)(s)\\
&\quad+
\sum_{h=1}^H
\left(
\prod_{t=1}^{h-1}
\mathbb J_t\mathbb E_t^\dagger
\right)
(\mathbb J_h\iota_h)(s)\\
&\quad+
\sum_{h=1}^H
\left(
\prod_{t=1}^{h-1}
\mathbb J_t\mathbb E_t^\dagger
\right)
\omega_h(s),
\end{align*}
where the products denote ordered composition of the operators, with the empty product interpreted as the identity operator. Since $V_{H+1}^{\pi^\star}=\widehat V_{H+1}\equiv0$, the first term vanishes.

By the definitions of $\mathbb J_h$ and $\mathbb E_h^\dagger$, we have
\begin{align*}
V_1^{\pi^\star}(s)-\widehat V_1(s)
&\leq
\sum_{h=1}^H
\mathbb E^{\pi^\star,P^{\pi^\star,\dagger}}
\left[
\iota_h(s_h,a_h)\mid s_1=s
\right]\\
&\quad+
\sum_{h=1}^H
\mathbb E^{\pi^\star,P^{\pi^\star,\dagger}}
\left[
\mathbb E_{a\sim\pi_h^\star(\cdot\mid s_h)}
[\widehat Q_h(s_h,a)]
-
\mathbb E_{a\sim\widehat\pi_h(\cdot\mid s_h)}
[\widehat Q_h(s_h,a)]
\mid s_1=s
\right],
\end{align*}
where $P^{\pi^\star,\dagger}=\{P_h^\dagger\}_{h=1}^H$ denotes the admissible transition-kernel sequence selected in the recursive upper bound above.

Next, we bound $\widehat V_1(s)-V_1^{\widehat\pi}(s)$. For any stage $h$,
\begin{align*}
\widehat V_h(s)-V_h^{\widehat\pi}(s)
&=
(\widehat{\mathbb J}_h\widehat Q_h)(s)
-
(\widehat{\mathbb J}_h Q_h^{\widehat\pi})(s)\\
&=
\widehat{\mathbb J}_h
\left(
\widehat Q_h-Q_h^{\widehat\pi}
\right)(s)\\
&=
\widehat{\mathbb J}_h
\left(
\mathbb B_h\widehat V_{h+1}
-
\mathbb B_hV_{h+1}^{\widehat\pi}
-
\iota_h
\right)(s).
\end{align*}
By the same argument as above,
\[
(\mathbb B_h\widehat V_{h+1})(s,a)
-
(\mathbb B_hV_{h+1}^{\widehat\pi})(s,a)
\leq
\bigl(
\mathbb E_h^\natural
[\widehat V_{h+1}-V_{h+1}^{\widehat\pi}]
\bigr)(s,a),
\]
where $\mathbb E_h^\natural$ is the expectation operator associated with the transition component $P_h^\natural$ of an admissible pair $(P_h^\natural,r_h^\natural)\in\mathcal U_h$ that simultaneously attains the corresponding supremum over all $(s,a)$. Hence,
\[
\widehat V_h(s)-V_h^{\widehat\pi}(s)
\leq
\bigl(
\widehat{\mathbb J}_h\mathbb E_h^\natural
[\widehat V_{h+1}-V_{h+1}^{\widehat\pi}]
\bigr)(s)
-
(\widehat{\mathbb J}_h\iota_h)(s).
\]
Recursively expanding this inequality and using $\widehat V_{H+1}=V_{H+1}^{\widehat\pi}\equiv0$, we obtain
\[
\widehat V_1(s)-V_1^{\widehat\pi}(s)
\leq
-\sum_{h=1}^H
\left(
\prod_{t=1}^{h-1}
\widehat{\mathbb J}_t\mathbb E_t^\natural
\right)
(\widehat{\mathbb J}_h\iota_h)(s).
\]
By the definitions of $\widehat{\mathbb J}_h$ and $\mathbb E_h^\natural$, this is equivalent to
\[
\widehat V_1(s)-V_1^{\widehat\pi}(s)
\leq
-\sum_{h=1}^H
\mathbb E^{\widehat\pi,P^{\widehat\pi,\natural}}
\left[
\iota_h(s_h,a_h)\mid s_1=s
\right],
\]
where $P^{\widehat\pi,\natural}=\{P_h^\natural\}_{h=1}^H$ denotes the admissible transition-kernel sequence selected in this recursive upper bound.

Combining the two bounds above, we obtain
\begin{align*}
\operatorname{SubOpt}(\widehat\pi;s)
&=
V_1^{\pi^\star}(s)-V_1^{\widehat\pi}(s)\\
&\leq
-\sum_{h=1}^H
\mathbb E^{\widehat\pi,P^{\widehat\pi,\natural}}
\left[
\iota_h(s_h,a_h)\mid s_1=s
\right]\\
&\quad+
\sum_{h=1}^H
\mathbb E^{\pi^\star,P^{\pi^\star,\dagger}}
\left[
\iota_h(s_h,a_h)\mid s_1=s
\right]\\
&\quad+
\sum_{h=1}^H
\mathbb E^{\pi^\star,P^{\pi^\star,\dagger}}
\left[
\mathbb E_{a\sim\pi_h^\star(\cdot\mid s_h)}
[\widehat Q_h(s_h,a)]
-
\mathbb E_{a\sim\widehat\pi_h(\cdot\mid s_h)}
[\widehat Q_h(s_h,a)]
\mid s_1=s
\right].
\end{align*}
\end{proof}

\subsection{Proof of Theorem~\ref{thm:suboptimality}}
\label{proof:suboptimality}
\begin{proof}[Proof of Theorem~\ref{thm:suboptimality}]
In order to characterize the suboptimality of Algorithm~\ref{alg:pess-source-ridge}, we first verify that our data-dependent penalty defines a valid $\xi$-uncertainty quantifier. 
\begin{lemma}[$\Gamma_h$ is a $\xi$-uncertainty quantifier]
\label{lem:uq}
In Algorithm~\ref{alg:pess-source-ridge}, we set
    $$
    \lambda=1, \beta=c \cdot d H \sqrt{\zeta}, \text { and } \zeta=\log (2 d K H N_{\max} / \xi)
    $$
Here $c>0$ is an absolute constant, $\xi \in(0,1)$ is the confidence parameter, and $N_{\max}=\max_{k\in[K]} N_k$. Then for all $h \in[H]$ and all $(s, a) \in \mathcal{S} \times \mathcal{A}$, it holds that
$$
\begin{aligned}
\left|\left(\mathbb{B}_h \widehat{V}_{h+1}\right)(s, a)-\left(\widehat{\mathbb{B}}_h \widehat{V}_{h+1}\right)(s, a)\right| & \leq \beta \sum_{i=1}^d \max _k\left\| \phi_i(s, a)\mathbf{e}_i\right\|_{\left(\Lambda_h^k\right)^{-1}} \\
& =\Gamma_h(s, a)
\end{aligned}
$$
with probability at least $1-\xi$, which is taken with respect to $\mathbb{P}_\mathcal{D}$. Equivalently, the event
\begin{equation}
\mathcal E=\Big\{\ \big|(\widehat{\mathbb B}_h \widehat V_{h+1})(s,a)-(\mathbb B_h \widehat V_{h+1})(s,a)\big|\ \le\ \Gamma_h(s,a),\ \forall(s,a),\ \forall h\in[H]\ \Big\}
\label{uncertainty-quantifier}
\end{equation}
satisfies $\mathbb P_{\mathcal D}(\mathcal E)\ge 1-\xi$.
\end{lemma}
\begin{proof}
We first define the expression of $\boldsymbol{\nu}_h^k$. Since
$$
\begin{aligned}
\left(\mathbb{B}_h \widehat{V}_{h+1}\right)(s, a) & =\sum_{i=1}^d \phi_i(s, a) \min _k\left[\theta_{h, i}^k+\left\langle\mu_{h, i}^k, \widehat{V}_{h+1}\right\rangle\right] \\
& =\phi(s, a)^{\top} \mathbf{w}_h\left(\widehat{V}_{h+1}\right)
\end{aligned}
$$
then
\begin{equation}
\boldsymbol{\nu}_h^k\left(\widehat{V}_{h+1}\right)=\boldsymbol{\theta}_{h}^k+\left\langle\boldsymbol{\mu}_{h}^k, \widehat{V}_{h+1}\right\rangle\label{eq:expression_nu}
\end{equation}
and
$$
\left(\mathbf{w}_h\right)_i=\min _{k \in[K]}\left(\boldsymbol{\nu}_h^k\right)_i, i=1, \ldots, d
$$
The following lemma upper bounds the norms of $\boldsymbol{\nu}_h^k\left(V\right)$ and $\widehat{\boldsymbol{\nu}}_h^k$, respectively.
 \begin{lemma}[Bounded Weights of Value Functions]
 \label{lemma-bounded-weights}
Let $V_{\max }>0$. For any measurable function $V:\mathcal S\to[0,V_{\max}]$, any $h \in[H]$ and any $k\in [K]$, we have
$$
\left\|\boldsymbol{\nu}_h^k\left(V\right)\right\|_2 \leq\left(1+V_{\max }\right) \sqrt{d}, \quad\left\|\widehat{\boldsymbol{\nu}}_h^k\right\|_2 \leq H \sqrt{N_k d / \lambda}
$$
where $\boldsymbol{\nu}_h^k$ and $\widehat{\boldsymbol{\nu}}_h^k$ are defined in Equations \eqref{eq:expression_nu} and \eqref{eq:estimate-nu}, respectively.
\end{lemma}  
\begin{proof}
For all $k\in[K]$ and all $h\in[H]$, we have
$$
\boldsymbol{\nu}_h^k\left(V\right)=\boldsymbol{\theta}_h^k+\int_{\mathcal{S}} V\left(s'\right) \boldsymbol{\mu}_h^k\left(ds'\right).
$$
By the triangle inequality, we have
\begin{align*}
\left\|\boldsymbol{\nu}_h^k(V)\right\|_2
&\le
\left\|\boldsymbol{\theta}_h^k\right\|_2
+
\left\|
\int_{\mathcal S} V(s')\,\boldsymbol{\mu}_h^k(ds')
\right\|_2 \\
&=
\left\|\boldsymbol{\theta}_h^k\right\|_2
+
\sqrt{
\sum_{i=1}^d
\left(
\int_{\mathcal S} V(s')\,\mu_{h,i}^k(ds')
\right)^2
}.
\end{align*}
Since each $\mu_{h,i}^k$ is a probability measure and
$0\le V(s')\le V_{\max}$ for all $s'\in\mathcal S$, we have, for every
$i\in[d]$,
\[
0
\le
\int_{\mathcal S} V(s')\,\mu_{h,i}^k(ds')
\le
V_{\max}.
\]
Moreover, since $\boldsymbol{\theta}_h^k\in[0,1]^d$, we have
$\|\boldsymbol{\theta}_h^k\|_2\le\sqrt d$. Therefore,
\begin{align*}
\left\|\boldsymbol{\nu}_h^k(V)\right\|_2
&\le
\sqrt d
+
\sqrt{\sum_{i=1}^d V_{\max}^2} \\
&=
(1+V_{\max})\sqrt d .
\end{align*}
For the $\widehat{\boldsymbol{\nu}}_h^k$, by the definition, we have
\begin{align*}
\left\|\widehat{\boldsymbol{\nu}}_h^k\right\|_2&=\left\|\left(\Lambda_h^k\right)^{-1}\left(\sum_{\tau=1}^{N_k} \phi\left(s_h^{(k, \tau)}, a_h^{(k, \tau)}\right) \cdot\left(r_h^{(k, \tau)}+\widehat{V}_{h+1}\left(s_{h+1}^{(k, \tau)}\right)\right)\right)\right\|_2\\
&\leq \sum_{\tau=1}^{N_k} \left\| \left(\Lambda_h^k\right)^{-1}\phi\left(s_h^{(k, \tau)}, a_h^{(k, \tau)}\right) \cdot\left(r_h^{(k, \tau)}+\widehat{V}_{h+1}\left(s_{h+1}^{(k, \tau)}\right)\right)\right\|_2
\end{align*}
Since $r_h^{(k,\tau)}\in[0,1]$ and $\widehat{V}_{h+1}\in[0,H-1]$, then $\left|r_h^{(k, \tau)}+\widehat{V}_{h+1}\left(s_{h+1}^{(k, \tau)}\right)\right|\in[0,H]$. Also, since $\Lambda_h^k \succeq \lambda\mathbf I_d$, we have
\begin{align*}
    \left\|\widehat{\boldsymbol{\nu}}_h^k\right\|_2&\leq H\sum_{\tau=1}^{N_k} \left\| \left(\Lambda_h^k\right)^{-1}\phi\left(s_h^{(k, \tau)}, a_h^{(k, \tau)}\right) \right\|_2\\
    &=H\sum_{\tau=1}^{N_k}\sqrt{\phi\left(s_h^{(k, \tau)}, a_h^{(k, \tau)}\right)^\top\left(\Lambda_h^k\right)^{-\frac{1}{2}}\left(\Lambda_h^k\right)^{-1}\left(\Lambda_h^k\right)^{-\frac{1}{2}} \phi\left(s_h^{(k, \tau)}, a_h^{(k, \tau)}\right)}\\
    &\leq \frac{H}{\sqrt{\lambda}}\sum_{\tau=1}^{N_k}\sqrt{\phi\left(s_h^{(k, \tau)}, a_h^{(k, \tau)}\right)^\top\left(\Lambda_h^k\right)^{-1} \phi\left(s_h^{(k, \tau)}, a_h^{(k, \tau)}\right)}
\end{align*}
By Cauchy-Schwarz inequality, we have
\begin{align*}
    \left\|\widehat{\boldsymbol{\nu}}_h^k\right\|_2&\leq H\sqrt{N_k/\lambda} \sqrt{\sum_{\tau=1}^{N_k}\phi\left(s_h^{(k, \tau)}, a_h^{(k, \tau)}\right)^\top\left(\Lambda_h^k\right)^{-1} \phi\left(s_h^{(k, \tau)}, a_h^{(k, \tau)}\right)}\\
    &=H\sqrt{N_k/\lambda} \sqrt{\mathrm{Tr}\left(\left(\Lambda_h^k\right)^{-1}\sum_{\tau=1}^{N_k}\phi\left(s_h^{(k, \tau)}, a_h^{(k, \tau)}\right) \phi\left(s_h^{(k, \tau)}, a_h^{(k, \tau)}\right)^\top\right)}\\
    &=H\sqrt{N_k/\lambda} \sqrt{\mathrm{Tr}\left(\left(\Lambda_h^k\right)^{-1}\left(\Lambda_h^k-\lambda\cdot \mathbf{I}_d\right)\right)}\\
    &\leq H\sqrt{N_k/\lambda} \sqrt{\mathrm{Tr}\left(\left(\Lambda_h^k\right)^{-1}\left(\Lambda_h^k\right)\right)}\\
    &=H\sqrt{N_kd/\lambda} 
\end{align*}
\end{proof}
Now, we begin to upper bound the difference between $\mathbb{B}_h \widehat{V}_{h+1}$ and $\widehat{\mathbb{B}}_h \widehat{V}_{h+1}$. In the following, for brevity, we write $w_{h,i}:=w_{h,i}(\widehat V_{h+1})$, $\nu_{h,i}^k:=\nu_{h,i}^k(\widehat V_{h+1})$,
$\mathbf{w}_{h}:=\mathbf{w}_{h}(\widehat V_{h+1})$ and
$\boldsymbol{\nu}_{h}^k:=\boldsymbol{\nu}_{h}^k(\widehat V_{h+1})$. For all $h \in[H]$, and all $(s, a) \in \mathcal{S} \times \mathcal{A}$, we have
\begin{align}
    \left|\left(\mathbb{B}_h \widehat{V}_{h+1}\right)(s, a)-\left(\widehat{\mathbb{B}}_h \widehat{V}_{h+1}\right)(s, a)\right|&=\left|\sum_{i=1}^d\phi_i(s, a)\left(w_{h,i}-\widehat{w}_{h,i}\right)\right|\nonumber\\
    &=\left|\sum_{i=1}^d\phi_i(s, a)\left(\min_{k\in[K]}\nu^k_{h,i}-\min_{k\in[K]}\widehat{\nu}^k_{h,i}\right)\right|\nonumber\\
    &\leq\sum_{i=1}^d\phi_i(s, a)\left|\min_{k\in[K]}\nu^k_{h,i}-\min_{k\in[K]}\widehat{\nu}^k_{h,i}\right|\nonumber\\
    &\leq \sum_{i=1}^d\phi_i(s, a)\max_{k\in[K]}\underbrace{\left|\mathbf{e}_i^\top\left(\boldsymbol{\nu}^k_{h}-\widehat{\boldsymbol{\nu}}^k_{h}\right)\right|}_{\text{(I)}}\label{eq:difference}.
\end{align}

For any bounded measurable function $V$, define the source-specific Bellman operator by
\[
(\mathbb B_h^k V)(s,a)=r_h^k(s,a)+\int_{\mathcal S}V(s')\,P_h^k(ds'\mid s,a)
=\boldsymbol{\phi}(s,a)^\top\boldsymbol{\nu}_h^k(V).
\]
Then we start to focus on (I):
{\footnotesize\begin{align*}
    \left|\mathbf{e}_i^\top\left(\boldsymbol{\nu}^k_{h}-\widehat{\boldsymbol{\nu}}^k_{h}\right)\right|=&\left|\mathbf{e}_i^\top\boldsymbol{\nu}^k_{h}-\mathbf{e}_i^\top\left(\Lambda_h^k\right)^{-1}\left(\sum_{\tau=1}^{N_k} \boldsymbol{\phi}\left(s_h^{(k, \tau)}, a_h^{(k, \tau)}\right) \cdot\left(r_h^{(k, \tau)}+\widehat{V}_{h+1}\left(s_{h+1}^{(k, \tau)}\right)\right)\right)\right|\\
    \le&\underbrace{\left|\mathbf{e}_i^\top\boldsymbol{\nu}^k_{h}-\mathbf{e}_i^\top\left(\Lambda_h^k\right)^{-1}\left(\sum_{\tau=1}^{N_k} \boldsymbol{\phi}\left(s_h^{(k, \tau)}, a_h^{(k, \tau)}\right) \cdot \left(\mathbb{B}_h^k\widehat{V}_{h+1}\right)\left(s_h^{(k, \tau)}, a_h^{(k, \tau)}\right)\right)\right|}_{\text{(i)}}\\
    &+\underbrace{\left|\mathbf{e}_i^\top\left(\Lambda_h^k\right)^{-1}\left(\sum_{\tau=1}^{N_k} \boldsymbol{\phi}\left(s_h^{(k, \tau)}, a_h^{(k, \tau)}\right) \cdot \left(r_h^{(k, \tau)}+\widehat{V}_{h+1}\left(s_{h+1}^{(k, \tau)}\right)-\left(\mathbb{B}_h^k\widehat{V}_{h+1}\right)\left(s_h^{(k, \tau)}, a_h^{(k, \tau)}\right)\right)\right)\right|}_{\text{(ii)}}
\end{align*}}

In the sequel, we begin to upper bound these two terms~(i) and term~(ii) for fixed $i\in[d]$. First we have $\widehat{V}_{h+1} \in[0, H-1]$, then by Lemma~\ref{lemma-bounded-weights}, we have $\left\|\boldsymbol{\nu}_h^k\right\|_2 \leq H\sqrt{d}$. Then the term~(i) can be bounded by
\begin{align}
    \left|\text{(i)}\right|&=\left|\mathbf{e}_i^\top\boldsymbol{\nu}^k_{h}-\mathbf{e}_i^\top\left(\Lambda_h^k\right)^{-1}\left(\sum_{\tau=1}^{N_k} \boldsymbol{\phi}\left(s_h^{(k, \tau)}, a_h^{(k, \tau)}\right) \cdot \left(\mathbb{B}_h^k\,\widehat{V}_{h+1}\right)\left(s_h^{(k, \tau)}, a_h^{(k, \tau)}\right)\right)\right|\nonumber\\
    &=\left|\mathbf{e}_i^\top\boldsymbol{\nu}^k_{h}-\mathbf{e}_i^\top\left(\Lambda_h^k\right)^{-1}\left(\sum_{\tau=1}^{N_k} \boldsymbol{\phi}\left(s_h^{(k, \tau)}, a_h^{(k, \tau)}\right) \cdot \boldsymbol{\phi}\left(s_h^{(k, \tau)}, a_h^{(k, \tau)}\right)^\top\boldsymbol{\nu}_h^k\right)\right|\nonumber\\
    &=\left|\mathbf{e}_i^\top\boldsymbol{\nu}^k_{h}-\mathbf{e}_i^\top\left(\Lambda_h^k\right)^{-1}\left(\Lambda_h^k-\lambda\cdot \mathbf{I}_d\right)\boldsymbol{\nu}_h^k\right|\nonumber\\
    &=\lambda\cdot\left|\mathbf{e}_i^\top\left(\Lambda_h^k\right)^{-1}\boldsymbol{\nu}_h^k\right|\nonumber\\
    &\leq\lambda\cdot\left\|\boldsymbol{\nu}_h^k\right\|_{\left(\Lambda_h^k\right)^{-1}} \cdot\left\|\mathbf{e}_i\right\|_{\left(\Lambda_h^k\right)^{-1}}\nonumber\\
    & = \lambda \sqrt{\left(\boldsymbol{\nu}_h^k\right)^{\top}\left(\Lambda_h^k\right)^{-1}\boldsymbol{\nu}_h^k} \cdot\left\|\mathbf{e}_i\right\|_{\left(\Lambda_h^k\right)^{-1}}\nonumber \\
    & \leq \lambda\left\|\left(\Lambda_h^k\right)^{-1}\right\|_{\mathrm{op}}^{1 / 2} \cdot\left\|\boldsymbol{\nu}_h^k\right\|_2 \cdot\left\|\mathbf{e}_i\right\|_{\left(\Lambda_h^k\right)^{-1}}\nonumber\\
    &\leq H\sqrt{d\lambda}\left\|\mathbf{e}_i\right\|_{\left(\Lambda_h^k\right)^{-1}}\label{eq:(i)}
\end{align}
where $\|\cdot\|_{\text {op }}$ denotes the matrix operator norm and we use the fact that $\left\|\left(\Lambda_h^k\right)^{-1}\right\|_{\mathrm{op}} \leq \lambda^{-1}$.

It remains to upper bound term~(ii). For fixed $k\in[K]$, any $h \in[H]$, any $\tau \in[N_k]$, and any measurable function $V: \mathcal{S} \rightarrow[0, H-1]$, we define the random variable
$$
\epsilon_{h}^{(k,\tau)}(V)=r_h^{(k, \tau)}+V\left(s_{h+1}^{(k, \tau)}\right)-\left(\mathbb{B}_h^k\,V\right)\left(s_h^{(k, \tau)}, a_h^{(k, \tau)}\right)
$$
Then by the Cauchy-Schwarz inequality, term~(ii) can be upper bounded by
\begin{align}
    \left|\text{(ii)}\right|&=\left|\mathbf{e}_i^{\top}\left(\Lambda_h^k\right)^{-1}\left(\sum_{\tau=1}^{N_k} \boldsymbol{\phi}\left(s_h^{(k, \tau)}, a_h^{(k, \tau)}\right) \cdot\epsilon_{h}^{(k,\tau)}(\widehat{V}_{h+1})\right)\right|\nonumber\\
    &\leq\underbrace{\left\|\sum_{\tau=1}^{N_k} \boldsymbol{\phi}\left(s_h^{(k, \tau)}, a_h^{(k, \tau)}\right) \cdot\epsilon_{h}^{(k,\tau)}(\widehat{V}_{h+1})\right\|_{\left(\Lambda_h^k\right)^{-1}} }_{\text{(iii)}}\cdot\|\mathbf{e}_i\|_{\left(\Lambda_h^k\right)^{-1}}\label{eq:(ii)}
\end{align}
In the sequel, we upper bound term~(iii) via concentration inequalities. An obstacle is that $\widehat V_{h+1}$ is constructed from the data at stages $h+1,\ldots,H$, including the next states $s_{h+1}^{(k,\tau)}$ that enter $\epsilon_h^{(k,\tau)}$, so $\widehat V_{h+1}$ and the noise terms are dependent. To this end, we resort to uniform concentration inequalities to upper bound
$$
\sup _{V \in \mathcal{V}_{h+1}(R,B,\lambda)}\left\|\sum_{\tau=1}^{N_k} \boldsymbol{\phi}\left(s_h^{(k, \tau)}, a_h^{(k, \tau)}\right) \cdot \epsilon_{h}^{(k,\tau)}\left(V\right)\right\|_{\left(\Lambda_h^k\right)^{-1}} 
$$
for each $h\in[H]$, where it holds that $\widehat{V}_{h+1} \in \mathcal{V}_{h+1}(R,B,\lambda)$. For all $h\in[H]$, we define the function class:
$$
\mathcal{V}_h(R, B, \lambda)=\left\{V_h(s ; \boldsymbol{\theta}, \beta,\Sigma_1,\cdots,\Sigma_K): \mathcal{S} \rightarrow[0, H] \text { with }\|\boldsymbol{\theta}\|_2 \leq R, \beta \in[0, B], \Sigma_1,\dots,\Sigma_K \succeq \lambda\mathbf I_d\right\}
$$
$$
\text { where } V_h(s ; \boldsymbol{\theta}, \beta,\Sigma_1,\cdots,\Sigma_K)=\max _{a \in \mathcal{A}}\left\{\min \left\{\boldsymbol{\phi}(s, a)^{\top} \boldsymbol{\theta}-\beta\sum_{i=1}^d\phi_i(s,a)\max_{k\in[K]}\|\mathbf{e}_i\|_{\left(\Sigma_k\right)^{-1}}, H-h+1\right\}^{+}\right\}
$$
For every admissible parameter tuple, the expression being maximized is measurable in $(s,a)$ and continuous in $a$ for every fixed $s$. Since $\mathcal A$ is a nonempty compact metric space, the maximum is attained. The measurable maximum theorem implies that every function in $\mathcal V_h(R,B,\lambda)$ is measurable. For convenience, define $\mathcal V_{H+1}(R,B,\lambda):=\{0\}$. For all $\varepsilon>0$ and all $h\in[H]$, let $\mathcal N_h(\varepsilon,R,B,\lambda)\subseteq\mathcal V_h(R,B,\lambda)$ be a fixed $\varepsilon$-cover of $\mathcal V_h(R,B,\lambda)$ with respect to the supremum norm that has minimal cardinality among all $\varepsilon$-covers contained in $\mathcal V_h(R,B,\lambda)$, i.e., for all $V \in \mathcal{V}_h(R,B,\lambda)$, there exists $V^{\dagger} \in \mathcal{N}_h(\varepsilon,R,B,\lambda)$ such that
\[
\sup _{s \in \mathcal{S}}\left|V(s)-V^{\dagger}(s)\right| \leq \varepsilon.
\]
In particular, every element of $\mathcal N_{h+1}(\varepsilon,R,B,\lambda)$ is measurable and takes values in $[0,H-h]\subseteq[0,H-1]$. We also set $\mathcal N_{H+1}(\varepsilon,R,B,\lambda):=\{0\}$. By Lemma~\ref{lemma-bounded-weights}, we have for any $k$, $\left\|\widehat{\boldsymbol{\nu}}_h^k\right\|_2 \leq H \sqrt{N_k d / \lambda}$. Then for any component $i\in[d]$,
\[
\left|\left(\widehat{\mathbf{w}}_h\right)_i\right|=\left|\min _k\left(\widehat{\boldsymbol{\nu}}_h^k\right)_i\right| \leq \max _k\left|\left(\widehat{\boldsymbol{\nu}}_h^k\right)_i\right| \leq \max _k\left\|\widehat{\boldsymbol{\nu}}_h^k\right\|_2
\]
and
\[
\left\|\widehat{\mathbf{w}}_h\right\|_2=\sqrt{\sum_{i=1}^d\left|\left(\widehat{\mathbf{w}}_h\right)_i\right|^2}\leq \sqrt{d}H \sqrt{N_{\max} d / \lambda}=Hd\sqrt{N_{\max}/\lambda},
\]
where $N_{\max}=\max _k N_k$. Hence, for all $h\in[H]$, we have
$$
\widehat{V}_{h+1} \in \mathcal{V}_{h+1}\left(R_0,B_0,\lambda\right), \quad \text { where } \quad R_0=Hd\sqrt{N_{\max}/\lambda}\quad \text { and } \quad B_0=2\beta
$$
For notational simplicity, we use $\mathcal{V}_{h+1}$ and $\mathcal{N}_{h+1}(\varepsilon)$ to denote $\mathcal{V}_{h+1}\left(R_0, B_0,\lambda\right)$ and $\mathcal{N}_{h+1}\left(\varepsilon ; R_0, B_0,\lambda\right)$, respectively. As it holds that $\widehat{V}_{h+1} \in \mathcal{V}_{h+1}$ and $\mathcal{N}_{h+1}(\varepsilon)$ is an $\varepsilon$-cover of $\mathcal{V}_{h+1}$, there exists a function $V_{h+1}^{\dagger} \in \mathcal{N}_{h+1}(\varepsilon)$ such that
\begin{equation}
    \sup _{s \in \mathcal{S}}\left|\widehat{V}_{h+1}(s)-V_{h+1}^{\dagger}(s)\right| \leq \varepsilon\label{eq:minimal cover}
\end{equation}
Now, we consider for fixed $k\in[K]$, all $\tau\in[N_k]$, and all $h\in[H]$,
\begin{align}
    &\left|\epsilon_{h}^{(k,\tau)}\left(\widehat{V}_{h+1}\right)-\epsilon_{h}^{(k,\tau)}\left(V_{h+1}^{\dagger}\right)\right|\nonumber\\
    =&\Bigl| r_h^{(k, \tau)}+\widehat{V}_{h+1}\left(s_{h+1}^{(k, \tau)}\right)-\left(\mathbb{B}_h^k\,\widehat{V}_{h+1}\right)\left(s_h^{(k, \tau)}, a_h^{(k, \tau)}\right) -r_h^{(k, \tau)}-V_{h+1}^{\dagger}\left(s_{h+1}^{(k, \tau)}\right)+\left(\mathbb{B}_h^k\,V_{h+1}^{\dagger}\right)\left(s_h^{(k, \tau)}, a_h^{(k, \tau)}\right) \Bigr|\nonumber\\
    \leq&\left|\widehat{V}_{h+1}\left(s_{h+1}^{(k, \tau)}\right) -V_{h+1}^{\dagger}\left(s_{h+1}^{(k, \tau)}\right) \right|+\left|\left(\mathbb{B}_h^k\,\widehat{V}_{h+1}\right)\left(s_h^{(k, \tau)}, a_h^{(k, \tau)}\right)-\left(\mathbb{B}_h^k\,V_{h+1}^{\dagger}\right)\left(s_h^{(k, \tau)}, a_h^{(k, \tau)}\right) \right|\nonumber\\
    =&\left|\widehat{V}_{h+1}\left(s_{h+1}^{(k, \tau)}\right)-V_{h+1}^{\dagger}\left(s_{h+1}^{(k, \tau)}\right)\right|+\left|
    \int_{\mathcal S}
    \left[\widehat V_{h+1}(s')-V^\dagger_{h+1}(s')\right]
    P_h^k\left(ds'\mid s_h^{(k,\tau)},a_h^{(k,\tau)}\right)
    \right|\nonumber\\
    \leq&\varepsilon+\varepsilon=2\varepsilon
    \label{eq:epsilon2}
\end{align}
Now, recall the expression of term~(iii). By the Cauchy-Schwarz inequality and $(x+y)^2\le 2x^2+2y^2$, for any two vectors $a, b \in \mathbb{R}^d$ and any positive definite matrix $\Lambda \in \mathbb{R}^{d \times d}$, it holds that $\|a+b\|_{\Lambda}^2 \leq 2 \cdot\|a\|_{\Lambda}^2+2 \cdot\|b\|_{\Lambda}^2$. Hence, for fixed $k\in[K]$ and all $h \in[H]$, we have
\begin{align}
\left|(\mathrm{iii})\right|^2= & \left\|\sum_{\tau=1}^{N_k} \boldsymbol{\phi}\left(s_h^{(k, \tau)}, a_h^{(k, \tau)}\right) \cdot \epsilon_{h}^{(k,\tau)}\left(\widehat{V}_{h+1}\right)\right\|_{(\Lambda_h^k)^{-1}}^2\nonumber \\
\leq & 2\left\|\sum_{\tau=1}^{N_k} \boldsymbol{\phi}\left(s_h^{(k, \tau)}, a_h^{(k, \tau)}\right) \cdot \epsilon_{h}^{(k,\tau)}\left(V_{h+1}^{\dagger}\right)\right\|_{\left(\Lambda_h^k\right)^{-1}}^2 \nonumber\\
& +2\left\|\sum_{\tau=1}^{N_k} \boldsymbol{\phi}\left(s_h^{(k, \tau)}, a_h^{(k, \tau)}\right) \cdot\left(\epsilon_{h}^{(k,\tau)}\left(\widehat{V}_{h+1}\right)-\epsilon_{h}^{(k,\tau)}\left(V_{h+1}^{\dagger}\right)\right)\right\|_{\left(\Lambda_h^k\right)^{-1}}^2\label{eq:(iii)^2}
\end{align}
The second term on the RHS of \eqref{eq:(iii)^2} can be upper bounded by
\begin{align}
& 2\cdot\left\|\sum_{\tau=1}^{N_k} \boldsymbol{\phi}\left(s_h^{(k, \tau)}, a_h^{(k, \tau)}\right) \cdot\left(\epsilon_{h}^{(k,\tau)}\left(\widehat{V}_{h+1}\right)-\epsilon_{h}^{(k,\tau)}\left(V_{h+1}^{\dagger}\right)\right)\right\|_{\left(\Lambda_h^k\right)^{-1}}^2 \nonumber\\
& =2 \sum_{\tau, \tau^{\prime}=1}^{N_k} \boldsymbol{\phi}\left(s_h^{(k, \tau)}, a_h^{(k, \tau)}\right)^{\top}\left(\Lambda_h^k\right)^{-1} \boldsymbol{\phi}\left(s_h^{\left(k, \tau^{\prime}\right)}, a_h^{\left(k, \tau^{\prime}\right)}\right) \nonumber\\
& \quad \times\left(\epsilon_{h}^{(k,\tau)}\left(\hat{V}_{h+1}\right)-\epsilon_{h}^{(k,\tau)}\left(V_{h+1}^{\dagger}\right)\right)\left(\epsilon_h^{(k,\tau')}\left(\hat{V}_{h+1}\right)-\epsilon_h^{(k,\tau')}\left(V_{h+1}^{\dagger}\right)\right) \nonumber\\
& \leq 8 \varepsilon^2 \sum_{\tau, \tau^{\prime}=1}^{N_k}\left|\boldsymbol{\phi}\left(s_h^{(k, \tau)}, a_h^{(k, \tau)}\right)^{\top}\left(\Lambda_h^k\right)^{-1} \boldsymbol{\phi}\left(s_h^{\left(k, \tau^{\prime}\right)}, a_h^{\left(k, \tau^{\prime}\right)}\right)\right| \nonumber\\
& \leq 8 \varepsilon^2 \cdot N_k^2 \cdot\left\|\left(\Lambda_h^k\right)^{-1}\right\|_{\mathrm{op}}\nonumber \\
& \leq \frac{8 \varepsilon^2 \cdot N_k^2}{\lambda}\label{eq:second_term_of_(iii)^2}
\end{align}
Now, combining \eqref{eq:(iii)^2} and \eqref{eq:second_term_of_(iii)^2}, for fixed $k\in[K]$ and all $h\in[H]$, we have 
\begin{equation}
\left|(\mathrm{iii})\right|^2 \leq 2 \sup _{V \in \mathcal{N}_{h+1}(\varepsilon)}\left\|\sum_{\tau=1}^{N_k} \boldsymbol{\phi}\left(s_h^{(k, \tau)}, a_h^{(k, \tau)}\right) \cdot \epsilon_{h}^{(k,\tau)}\left(V\right)\right\|_{\left(\Lambda_h^k\right)^{-1}}^2+\frac{8 \varepsilon^2 \cdot N_k^2}{\lambda} \label{eq:(iii)^2(2)}
\end{equation}
Since the right-hand side of Equation~\eqref{eq:(iii)^2(2)} does not depend on the data-driven estimators $\widehat{Q}_h$ and $\widehat{V}_{h+1}$, the first term can be bounded above by applying uniform concentration inequalities. We use the following lemma.
\begin{lemma}[Concentration of self-normalized processes; \citealp{abbasi2011improved}]
\label{le:concentration}
    Let $\left\{\mathcal{F}_t\right\}_{t=0}^{\infty}$ be a filtration and $\left\{\epsilon_t\right\}_{t=1}^{\infty}$ be an $\mathbb{R}$-valued stochastic process such that $\epsilon_t$ is $\mathcal{F}_t$-measurable for all $t \geq 1$. Moreover, suppose that conditioning on $\mathcal{F}_{t-1}, \epsilon_t$ is a zero-mean and $\Sigma$-sub-Gaussian random variable for all $t \geq 1$, that is,
$$
\mathbb{E}\left[\epsilon_t \mid \mathcal{F}_{t-1}\right]=0, \quad \mathbb{E}\left[\exp \left(\lambda \epsilon_t\right) \mid \mathcal{F}_{t-1}\right] \leq \exp \left(\lambda^2 \Sigma^2 / 2\right), \quad \forall \lambda \in \mathbb{R}
$$
Meanwhile, let $\left\{\boldsymbol{\phi}_t\right\}_{t=1}^{\infty}$ be an $\mathbb{R}^d$-valued stochastic process such that $\boldsymbol{\phi}_t$ is $\mathcal{F}_{t-1}$-measurable for all $t \geq 1$. Also, let $M_0 \in \mathbb{R}^{d \times d}$ be a deterministic positive-definite matrix and
$$
M_t=M_0+\sum_{s=1}^t \boldsymbol{\phi}_s \boldsymbol{\phi}_s^{\top}
$$
for all $t \geq 1$. For all $\delta\in(0,1)$, it holds that
$$
\left\|\sum_{s=1}^t \boldsymbol{\phi}_s \epsilon_s\right\|_{M_t^{-1}}^2 \leq 2 \Sigma^2 \cdot \log \left(\frac{\operatorname{det}\left(M_t\right)^{1 / 2} \cdot \operatorname{det}\left(M_0\right)^{-1 / 2}}{\delta}\right)
$$
with probability at least $1-\delta$, simultaneously for all $t\ge1$.
\end{lemma}
Recall the definition of $\epsilon_{h}^{(k,\tau)}(V)$:
\begin{align*}
\epsilon_{h}^{(k,\tau)}(V)&=r_h^{(k, \tau)}+V\left(s_{h+1}^{(k, \tau)}\right)-\left(\mathbb{B}_h^k V\right)\left(s_h^{(k, \tau)}, a_h^{(k, \tau)}\right)\\
&=r_h^{(k, \tau)}+V\left(s_{h+1}^{(k, \tau)}\right)-\boldsymbol{\phi}\left(s_h^{(k, \tau)}, a_h^{(k, \tau)}\right)^{\top} \left(\boldsymbol{\theta}_h^k+\int_{\mathcal{S}} V\left(s^{\prime}\right) \boldsymbol{\mu}_h^k\left(ds^{\prime}\right)\right)
\end{align*}
For fixed $h\in[H]$ and $k\in[K]$, define the filtration $\{\mathcal F_{h,k}^{\tau}\}_{\tau=0}^{N_k}$ by
\[
\mathcal F_{h,k}^{\tau}
=
\sigma\left(
\left\{(s_h^{(k,j)},a_h^{(k,j)})\right\}_{j=1}^{(\tau+1)\wedge N_k},
\left\{(r_h^{(k,j)},s_{h+1}^{(k,j)})\right\}_{j=1}^{\tau}
\right),
\]
where $(\tau+1)\wedge N_k:=\min\{\tau+1,N_k\}$ and the second collection is understood to be empty when $\tau=0$. By construction, for each $\tau\in[N_k]$, $\boldsymbol{\phi}(s_h^{(k,\tau)},a_h^{(k,\tau)})$ is $\mathcal F_{h,k}^{\tau-1}$-measurable, whereas $\epsilon_h^{(k,\tau)}(V)$ is $\mathcal F_{h,k}^{\tau}$-measurable. By the source-wise compliance condition in Section~\ref{sec:linear-mdp}, for any fixed bounded measurable function $V$, we have
\begin{align*}
\mathbb E_{\mathcal D_k}\left[
\epsilon_h^{(k,\tau)}(V)
\mid \mathcal F_{h,k}^{\tau-1}
\right]
&=
\mathbb E_{\mathcal D_k}\left[
r_h^{(k,\tau)}+V(s_{h+1}^{(k,\tau)})
\mid \mathcal F_{h,k}^{\tau-1}
\right]-
(\mathbb B_h^k V)
\left(s_h^{(k,\tau)},a_h^{(k,\tau)}\right)\\
&=
r_h^k\left(s_h^{(k,\tau)},a_h^{(k,\tau)}\right)
+
\int_{\mathcal S}V(s')\,
P_h^k\left(ds'\mid s_h^{(k,\tau)},a_h^{(k,\tau)}\right)\\
&\quad-
(\mathbb B_h^k V)
\left(s_h^{(k,\tau)},a_h^{(k,\tau)}\right)\\
&=0.
\end{align*}
Moreover, as it holds that $r_{h}^{(k,\tau)} \in[0,1]$ and $V \in[0, H-1]$, we have $r_h^{(k, \tau)}+V\left(s_{h+1}^{(k, \tau)}\right)\in[0, H]$. Meanwhile, we have $\left(\mathbb{B}_h^k V\right)\left(s_h^{(k, \tau)}, a_h^{(k, \tau)}\right)\in[0, H]$, which implies $\left|\epsilon_{h}^{(k,\tau)}(V)\right| \leq H$. Hence, for the fixed $h \in[H]$, fixed $k\in[K]$, and all $\tau \in[N_k]$, the random variable $\epsilon_{h}^{(k,\tau)}(V)$ is mean-zero and $H$-sub-Gaussian conditioning on $\mathcal{F}_{h, k}^{\tau-1}$.


By using Lemma~\ref{le:concentration} with $M_0=\lambda\cdot \mathbf{I}_d$ and $M_{N_k}=\sum_{\tau=1}^{N_k} \boldsymbol{\phi}\left(s_h^{(k, \tau)}, a_h^{(k, \tau)}\right) \boldsymbol{\phi}\left(s_h^{(k, \tau)}, a_h^{(k, \tau)}\right)^{\top}+\lambda\cdot \mathbf{I}_d$, for fixed $k\in$[K], fixed $h\in[H]$, and the fixed measurable function $V:\mathcal{S}\rightarrow[0,H-1]$, we have 
$$
\mathbb{P}_{\mathcal{D}_k}\left(\left\|\sum_{\tau=1}^{N_k} \boldsymbol{\phi}\left(s_h^{(k, \tau)}, a_h^{(k, \tau)}\right) \cdot \epsilon_{h}^{(k,\tau)}(V)\right\|_{\left(\Lambda_h^k\right)^{-1}}^2>2 H^2 \cdot \log \left(\frac{\operatorname{det}\left(\Lambda_h^k\right)^{1 / 2}}{\delta \cdot \operatorname{det}(\lambda \cdot \mathbf{I}_d)^{1 / 2}}\right)\right) \leq \delta
$$
for all $\delta \in(0,1)$. Moreover,
$$
\left\|\Lambda_h^k\right\|_{\mathrm{op}}=\left\|\lambda\cdot\mathbf{I}_d+\sum_{\tau=1}^{N_k} \boldsymbol{\phi}\left(s_h^{(k, \tau)}, a_h^{(k, \tau)}\right) \boldsymbol{\phi}\left(s_h^{(k, \tau)}, a_h^{(k, \tau)}\right)^{\top}\right\|_{\mathrm{op}} \leq \lambda+N_k .
$$
Thus,
$$
\operatorname{det}\left(\Lambda_h^k\right) \leq\left(\lambda+N_k\right)^d, \quad \operatorname{det}\left(\lambda\cdot\mathbf{I}_d\right)=\lambda^d
$$
Since
$$
\begin{aligned}
2 H^2 \log \left(\frac{\operatorname{det}\left(\Lambda_h^k\right)^{1 / 2}}{\delta \operatorname{det}\left(\lambda\cdot\mathbf{I}_d\right)^{1 / 2}}\right) & \leq 2 H^2 \log \left(\frac{\left(\lambda+N_k\right)^{d / 2}}{\delta \lambda^{d / 2}}\right) \\
& =2 H^2\left[\log \left(\frac{1}{\delta}\right)+\frac{d}{2} \log \left(1+\frac{N_k}{\lambda}\right)\right] \\
& =H^2\left[2 \log \left(\frac{1}{\delta}\right)+d \log \left(1+\frac{N_k}{\lambda}\right)\right]
\end{aligned}
$$
then for any fixed function $V$, fixed $k\in[K]$ and fixed $h\in[H]$, and any $\delta\in(0,1)$, we have
$$
\mathbb{P}_{\mathcal{D}_k}\left(\left\|\sum_{\tau=1}^{N_k} \boldsymbol{\phi}\left(s_h^{(k, \tau)}, a_h^{(k, \tau)}\right) \cdot \epsilon_{h}^{(k,\tau)}(V)\right\|_{\left(\Lambda_h^k\right)^{-1}}^2>H^2\left[2 \log \left(\frac{1}{\delta}\right)+d \log \left(1+\frac{N_k}{\lambda}\right)\right]\right) \leq \delta
$$
Then by applying the union bound, for any fixed $k\in[K]$ and any fixed $h\in[H]$, we have
\begin{equation*}
\begin{split}
    \mathbb{P}_{\mathcal{D}_k}\Bigg( & \sup _{V \in \mathcal{N}_{h+1}(\varepsilon)} \left\|\sum_{\tau=1}^{N_k} \boldsymbol{\phi}\left(s_h^{(k, \tau)}, a_h^{(k, \tau)}\right) \cdot \epsilon_{h}^{(k,\tau)}(V)\right\|_{\left(\Lambda_h^k\right)^{-1}}^2 \\
    & > H^2\left[2 \log \left(\frac{1}{\delta}\right)+d \log \left(1+\frac{N_k}{\lambda}\right)\right] \Bigg) \leq \delta \left| \mathcal{N}_{h+1}(\varepsilon)\right|
\end{split}
\end{equation*}
For all $\xi \in(0,1)$ and all $\varepsilon>0$, we set $\delta=\xi /\left(K\cdot H \cdot\left|\mathcal{N}_{h+1}(\varepsilon)\right|\right)$ (union bound over $k,h$ and the cover). Hence, for any fixed $k\in[K]$ and any fixed $h \in[H]$, it holds that
\begin{equation} \label{eq:sup}
\begin{split}
    \sup _{V \in \mathcal{N}_{h+1}(\varepsilon)} & \left\|\sum_{\tau=1}^{N_k} \boldsymbol{\phi}\left(s_h^{(k, \tau)}, a_h^{(k, \tau)}\right) \epsilon_{h}^{(k,\tau)}(V)\right\|_{\left(\Lambda_h^k\right)^{-1}}^2 \\
    & \leq H^2 \left[ 2 \log \left( \frac{K H \left|\mathcal{N}_{h+1}(\varepsilon)\right|}{\xi} \right) + d \log \left(1 + \frac{N_k}{\lambda}\right) \right]
\end{split}
\end{equation}
with probability at least $1-\xi/\left(K\cdot H\right)$, which is taken with respect to $\mathbb{P}_\mathcal{D}$.
Now, combining \eqref{eq:(iii)^2(2)} and \eqref{eq:sup}, we have for any fixed $k\in[K]$,
\begin{equation}
\begin{split}
\mathbb{P}_{\mathcal{D}}\Bigg( \bigcap_{h \in [H]} \Bigg\{ & \left\| \sum_{\tau=1}^{N_k} \boldsymbol{\phi}\left(s_h^{(k, \tau)}, a_h^{(k, \tau)}\right) \epsilon_{h}^{(k,\tau)}(\widehat{V}_{h+1}) \right\|_{(\Lambda_h^k)^{-1}}^2 \\
& \leq 2H^2 \left[ 2 \log\left(\frac{K H |\mathcal{N}_{h+1}(\varepsilon)|}{\xi}\right) + d \log\left(1+\frac{N_k}{\lambda}\right) \right] + \frac{8\varepsilon^2 N_k^2}{\lambda} \Bigg\} \Bigg) \geq 1 - \frac{\xi}{K}
\end{split}
\end{equation}
which is derived from the union bound. Since $N_{\max}=\max_{k\in[K]} N_k$, then for fixed $k \in[K]$, for all $h \in[H]$, it holds that
\begin{equation*}
\begin{split}
    & \left\|\sum_{\tau=1}^{N_k} \boldsymbol{\phi}\left(s_h^{(k, \tau)}, a_h^{(k, \tau)}\right) \epsilon_{h}^{(k,\tau)}\left(\widehat{V}_{h+1}\right)\right\|_{\left(\Lambda_h^k\right)^{-1}}^2 \\
    & \quad \leq 2 H^2\left[ 2 \log \left(\frac{KH}{\xi}\right) + 2 \log \left|\mathcal{N}_{h+1}(\varepsilon)\right| + d \log \left(1+\frac{N_{\max}}{\lambda}\right)\right] + \frac{8 \varepsilon^2 N_{\max}^2}{\lambda}
\end{split}
\end{equation*}
with probability at least $1-\xi/K$.

In the sequel, we set $\varepsilon=dH/N_{\max}$ and $\lambda=1$. For any fixed $k\in[K]$ and all $h\in[H]$, it holds that
\begin{equation}
    \begin{split}
        &\left\|\sum_{\tau=1}^{N_k} \boldsymbol{\phi}\left(s_h^{(k, \tau)}, a_h^{(k, \tau)}\right) \cdot \epsilon_{h}^{(k,\tau)}\left(\widehat{V}_{h+1}\right)\right\|_{\left(\Lambda_h^k\right)^{-1}}^2 \\
        &\quad \leq 2 H^2\left[ 2 \log \left(\frac{KH}{\xi}\right) + 2 \log \left|\mathcal{N}_{h+1}(\varepsilon)\right| + d \log \left(1+N_{\max}\right)+4d^2\right] 
    \end{split}
\label{eq:square}
\end{equation}
with probability at least $1-\xi/K$. Recall the definition of function class:
$$
\mathcal{V}_h(R, B, \lambda)=\left\{V_h(s ; \boldsymbol{\theta}, \beta,\Sigma_1,\cdots,\Sigma_K): \mathcal{S} \rightarrow[0, H] \text { with }\|\boldsymbol{\theta}\|_2 \leq R, \beta \in[0, B], \Sigma_1,\dots,\Sigma_K \succeq \lambda\mathbf I_d\right\}
$$
$$
\text { where } V_h(s ; \boldsymbol{\theta}, \beta,\Sigma_1,\cdots,\Sigma_K)=\max _{a \in \mathcal{A}}\left\{\min \left\{\boldsymbol{\phi}(s, a)^{\top} \boldsymbol{\theta}-\beta\sum_{i=1}^d\phi_i(s,a)\max_{k\in[K]}\|\mathbf{e}_i\|_{\left(\Sigma_k\right)^{-1}}, H-h+1\right\}^{+}\right\}
$$
In order to upper bound the $\varepsilon$-covering number of this function class, we use the following lemma:
\begin{lemma}[$\varepsilon$-Covering Number]
\label{lemma:covering-number-used}
For all $h \in[H]$ and all $\varepsilon>0$, we have
\[
\log \left|\mathcal{N}_h(\varepsilon, R, B, \lambda)\right| \leq d \cdot \log \left(1+\frac{6 R}{\varepsilon}\right)+\log \left(1+\frac{6 B}{\varepsilon\sqrt{\lambda} }\right)+d \cdot \log \left(1+\frac{6 B}{\varepsilon\sqrt{\lambda}}\right)
\]
\end{lemma}
\begin{proof}
Define the vector function $t(\boldsymbol{\Sigma}) \in \mathbb{R}^d$ element-wise by:
\[
t_i(\boldsymbol{\Sigma})=\max_{k\in[K]} \|\mathbf{e}_i\|_{(\Sigma_k)^{-1}}.
\]
The value function takes the form:
\[
V_h(s ; \boldsymbol{\theta}, \beta,\boldsymbol{\Sigma})=\max _{a \in \mathcal{A}}\left\{\min \left\{\boldsymbol{\phi}(s, a)^{\top} \boldsymbol{\theta}-\beta\boldsymbol{\phi}(s, a)^{\top}t(\boldsymbol{\Sigma}), H-h+1\right\}^{+}\right\}.
\]
Since both $\min \{\cdot, C\}^{+}$ and $\max _{a \in \mathcal{A}}$ are non-expansive (Lipschitz continuous with constant 1), for any two functions $V_1, V_2 \in \mathcal{V}_h$ parameterized by $(\boldsymbol{\theta}_1, \beta_1, \boldsymbol{\Sigma}_1)$ and $(\boldsymbol{\theta}_2, \beta_2, \boldsymbol{\Sigma}_2)$, we have:
{\small
\begin{align*}
\sup_{s\in\mathcal S}|V_1(s)-V_2(s)|
&\le \sup_{s,a} \left| \boldsymbol{\phi}(s,a)^\top(\boldsymbol{\theta}_1-\boldsymbol{\theta}_2) - \left(\beta_1\boldsymbol{\phi}(s,a)^\top t(\boldsymbol{\Sigma}_1) - \beta_2\boldsymbol{\phi}(s,a)^\top t(\boldsymbol{\Sigma}_2)\right) \right| \\
&\le \sup_{s,a} \left( \|\boldsymbol{\phi}(s,a)\|_2 \|\boldsymbol{\theta}_1-\boldsymbol{\theta}_2\|_2 + \left| \boldsymbol{\phi}(s,a)^\top \left( \beta_1 t(\boldsymbol{\Sigma}_1) - \beta_2 t(\boldsymbol{\Sigma}_2) \right) \right| \right) \\
&\leq \|\boldsymbol{\theta}_1-\boldsymbol{\theta}_2\|_2 + \left|\beta_1-\beta_2\right|\|t(\boldsymbol{\Sigma}_1)\|_\infty + B\|t(\boldsymbol{\Sigma}_1)-t(\boldsymbol{\Sigma}_2)\|_\infty.
\end{align*}
}

For all $k\in[K]$, since $\Sigma_k \succeq \lambda I$, we have $(\Sigma_k)^{-1} \preceq \lambda^{-1} I$. Thus, for each coordinate $i$:
\[
0 \leq t_i(\boldsymbol{\Sigma}) = \max_{k\in[K]} \sqrt{\mathbf{e}_i^\top (\Sigma_k)^{-1} \mathbf{e}_i} \leq \sqrt{\lambda^{-1}} = \lambda^{-1/2}.
\]
This implies $\|t(\boldsymbol{\Sigma})\|_\infty \leq \lambda^{-1/2}$. We now construct $\varepsilon/3$-covers for each component:
\begin{enumerate}
    \item For $\theta$: Let $\mathcal{C}_{\theta}$ be an $\varepsilon / 3$-cover of the Euclidean ball $\{\theta \in \mathbb{R}^d \mid \| \theta\|_2 \leq R\}$ with centers in the ball. Its size is bounded by:
    \[ |\mathcal{C}_{\theta}| \leq \left(1+\frac{6R}{\varepsilon}\right)^d. \] 
    \item For $\beta$: Since the Lipschitz constant for $\beta$ is bounded by $\|t(\boldsymbol{\Sigma})\|_\infty \le \lambda^{-1/2}$, we need an $\varepsilon/3$-cover of the interval $[0, B]$ with respect to the scaled metric $\lambda^{-1/2}|\cdot|$. This is equivalent to covering with precision $\frac{\varepsilon\sqrt{\lambda}}{3}$. Let $\mathcal{C}_{\beta}\subseteq[0,B]$ be such a cover:
    \[ |\mathcal{C}_{\beta}| \leq 1+\frac{2B}{\varepsilon\sqrt{\lambda}/3} = 1+\frac{6B}{\varepsilon\sqrt{\lambda}}. \]

    \item For $t(\boldsymbol{\Sigma})$: The term $t(\boldsymbol{\Sigma})$ lies in the hypercube $[0, \lambda^{-1/2}]^d$. The relevant Lipschitz constant is $B$. We need a cover $\mathcal{C}_t$ with respect to the $\ell_\infty$-norm with precision $\frac{\varepsilon}{3B}$.
    \[ |\mathcal{C}_t| \leq \left(1+\frac{\lambda^{-1/2}}{\varepsilon/(3B)}\right)^d = \left(1+\frac{3B}{\varepsilon\sqrt{\lambda}}\right)^d \le \left(1+\frac{6B}{\varepsilon\sqrt{\lambda}}\right)^d. \]
    The grid points can be chosen in $(0,\lambda^{-1/2}]^d$. Each such point $t$ is realized by taking $\Sigma_k=\operatorname{diag}(t_1^{-2},\ldots,t_d^{-2})$ for every $k\in[K]$.
\end{enumerate}

Combining these, the set
\[
\left\{V_h(\cdot;\boldsymbol\theta,\beta,\boldsymbol\Sigma):\ \boldsymbol\theta\in\mathcal C_\theta,\ \beta\in\mathcal C_\beta,\ \Sigma_k=\operatorname{diag}(t_1^{-2},\ldots,t_d^{-2})\ \text{for all }k\in[K],\ t\in\mathcal C_t\right\}
\]
is an $\varepsilon$-cover of $\mathcal V_h(R,B,\lambda)$. Each of its elements satisfies $\|\boldsymbol\theta\|_2\le R$, $\beta\in[0,B]$, and $\Sigma_k\succeq\lambda\mathbf I_d$, so the cover is contained in $\mathcal V_h(R,B,\lambda)$. Since $\mathcal N_h(\varepsilon,R,B,\lambda)$ has minimal cardinality among such covers, its log-cardinality is bounded by the sum of the logs of the sizes:
\[
\log |\mathcal{N}_h| \leq d \log \left(1+\frac{6R}{\varepsilon}\right) + \log \left(1+\frac{6 B}{\varepsilon\sqrt{\lambda} }\right) + d \log \left(1+\frac{6 B}{\varepsilon\sqrt{\lambda}}\right).
\]
\end{proof}
Now, recall that $ \widehat{V}_{h+1} \in \mathcal{V}_{h+1}(R_0, B_0, \lambda)$ and $\varepsilon=dH/N_{\max}$, where
\[
\lambda=1, \quad 
    R_0=Hd \sqrt{\frac{N_{\max}}{\lambda}}, \quad 
    B_0=2\beta, \quad 
    \beta=c d H\sqrt{\zeta}, \quad 
    \zeta=\log\left(\frac{2dKHN_{\max}}{\xi}\right).
\]
For $h=H$, $\left|\mathcal{N}_{H+1}(\varepsilon)\right|=1$, so the desired bound is immediate. For $h\in[H-1]$, substituting these quantities into Lemma~\ref{lemma:covering-number-used} gives the same bound. Therefore, for all $h\in[H]$,
\begin{align}
\log \left|\mathcal{N}_{h+1}(\varepsilon)\right| &\leq d\cdot\log\left(1+\frac{6R_0}{\varepsilon}\right)+\log\left(1+\frac{6B_0}{\varepsilon\sqrt{\lambda}} \right)+d\cdot\log\left(1+\frac{6B_0}{\varepsilon\sqrt{\lambda}} \right)\nonumber\\
  &= d\cdot\log (1+6N_{\max}^\frac{3}{2})+(d+1)\log\left(1+12cN_{\max}\sqrt{\zeta}\right)\nonumber\\
    &\leq d \cdot \log (1+6N_{\max}^\frac{3}{2})+2d^2 \cdot \log \left(1+32d^\frac{1}{2}c^2N_{\max}^2\zeta\right),\label{eq:covering}
\end{align}
where the last inequality uses that, for $d \geq 1$ and $c \geq 1$, we have $d+1 \leq 2 d^2$ and $12 c N_{\max } \sqrt{\zeta} \leq 32 d^{1 / 2} c^2 N_{\max }^2 \zeta$. Moreover, since $\zeta>\log 2$ and we take $c \geq 1$, it follows that $32 d^{1 / 2} c^2 N_{\max }^2 \zeta \geq 1$. Hence, we have
\begin{align}
     \log \left|\mathcal{N}_{h+1}(\varepsilon)\right|&\leq d \cdot \log (1+6N_{\max}^\frac{3}{2})+2d^2 \cdot \log \left(1+32d^\frac{1}{2}c^2N_{\max}^2\zeta\right)\nonumber\\
     &\leq d^2\log \left(1+32d^\frac{1}{2}c^2N_{\max}^2\zeta\right)+2d^2 \cdot \log \left(1+32d^\frac{1}{2}c^2N_{\max}^2\zeta\right)\nonumber\\
     &\leq 3d^2 \cdot \log \left(64d^\frac{1}{2}c^2N_{\max}^2\zeta\right)\label{eq:log_cover}
\end{align}
Now, combining \eqref{eq:square} and \eqref{eq:log_cover}, for fixed $k\in[K]$ and all $h\in[H]$, it holds that
\begin{equation}
    \begin{split}
        &\left\|\sum_{\tau=1}^{N_k} \boldsymbol{\phi}\left(s_h^{(k, \tau)}, a_h^{(k, \tau)}\right) \cdot \epsilon_{h}^{(k,\tau)}\left(\widehat{V}_{h+1}\right)\right\|_{\left(\Lambda_h^k\right)^{-1}}^2 \\
        &\quad\leq 2 H^2\left(2 \log (KH / \xi)+ 6d^2 \cdot \log \left(64d^\frac{1}{2}c^2N_{\max}^2\zeta\right) +d \cdot \log \left(1+N_{\max}\right)+4d^2\right)
    \end{split}
\label{eq:square(2)}
\end{equation}
with probability at least $1-\xi/K$. Note that $\log (1+N_{\max}) \leq \log (2 N_{\max}) \leq \zeta$ and $\log \zeta \leq \zeta$. Hence, we have
$$
\begin{aligned}
& 2 \cdot \log (KH / \xi)+6 d^2 \cdot \log \left(d^{1 / 2} N_{\max}^2 \zeta\right)+d \cdot \log (1+N_{\max})+4 d^2 \\
& \quad \leq 3 d^2 \cdot \log \left(d KH N_{\max}^4 / \xi\right)+d \zeta+10 d^2 \zeta \leq 24 d^2 \zeta
\end{aligned}
$$
Using the bound above and the fact that $\zeta>\log 2$, \eqref{eq:square(2)} implies that for fixed $k\in[K]$ and all $h\in[H]$
$$
\left\|\sum_{\tau=1}^{N_k} \boldsymbol{\phi}\left(s_h^{(k, \tau)}, a_h^{(k, \tau)}\right) \cdot \epsilon_{h}^{(k,\tau)}\left(\widehat{V}_{h+1}\right)\right\|_{\left(\Lambda_h^k\right)^{-1}}^2 \leq d^2 H^2 \zeta \cdot\left(48+ \frac{12}{\log 2} \cdot \log \left(64 c^2\right)\right) 
$$
holds with probability at least $1-\xi/K$. We can set $c\geq 1$ to be sufficiently large, which guarantees that  $48+ \frac{12}{\log 2}\cdot \log \left(64 c^2\right)\leq\frac{c^2}{4}$, then for fixed $k\in[K]$ and all $h\in[H]$, it holds that
\begin{equation}
    \left\|\sum_{\tau=1}^{N_k} \boldsymbol{\phi}\left(s_h^{(k, \tau)}, a_h^{(k, \tau)}\right) \cdot \epsilon_{h}^{(k,\tau)}\left(\widehat{V}_{h+1}\right)\right\|_{\left(\Lambda_h^k\right)^{-1}}\leq \frac{c}{2}dH\sqrt{\zeta}\label{eq:sqrt}
\end{equation}
Now, by equation \eqref{eq:(ii)} and \eqref{eq:sqrt}, for fixed $k\in[K]$, for all $i\in[d]$ and all $h\in[H]$, it holds that
\begin{equation}
    |(\mathrm{ii})|\leq \frac{c}{2}dH\sqrt{\zeta}\left\|\mathbf{e}_i\right\|_{\left(\Lambda_h^k\right)^{-1}}=\frac{\beta}{2}\left\|\mathbf{e}_i\right\|_{\left(\Lambda_h^k\right)^{-1}}
\end{equation}
with probability at least $1-\xi/K$. Combining with equation \eqref{eq:(i)}, for fixed $k\in[K]$, for all $i\in[d]$, and all $h\in[H]$, it holds that
\begin{align}
    \left|\mathbf{e}_i^{\top}\left(\boldsymbol{\nu}_h^k-\widehat{\boldsymbol{\nu}}_h^k\right)\right|\leq \left(H\sqrt{d}+\frac{\beta}{2}\right)\left\|\mathbf{e}_i\right\|_{\left(\Lambda_h^k\right)^{-1}}\leq \beta\left\|\mathbf{e}_i\right\|_{\left(\Lambda_h^k\right)^{-1}}\leq\max_{k\in[K]} \beta\left\|\mathbf{e}_i\right\|_{\left(\Lambda_h^k\right)^{-1}}
\end{align}
with probability at least $1-\xi/K$. Taking a union bound over $k\in[K]$, with probability at least $1-\xi$, for all $i\in[d]$ and all $h\in[H]$,
\begin{align}
\max_{k\in[K]}\left|\mathbf{e}_i^{\top}\left(\boldsymbol{\nu}_h^k-\widehat{\boldsymbol{\nu}}_h^k\right)\right|\leq \max_{k\in[K]} \beta\left\|\mathbf{e}_i\right\|_{\left(\Lambda_h^k\right)^{-1}}
\end{align}
Finally, according to equation \eqref{eq:difference}, for all $h\in[H]$ and all $(s,a)\in\mathcal{S}\times\mathcal{A}$, it holds that
\begin{align*}
    \left|\left(\mathbb{B}_h \widehat{V}_{h+1}\right)(s, a)-\left(\widehat{\mathbb{B}}_h \widehat{V}_{h+1}\right)(s, a)\right|
    &\leq \sum_{i=1}^d \phi_i(s, a) \max _k \left|\mathbf{e}_i^{\top}\left(\boldsymbol{\nu}_h^k-\widehat{\boldsymbol{\nu}}_h^k\right)\right|\\
    &\leq \beta\sum_{i=1}^d\phi_i(s,a)\max_{k\in[K]} \left\|\mathbf{e}_i\right\|_{\left(\Lambda_h^k\right)^{-1}}\\
    & = \beta\sum_{i=1}^d\max_{k\in[K]} \left\|\phi_i(s,a)\mathbf{e}_i\right\|_{\left(\Lambda_h^k\right)^{-1}}\\
    &=\Gamma_h(s,a)
\end{align*}
with probability at least $1-\xi$. This completes the proof that $\{\Gamma_h\}_{h=1}^H$ is a $\xi$-uncertainty quantifier.
\end{proof}
The following lemma bounds the model evaluation error $\iota_h(s,a)$.
\begin{lemma}[Bounded evaluation error \citep{jin2021pessimism}]
\label{le:pessimism}
Suppose $\{\Gamma_h\}_{h=1}^H$ in Algorithm~\ref{alg:pess-source-ridge} is a $\xi$-uncertainty quantifier. On the event $\mathcal E$ in \eqref{uncertainty-quantifier}, which has probability at least $1-\xi$, we have for all $(s,a)$ and $h\in[H]$,
\[
0\ \le\ \iota_h(s,a) \le\ 2\,\Gamma_h(s,a),
\]
where $\iota_h(s,a)$ is defined in Equation~\eqref{eq:iota-def}.
\end{lemma}
On the event $\mathcal E$, Lemma~\ref{le:pessimism} implies that $\iota_h(s,a)\ge0$, so term~(i) in Lemma~\ref{lemma:decomposition} is non-positive. Moreover, since $\widehat\pi_h$ is greedy with respect to $\widehat Q_h$, term~(iii) is also non-positive. Therefore,
$$
\operatorname{SubOpt}(\widehat{\pi} ; s)=V_1^{\pi^\star}(s)-V_1^{\widehat{\pi}}(s)
    \leq \sum_{h=1}^H \mathbb{E}^{\pi^\star, P^{\pi^\star, \dagger}}\left[\iota_h\left(s_h, a_h\right) \mid s_1=s\right]
$$
Recall that 
$$
\Gamma_h(s,a)=\sum_{i=1}^d \beta \max _k\left\|\phi_i(s, a) \mathbf{e}_i\right\|_{\left(\Lambda_h^k\right)^{-1}}
$$
and 
$$
0\leq \iota_h\left(s_h, a_h\right)\leq 2\Gamma_h(s_h,a_h)
$$
Then we have
$$
\operatorname{SubOpt}(\widehat{\pi} ; s)
    \leq 2\beta\sum_{h=1}^H \sum_{i=1}^d \mathbb{E}^{\pi^\star, P^{\pi^\star, \dagger}}\left[ \max _k\left\|\phi_i(s_h, a_h) \mathbf{e}_i\right\|_{\left(\Lambda_h^k\right)^{-1}}\mid s_1=s\right]
$$
This proves Theorem~\ref{thm:suboptimality}.
\end{proof}

\subsection{Proof of Corollary~\ref{corollary-suboptimality}}
\label{proof:corollary}
\begin{proof}[Proof of Corollary~\ref{corollary-suboptimality}]
Work on the event in Theorem~\ref{thm:suboptimality}, which occurs with probability at least $1-\xi$. Fix a realized dataset $\mathcal D$ satisfying Assumption~\ref{partial-coverage} for the fixed initial state $s$. For each $i\in[d]$ and $h\in[H]$, choose 
\[
k_{i,h}^\star
\in
\arg\max_{k\in[K]}
\mathbf e_i^\top(\Lambda_h^k)^{-1}\mathbf e_i .
\]
Since $\phi_i(s,a)\ge 0$, this choice satisfies, for every $(s,a)$,
\[
\max_{k\in[K]}
\|\phi_i(s,a)\mathbf e_i\|_{(\Lambda_h^k)^{-1}}
=
\|\phi_i(s,a)\mathbf e_i\|_{(\Lambda_h^{k_{i,h}^\star})^{-1}} .
\]

We first consider the summation over $i\in[d]$ at a fixed step $h$:
\begin{align*}
&\sum_{i=1}^d
\mathbb{E}^{\pi^\star, P^{\pi^\star,\dagger}}
\left[
\max_{k\in[K]}
\left\|\phi_i(s_h,a_h)\mathbf e_i\right\|_{(\Lambda_h^k)^{-1}}
\,\middle|\,
s_1=s
\right] \\
&=
\sum_{i=1}^d
\mathbb{E}_{(s_h,a_h)\sim
d_{P^{\pi^\star,\dagger},h}^{\pi^\star,s}}
\left[
\left\|\phi_i(s_h,a_h)\mathbf e_i\right\|_
{(\Lambda_h^{k_{i,h}^\star})^{-1}}
\right] \\
&=
\sum_{i=1}^d
\mathbb{E}_{(s_h,a_h)\sim
d_{P^{\pi^\star,\dagger},h}^{\pi^\star,s}}
\left[
\sqrt{
\left(\phi_i(s_h,a_h)\mathbf e_i\right)^\top
(\Lambda_h^{k_{i,h}^\star})^{-1}
\left(\phi_i(s_h,a_h)\mathbf e_i\right)
}
\right] \\
&\le
\sum_{i=1}^d
\sqrt{
\mathbb{E}_{(s_h,a_h)\sim
d_{P^{\pi^\star,\dagger},h}^{\pi^\star,s}}
\left[
\left(\phi_i(s_h,a_h)\mathbf e_i\right)^\top
(\Lambda_h^{k_{i,h}^\star})^{-1}
\left(\phi_i(s_h,a_h)\mathbf e_i\right)
\right]
} \\
&=
\sum_{i=1}^d
\sqrt{
\operatorname{Tr}
\left(
\Sigma_{P^{\pi^\star,\dagger},h,i}
(\Lambda_h^{k_{i,h}^\star})^{-1}
\right)
},
\end{align*}
where
\[
\Sigma_{P^{\pi^\star,\dagger},h,i}
:=
\mathbb{E}_{(s_h,a_h)\sim
d_{P^{\pi^\star,\dagger},h}^{\pi^\star,s}}
\left[
\left(\phi_i(s_h,a_h)\mathbf e_i\right)
\left(\phi_i(s_h,a_h)\mathbf e_i\right)^\top
\right].
\]
Since $P^{\pi^\star,\dagger}$ is an admissible transition-kernel sequence, Assumption~\ref{partial-coverage} applies to $P=P^{\pi^\star,\dagger}$. Thus, for every $k\in[K]$,
\[
\frac{1}{N_k}\Lambda_h^k
\succeq
\frac{\lambda}{N_k}\mathbf I_d
+
c^\dagger
\Sigma_{P^{\pi^\star,\dagger},h,i},
\]
and hence
\[
\Lambda_h^k
\succeq
\lambda \mathbf I_d
+
c^\dagger N_k
\Sigma_{P^{\pi^\star,\dagger},h,i}.
\]
Since we set $\lambda=1$, for $k=k_{i,h}^\star$ we have
\[
\Lambda_h^{k_{i,h}^\star}
\succeq
\mathbf I_d
+
c^\dagger N_{k_{i,h}^\star}
\Sigma_{P^{\pi^\star,\dagger},h,i}.
\]
By the monotonicity of the matrix inverse,
\[
(\Lambda_h^{k_{i,h}^\star})^{-1}
\preceq
\left(
\mathbf I_d
+
c^\dagger N_{k_{i,h}^\star}
\Sigma_{P^{\pi^\star,\dagger},h,i}
\right)^{-1}.
\]

Define
\[
m_{h,i}
:=
\mathbb{E}_{(s_h,a_h)\sim
d_{P^{\pi^\star,\dagger},h}^{\pi^\star,s}}
\left[
\phi_i(s_h,a_h)^2
\right].
\]
Then
\[
\Sigma_{P^{\pi^\star,\dagger},h,i}
=
m_{h,i}\mathbf e_i\mathbf e_i^\top .
\]
Therefore,
\begin{align*}
&\operatorname{Tr}
\left(
\Sigma_{P^{\pi^\star,\dagger},h,i}
(\Lambda_h^{k_{i,h}^\star})^{-1}
\right) \\
&\le
\operatorname{Tr}
\left(
\Sigma_{P^{\pi^\star,\dagger},h,i}
\left(
\mathbf I_d
+
c^\dagger N_{k_{i,h}^\star}
\Sigma_{P^{\pi^\star,\dagger},h,i}
\right)^{-1}
\right) \\
&=
\frac{m_{h,i}}
{1+c^\dagger N_{k_{i,h}^\star}m_{h,i}} \\
&\le
\frac{m_{h,i}}
{1+c^\dagger N_{\min}m_{h,i}}
\le
\frac{1}{c^\dagger N_{\min}},
\end{align*}
where $N_{\min}=\min_{k\in[K]} N_k$. Consequently,
\[
\sum_{i=1}^d
\mathbb{E}^{\pi^\star, P^{\pi^\star,\dagger}}
\left[
\max_{k\in[K]}
\left\|\phi_i(s_h,a_h)\mathbf e_i\right\|_{(\Lambda_h^k)^{-1}}
\,\middle|\,
s_1=s
\right]
\le
\frac{d}{\sqrt{c^\dagger N_{\min}}}.
\]

Finally, by Theorem~\ref{thm:suboptimality}, we have
\begin{align*}
\operatorname{SubOpt}(\widehat{\pi};s)
&\le
2\beta
\sum_{h=1}^H
\sum_{i=1}^d
\mathbb{E}^{\pi^\star, P^{\pi^\star,\dagger}}
\left[
\max_{k\in[K]}
\left\|\phi_i(s_h,a_h)\mathbf e_i\right\|_{(\Lambda_h^k)^{-1}}
\,\middle|\,
s_1=s
\right] \\
&\le
2c\cdot dH\sqrt{\log(2dKHN_{\max}/\xi)}
\cdot
\frac{Hd}{\sqrt{c^\dagger N_{\min}}} \\
&=
c^\ddagger d^2H^2
\sqrt{
\frac{\log(2dKHN_{\max}/\xi)}{N_{\min}}
}.
\end{align*}
Thus, on the event in Theorem~\ref{thm:suboptimality}, whenever the realized offline dataset satisfies Assumption~\ref{partial-coverage} for the fixed initial state $s$, the stated bound holds. Since this event has probability at least $1-\xi$, the desired implication follows.
\end{proof}

\subsection{Proof of Corollary~\ref{cor:cluster-level-subopt}}
\label{proof:corollary-cluster-level}
\begin{proof}[Proof of Corollary~\ref{cor:cluster-level-subopt}]
Work on the event on which the cluster-level concentration bounds derived below hold simultaneously, which occurs with probability at least $1-\xi$. Fix a realized dataset satisfying Assumption~\ref{partial-coverage-cluster} for the fixed initial state $s$. Throughout this proof, we suppress the superscript
``$\mathrm{pool}$'' and write
$\widehat V_h:=\widehat V_h^{\mathrm{pool}}$,
$\widehat Q_h:=\widehat Q_h^{\mathrm{pool}}$, $\overline Q_h:=\overline Q_h^{\mathrm{pool}}$ and
$\widehat\pi_h:=\widehat\pi_h^{\mathrm{pool}}$. 

For each cluster $m\in[M]$, fix an arbitrary reference source $k_m^\circ\in\mathcal C_m$, and define, for any bounded measurable $V$, $\boldsymbol{\nu}_h^m(V)
:=
\boldsymbol{\nu}_h^{k_m^\circ}(V)$. This reference is introduced only for the analysis. We further define 
\[
\eta_{h,i}(V):=\min_{m\in[M]}\nu_{h,i}^m(V),
\qquad
(\mathbb B_h^{\mathrm{pool}}V)(s,a)
:=
\sum_{i=1}^d\phi_i(s,a)\eta_{h,i}(V),
\]
which serves as an auxiliary reference-source Bellman operator for the analysis.

Since $\boldsymbol{\nu}_h^m(V)$ is the Bellman coefficient of an actual source, for $V: \mathcal{S} \rightarrow\left[0, V_{\max }\right]$, Lemma~\ref{lemma-bounded-weights} gives
\[
\|\boldsymbol{\nu}_h^m(V)\|_2
\le (1+V_{\max})\sqrt d.
\]
For each cluster $m \in[M]$, let $\widehat{\boldsymbol{\nu}}_h^{\mathcal C_m}$ denote the ridge estimator obtained from the pooled dataset $\mathcal{D}_{\mathcal{C}_m}$ of size $N_m$. By the same ridge-regression argument as in Lemma~\ref{lemma-bounded-weights},
\[
\|\widehat{\boldsymbol{\nu}}_h^{\mathcal C_m}\|_2
\le
H\sqrt{\frac{N_m d}{\lambda}}.
\]

To analyze the suboptimality, we decompose the Bellman update error. For brevity, let
$\boldsymbol{\nu}_h^m
:=\boldsymbol{\nu}_h^m(\widehat V_{h+1})$,
$w_{h,i}:=w_{h,i}(\widehat V_{h+1})$,
$\eta_{h,i}:=\eta_{h,i}(\widehat V_{h+1})$, and $\widehat\eta_{h,i}
:=\min_{m\in[M]}\widehat\nu_{h,i}^{\mathcal C_m}$. Accordingly, define the pooled empirical Bellman backup by $(\widehat{\mathbb B}_h^{\mathrm{pool}}\widehat V_{h+1})(s,a)
:=
\sum_{i=1}^d
\phi_i(s,a)\widehat\eta_{h,i}$. The discrepancy between the true robust Bellman operator and the pooled empirical operator is:
\begin{align*}
\Big|(\mathbb{B}_h \widehat{V}_{h+1})(s, a)-(\widehat{\mathbb{B}}_h^\mathrm{pool}\widehat{V}_{h+1})(s, a)\Big|
&= \Big|\sum_{i=1}^d \phi_i(s, a)\big(w_{h,i}-\widehat{\eta}_{h,i}\big)\Big| \\
&= \Big|\sum_{i=1}^d \phi_i(s, a)\big[(w_{h,i}-\eta_{h,i})+(\eta_{h,i}-\widehat{\eta}_{h,i})\big]\Big| \\
&\le \underbrace{\Big|\sum_{i=1}^d \phi_i(s, a)\big(\eta_{h,i}-\widehat{\eta}_{h,i}\big)\Big|}_{\text{(a)}}
   +\underbrace{ \Big|\sum_{i=1}^d \phi_i(s, a)\big(w_{h,i}-\eta_{h,i}\big)\Big|}_{\text{(b)}}.
\end{align*}
For term~(a), the coordinate-wise minimum inequality gives
\begin{equation}
\text{(a)}
\le
\sum_{i=1}^d\phi_i(s,a)
\max_{m\in[M]}
\underbrace{
\left|
\mathbf e_i^\top
\left(
\boldsymbol\nu_h^m
-\widehat{\boldsymbol\nu}_h^{\mathcal C_m}
\right)
\right|
}_{\text{(I)}}.
\label{eq:difference-cluster}
\end{equation}

To bound term~(I), fix $h\in[H]$ and $m\in[M]$. Order the sources in $\mathcal C_m$ deterministically and concatenate their trajectories, preserving the original trajectory order within each source. Let $k(\tau)$ and $j(\tau)$ denote the source label and within-source trajectory index of the $\tau$-th pooled trajectory, respectively. We write $(s_h^{(m,\tau)},a_h^{(m,\tau)},r_h^{(m,\tau)}, s_{h+1}^{(m,\tau)})$ for the corresponding observation and set
\[
\boldsymbol{\phi}_\tau
:=
\boldsymbol\phi(s_h^{(m,\tau)},a_h^{(m,\tau)}).
\]

For any measurable function $V:\mathcal S\to[0,H-1]$, recall the source-specific residual
\[
\epsilon_h^{(k,j)}(V)
=
r_h^{(k,j)}+V(s_{h+1}^{(k,j)})
-\boldsymbol\phi(s_h^{(k,j)},a_h^{(k,j)})^\top
\boldsymbol\nu_h^k(V),
\]
and define the residual centered at the reference source by
\[
\tilde\epsilon_h^{(m,\tau)}(V)
=
r_h^{(m,\tau)}+V(s_{h+1}^{(m,\tau)})
-\boldsymbol{\phi}_\tau^\top\boldsymbol\nu_h^m(V).
\]
These residuals satisfy
\[
\tilde\epsilon_h^{(m,\tau)}(V)
=
\epsilon_h^{(k(\tau),j(\tau))}(V)
+
\boldsymbol{\phi}_\tau^\top
\left(
\boldsymbol\nu_h^{k(\tau)}(V)
-\boldsymbol\nu_h^m(V)
\right).
\]
Thus, centering at the reference source introduces a within-cluster discrepancy in addition to the source-specific residual.

Using $\boldsymbol\Lambda_h^{\mathcal C_m}=\lambda\mathbf I_d+\sum_{\tau=1}^{N_m}\boldsymbol{\phi}_\tau\boldsymbol{\phi}_\tau^\top$ and the definition of the pooled ridge estimator, we obtain
\begin{align*}
\text{(I)}
&\le
\underbrace{
\lambda
\left|
\mathbf e_i^\top
(\boldsymbol\Lambda_h^{\mathcal C_m})^{-1}
\boldsymbol\nu_h^m
\right|
}_{\text{(i)}}\\
&\quad+
\underbrace{
\left|
\mathbf e_i^\top
(\boldsymbol\Lambda_h^{\mathcal C_m})^{-1}
\sum_{\tau=1}^{N_m}
\boldsymbol{\phi}_\tau\,
\tilde\epsilon_h^{(m,\tau)}(\widehat V_{h+1})
\right|
}_{\text{(ii)}}.
\end{align*}
Since $\|\boldsymbol\nu_h^m\|_2\le H\sqrt d$ and $\boldsymbol\Lambda_h^{\mathcal C_m}\succeq\lambda\mathbf I_d$, the ridge-bias term satisfies
\[
\text{(i)}
\le
H\sqrt{d\lambda}\,
\|\mathbf e_i\|_{(\boldsymbol\Lambda_h^{\mathcal C_m})^{-1}}.
\]
For term~(ii), the residual decomposition above and Cauchy--Schwarz yield
\begin{align*}
\text{(ii)}
&\le
\|\mathbf e_i\|_{(\boldsymbol\Lambda_h^{\mathcal C_m})^{-1}}
\underbrace{
\left\|
\sum_{\tau=1}^{N_m}
\boldsymbol{\phi}_\tau\,
\epsilon_h^{(k(\tau),j(\tau))}(\widehat V_{h+1})
\right\|_{(\boldsymbol\Lambda_h^{\mathcal C_m})^{-1}}
}_{\text{(iii)}}\\
&\quad+
\|\mathbf e_i\|_{(\boldsymbol\Lambda_h^{\mathcal C_m})^{-1}}
\underbrace{
\left\|
\sum_{\tau=1}^{N_m}
\boldsymbol{\phi}_\tau\boldsymbol{\phi}_\tau^\top
\left(
\boldsymbol\nu_h^{k(\tau)}(\widehat V_{h+1})
-\boldsymbol\nu_h^m(\widehat V_{h+1})
\right)
\right\|_{(\boldsymbol\Lambda_h^{\mathcal C_m})^{-1}}
}_{\text{(iv)}}.
\end{align*}
Term~(iii) is controlled by concentration, whereas term~(iv) is bounded using the within-cluster discrepancy.

To control term~(iii), define the filtration $\{\mathcal F_\tau\}_{\tau=0}^{N_m}$ by
\[
\mathcal F_\tau
:=
\sigma\left(
\left\{(s_h^{(m,\ell)},a_h^{(m,\ell)})\right\}
_{\ell=1}^{(\tau+1)\wedge N_m},
\left\{(r_h^{(m,\ell)},s_{h+1}^{(m,\ell)})\right\}
_{\ell=1}^{\tau}
\right),
\]
where the second collection is empty when $\tau=0$. For each $\tau\in[N_m]$, $\boldsymbol{\phi}_\tau$ is $\mathcal F_{\tau-1}$-measurable. For any fixed measurable function $V:\mathcal S\to[0,H-1]$, the source-wise compliance condition and the independence of the source datasets imply
\[
\mathbb E\left[
\epsilon_h^{(k(\tau),j(\tau))}(V)
\mid\mathcal F_{\tau-1}
\right]=0.
\]
Moreover, this residual is $\mathcal F_\tau$-measurable and bounded in absolute value by $H$. Therefore, the self-normalized concentration inequality applies to the pooled residual sum for each fixed $V$. To handle the data-dependent function $\widehat V_{h+1}$, we make this bound uniform over a deterministic value-function class.

Let $\mathcal V_h^{\mathrm{pool}}(R,B,\lambda)$ denote the cluster-level analogue of the value-function class introduced in the proof of Lemma~\ref{lem:uq}. Specifically, define
$$
\begin{aligned}
& \mathcal{V}_h^{\mathrm{pool}}(R, B, \lambda)=\left\{V_h\left(s ; \boldsymbol{\theta}, \beta, \Sigma_1, \cdots, \Sigma_M\right): \mathcal{S} \rightarrow[0, H] \text { with }\|\boldsymbol{\theta}\|_2 \leq R, \beta \in[0, B], \Sigma_1, \cdots, \Sigma_M \succeq \lambda\mathbf I_d\right\}, \\
&V_h\left(s ; \boldsymbol{\theta}, \beta, \Sigma_1, \cdots, \Sigma_M\right)=\max _{a \in \mathcal{A}}\left\{\min \left\{\boldsymbol{\phi}(s, a)^{\top} \boldsymbol{\theta}-\beta \sum_{i=1}^d \max _{m\in[M]}\left\|\phi_i(s, a) \mathbf{e}_i\right\|_{\left(\Sigma_m\right)^{-1}}, H-h+1\right\}^{+}\right\}.
\end{aligned}
$$
By the same argument as in the proof of Lemma~\ref{lem:uq}, the maximum over actions is attained, and every function in $\mathcal V_h^{\mathrm{pool}}(R,B,\lambda)$ is measurable. For convenience, define $\mathcal V_{H+1}^{\mathrm{pool}}(R,B,\lambda):=\{0\}$. Let
\[
\beta^{\mathrm{pool}}
:=
c\,dH\sqrt{\zeta^{\mathrm{pool}}},
\qquad
\zeta^{\mathrm{pool}}
:=
\log\left(
\frac{2dMHN_{\max}^{\mathrm{pool}}}{\xi}
\right),
\qquad
N_{\max}^{\mathrm{pool}}
:=
\max_{m\in[M]}N_m.
\]
By the same boundedness argument as in the proof of
Lemma~\ref{lem:uq}, with the source index replaced by the cluster
index, the cluster-level value estimates satisfy $\widehat V_h
\in\mathcal V_h^{\mathrm{pool}}
\left(R_0^{\mathrm{pool}},B_0^{\mathrm{pool}},\lambda\right)$, where $R_0^{\mathrm{pool}}:=
Hd\sqrt{\frac{N_{\max}^{\mathrm{pool}}}{\lambda}}$ and $B_0^{\mathrm{pool}}
:=2\beta^{\mathrm{pool}}$. For brevity, write $\mathcal V_h^{\mathrm{pool}}
:=\mathcal V_h^{\mathrm{pool}}
\left(R_0^{\mathrm{pool}},B_0^{\mathrm{pool}},\lambda\right)$.

By choosing $\varepsilon = dH / N_{\max}^{\mathrm{pool}}$ for the covering of $\mathcal{V}_{h+1}^{\mathrm{pool}}$ and applying the self-normalized concentration inequality (Lemma~\ref{le:concentration}) as in the proof of Lemma~\ref{lem:uq}, we obtain, for each fixed cluster $m$, $\text{(iii)}\le \beta^{\mathrm{pool}}/2$ simultaneously for all $h\in[H]$, with probability at least $1-\xi/M$.

To control term~(iv), note that the clipping step ensures that
\[
0\leq \widehat V_{h+1}(s)\leq H-h
\qquad\text{for all }s\in\mathcal S,
\]
and hence $\widehat V_{h+1}\in\mathcal B_{h+1}$. Therefore, by the definition of $\Delta_{h,m}$ and $\boldsymbol\nu_h^m=\boldsymbol\nu_h^{k_m^\circ}$,
\[
\begin{aligned}
\max_{k\in\mathcal C_m}
\left\|
\boldsymbol\nu_h^k(\widehat V_{h+1})
-
\boldsymbol\nu_h^m(\widehat V_{h+1})
\right\|_2
&=
\max_{k\in\mathcal C_m}
\left\|
\boldsymbol\nu_h^k(\widehat V_{h+1})
-
\boldsymbol\nu_h^{k_m^\circ}(\widehat V_{h+1})
\right\|_2\\
&\le
\max_{k,k'\in\mathcal C_m}
\left\|
\boldsymbol\nu_h^k(\widehat V_{h+1})
-
\boldsymbol\nu_h^{k'}(\widehat V_{h+1})
\right\|_2\\
&\le
\Delta_{h,m}.
\end{aligned}
\]

Let $\mathbf{g}:=\sum_{\tau=1}^{N_m}\boldsymbol{\phi}_\tau\,\boldsymbol{\phi}_\tau^\top\big(\boldsymbol{\nu}_h^{k(\tau)}(\widehat V_{h+1}) - \boldsymbol{\nu}_h^m(\widehat V_{h+1})\big)$, so that $\left|\text{(iv)}\right|=\left\|\mathbf{g}\right\|_{(\boldsymbol{\Lambda}_h^{\mathcal C_m})^{-1}}$. By the Cauchy--Schwarz inequality,
\[
\begin{aligned}
\left\|\mathbf{g}\right\|^2_{(\boldsymbol{\Lambda}_h^{\mathcal C_m})^{-1}}
&=\mathbf{g}^\top(\boldsymbol{\Lambda}_h^{\mathcal C_m})^{-1}\mathbf{g}
=\sum_{\tau=1}^{N_m}\boldsymbol{\phi}_\tau^\top\big(\boldsymbol{\nu}_h^{k(\tau)}(\widehat V_{h+1}) - \boldsymbol{\nu}_h^m(\widehat V_{h+1})\big)\cdot\boldsymbol{\phi}_\tau^\top(\boldsymbol{\Lambda}_h^{\mathcal C_m})^{-1}\mathbf{g}\\
&\le\left(\sum_{\tau=1}^{N_m}\Big(\boldsymbol{\phi}_\tau^\top\big(\boldsymbol{\nu}_h^{k(\tau)}(\widehat V_{h+1}) - \boldsymbol{\nu}_h^m(\widehat V_{h+1})\big)\Big)^2\right)^{1/2}
\left(\sum_{\tau=1}^{N_m}\Big(\boldsymbol{\phi}_\tau^\top(\boldsymbol{\Lambda}_h^{\mathcal C_m})^{-1}\mathbf{g}\Big)^2\right)^{1/2}.
\end{aligned}
\]
Since $\|\boldsymbol\phi_\tau\|_2\le\|\boldsymbol\phi_\tau\|_1=1$, the definition of $\Delta_{h,m}$ gives
$\big|\boldsymbol{\phi}_\tau^\top\big(\boldsymbol{\nu}_h^{k(\tau)}(\widehat V_{h+1}) - \boldsymbol{\nu}_h^m(\widehat V_{h+1})\big)\big|\le\Delta_{h,m}$ for every $\tau$, so the first factor is at most $\sqrt{N_m}\,\Delta_{h,m}$. For the second factor, since $\sum_{\tau=1}^{N_m}\boldsymbol\phi_\tau\boldsymbol\phi_\tau^\top=\boldsymbol\Lambda_h^{\mathcal C_m}-\lambda\mathbf I_d\preceq\boldsymbol\Lambda_h^{\mathcal C_m}$,
\[
\sum_{\tau=1}^{N_m}\Big(\boldsymbol{\phi}_\tau^\top(\boldsymbol{\Lambda}_h^{\mathcal C_m})^{-1}\mathbf{g}\Big)^2
=\mathbf g^\top(\boldsymbol{\Lambda}_h^{\mathcal C_m})^{-1}\big(\boldsymbol{\Lambda}_h^{\mathcal C_m}-\lambda\mathbf I_d\big)(\boldsymbol{\Lambda}_h^{\mathcal C_m})^{-1}\mathbf g
\le\left\|\mathbf{g}\right\|^2_{(\boldsymbol{\Lambda}_h^{\mathcal C_m})^{-1}} .
\]
Combining the two bounds yields $\left\|\mathbf{g}\right\|^2_{(\boldsymbol{\Lambda}_h^{\mathcal C_m})^{-1}}\le\sqrt{N_m}\,\Delta_{h,m}\left\|\mathbf{g}\right\|_{(\boldsymbol{\Lambda}_h^{\mathcal C_m})^{-1}}$, and therefore
\[
\left|\text{(iv)}\right|=\left\|\mathbf{g}\right\|_{(\boldsymbol{\Lambda}_h^{\mathcal C_m})^{-1}}\le\sqrt{N_m}\,\Delta_{h,m}.
\]

Let $\sigma_{h,i}^2$ be defined as in Corollary~\ref{cor:cluster-level-subopt}. For any $P\in\prod_{t=1}^H\mathcal P_t$, write $\sigma_{h,i}^2(P):=\mathbb{E}_{(s_h,a_h)\sim d_{P,h}^{\pi^\star,s}}\left[\phi_i(s_h,a_h)^2\right]$,
so that $\sigma_{h,i}^2=\sup_{P\in\prod_{t=1}^H\mathcal P_t}\sigma_{h,i}^2(P)$. Since 
\[
\mathbb E_{(s_h,a_h)\sim d^{\pi^\star,s}_{P,h}}
\Big[\big(\phi_i(s_h,a_h)\mathbf e_i\big)\big(\phi_i(s_h,a_h)\mathbf e_i\big)^\top\Big]
=\sigma_{h,i}^2(P)\,\mathbf e_i\mathbf e_i^\top,
\]
Assumption~\ref{partial-coverage-cluster} gives, for every $P\in\prod_{t=1}^H\mathcal P_t$,
\[
\boldsymbol{\Lambda}_h^{\mathcal C_m} \succeq \mathbf{D}_P
:= \lambda \mathbf{I}_d + N_m c' \sigma_{h,i}^2(P)\, \mathbf{e}_i \mathbf{e}_i^\top \succ 0 .
\]
Note that $\mathbf{D}_P$ is a diagonal matrix. By the order-reversing property of the matrix inverse, we have $(\boldsymbol{\Lambda}_h^{\mathcal C_m})^{-1} \preceq \mathbf{D}_P^{-1}$. Therefore,
\begin{align*}
\|\mathbf{e}_i\|_{(\boldsymbol{\Lambda}_h^{\mathcal C_m})^{-1}}^2
&= \mathbf{e}_i^\top (\boldsymbol{\Lambda}_h^{\mathcal C_m})^{-1} \mathbf{e}_i
\le \mathbf{e}_i^\top \mathbf{D}_P^{-1} \mathbf{e}_i
= (\mathbf{D}_P^{-1})_{ii}
= \frac{1}{\lambda + N_m c' \sigma_{h,i}^2(P)} .
\end{align*}
Since the left-hand side does not depend on $P$, taking the infimum of the right-hand side over $P\in\prod_{t=1}^H\mathcal P_t$ yields
\[
\|\mathbf{e}_i\|_{(\boldsymbol{\Lambda}_h^{\mathcal C_m})^{-1}}
\le \frac{1}{\sqrt{\lambda + N_m c'  \sigma_{h,i}^2}},
\]
and
\begin{align*}
\|\mathbf e_i\|_{(\boldsymbol{\Lambda}_h^{\mathcal C_m})^{-1}}\,|\mathrm{(iv)}|
&\le
\Delta_{h,m}
\sqrt{\frac{N_m}
{\lambda+N_m c'\sigma_{h,i}^2}}\\
&=
\Delta_{h,m}
\sqrt{
\frac{1}
{\lambda/N_m+c'\sigma_{h,i}^2}
}\\
&\le
\Delta_{h,m}
\sqrt{
\frac{1}
{\lambda/N_{\max}^{\mathrm{pool}}+c'\sigma_{h,i}^2}
}.
\end{align*}
Then for a fixed cluster $m$, all $h\in[H]$, and all $i\in[d]$, it holds that
\[
\text{Term (ii)} \leq \frac{\beta^{\mathrm{pool}}}{2}\|\mathbf{e}_i\|_{(\boldsymbol{\Lambda}_h^{\mathcal C_m})^{-1}}+\Delta_{h, m} \sqrt{
\frac{1}{\lambda/N_{\max}^{\mathrm{pool}}+c'\sigma_{h,i}^2}
}
\]
with probability at least $1-\xi/M$. Since $\lambda=1$ and the absolute constant $c$ in
$\beta^{\mathrm{pool}}=cdH\sqrt{\zeta^{\mathrm{pool}}}$ is chosen sufficiently large, we have
\[
H\sqrt{d\lambda}\le \frac{\beta^{\mathrm{pool}}}{2}.
\]
Therefore, the ridge-bias term (i) and the statistical part of term (ii) are
absorbed into $\beta^{\mathrm{pool}}\|\mathbf e_i\|_{(\boldsymbol{\Lambda}_h^{\mathcal C_m})^{-1}}$. For a fixed $m\in[M]$, all $h\in[H]$, and all $i\in[d]$, it holds that 
\[
\text{Term (I)} \leq \beta^{\mathrm{pool}} \|\mathbf{e}_i\|_{(\boldsymbol{\Lambda}_h^{\mathcal C_m})^{-1}}+ \Delta_{h, m}\sqrt{
\frac{1}{\lambda/N_{\max}^{\mathrm{pool}}+c'\sigma_{h,i}^2}
} 
\]
with probability at least $1-\xi/M$. Taking a union bound over $m\in[M]$, the above bound holds simultaneously for all clusters with probability at least $1-\xi$. On this event,
\begin{align*}
    \text{Term (a)} &\leq \sum_{i=1}^d \phi_i(s, a) \max _m \underbrace{\left|\mathbf{e}_i^{\top}\left(\boldsymbol{\nu}_h^m-\widehat{\boldsymbol{\nu}}_h^{\mathcal C_m}\right)\right|}_{(\mathrm{I})}\\
    &\leq \sum_{i=1}^d \phi_i(s, a) \max _m \left[\beta^{\mathrm{pool}} \|\mathbf{e}_i\|_{(\boldsymbol{\Lambda}_h^{\mathcal C_m})^{-1}}+ \Delta_{h, m} \sqrt{
\frac{1}{\lambda/N_{\max}^{\mathrm{pool}}+c'\sigma_{h,i}^2}
} \right]\\
    &\leq \sum_{i=1}^d \phi_i(s, a) \max _m \beta^{\mathrm{pool}} \|\mathbf{e}_i\|_{(\boldsymbol{\Lambda}_h^{\mathcal C_m})^{-1}} + C_h\Delta_h\\
    & = \Gamma_h^{\mathrm{pool}}(s,a) + C_h\Delta_h,
\end{align*}
where $\Gamma_h^{\mathrm{pool}}(s,a):=\beta^{\mathrm{pool}}\sum_{i=1}^d\max_{m\in[M]}\|\phi_i(s,a)\mathbf e_i\|_{(\boldsymbol\Lambda_h^{\mathcal C_m})^{-1}}$ and 
\[
C_h
=
\sup_{(\tilde s,a)\in\mathcal S\times\mathcal A}
\sum_{i=1}^d
\phi_i(\tilde s,a)
\sqrt{
\frac{1}{
\lambda/N_{\max}^{\mathrm{pool}}+c'\sigma_{h,i}^2
}
}.
\]

Term (b) captures the discrepancy induced by replacing all sources with one reference source from each cluster.
\begin{align*}
\left|\sum_{i=1}^d \phi_i(s, a)\left(w_{h, i}-\eta_{h,i}\right)\right|
&= \left|\sum_{i=1}^d \phi_i(s, a)\left(\min_{k\in[K]} \nu^k_{h, i}-\min_{m\in[M]} \nu^m_{h, i} \right)\right|\\
&\le \sum_{i=1}^d \phi_i(s, a)\left|\min_{k\in[K]} \nu^k_{h, i}-\min_{m\in[M]} \nu^m_{h, i} \right|.
\end{align*}
Fix $i\in[d]$. Since each $\boldsymbol{\nu}_h^m$ corresponds to the reference source $k_m^\circ\in\mathcal C_m$, the set of reference sources is a subset of the original $K$ sources. Hence, 
\[
w_{h,i}
=
\min_{k\in[K]}\nu_{h,i}^k
\le
\min_{m\in[M]}\nu_{h,i}^m
=
\eta_{h,i}.
\]
Let $k^\star\in\arg\min_{k\in[K]}\nu_{h,i}^k$, and let $m^\star$ be the cluster such that $k^\star\in\mathcal C_{m^\star}$. Then
\[
\eta_{h,i}
\le
\nu_{h,i}^{m^\star}
=
\nu_{h,i}^{k_{m^\star}^\circ},
\]
and therefore
\[
0
\le
\eta_{h,i}-w_{h,i}
\le
\nu_{h,i}^{k_{m^\star}^\circ}-\nu_{h,i}^{k^\star}
\le
\left\|
\boldsymbol{\nu}_h^{k_{m^\star}^\circ}
-
\boldsymbol{\nu}_h^{k^\star}
\right\|_2
\le
\Delta_{h,m^\star}
\le
\Delta_h.
\]
Thus,
\[
|w_{h,i}-\eta_{h,i}|
=
\eta_{h,i}-w_{h,i}
\le
\Delta_h.
\]
Consequently,
\[
\text{Term (b)}
\le
\sum_{i=1}^d
\phi_i(s,a)\,|w_{h,i}-\eta_{h,i}|
\le
\Delta_h\sum_{i=1}^d\phi_i(s,a)
=
\Delta_h.
\]

Combining the bounds for (a) and (b), it holds that
\begin{equation}
    \left|\left(\mathbb{B}_h \widehat{V}_{h+1}\right)(s, a)-\left(\widehat{\mathbb{B}}_h^{\mathrm{pool}} \widehat{V}_{h+1}\right)(s, a)\right|\leq \Gamma_h^{\mathrm{pool}}(s, a)+\left(C_h +1\right)\Delta_h
\label{eq:Pool-difference}
\end{equation}
with probability at least $1-\xi$. Recall in our algorithm, 
\[
\overline{Q}_h(s,a)=\left(\widehat{\mathbb{B}}^{\mathrm{pool}}_h \widehat{V}_{h+1}\right)(s, a)-\Gamma_h^{\mathrm{pool}}(s,a),\quad \widehat{Q}_h(s, a)=\min \left\{\overline{Q}_h(s, a), H-h+1\right\}^{+}.
\]
\begin{lemma}[Bounded evaluation error, cluster level]
\label{le:bounded-evaluation-error-cluster}
If \eqref{eq:Pool-difference} holds, then for all $(s,a)$ and $h\in[H]$,
\[
-\left(C_h +1\right)\Delta_h\leq\iota_h(s, a)\leq 2\Gamma_h^{\mathrm{pool}}(s,a)+\left(C_h +1\right)\Delta_h
\]
where $\iota_h(s,a)$ is defined in main-text Equation~\eqref{eq:iota-def}.
\end{lemma}
\begin{proof}
Recall that $\iota_h(s,a):=(\mathbb{B}_h\widehat V_{h+1})(s,a)-\widehat Q_h(s,a)$.
From \eqref{eq:Pool-difference} and $\overline Q_h := (\widehat{\mathbb{B}}^{\rm pool}_h \widehat V_{h+1}) - \Gamma_h^{\rm pool}$,
we have for all $(s,a)$,
\[
-(C_h+1)\Delta_h \le (\mathbb{B}_h \widehat V_{h+1})(s,a) - \overline Q_h(s,a)
\le 2\Gamma_h^{\rm pool}(s,a) + (C_h+1)\Delta_h.
\]

\emph{Upper bound.}
If $\overline Q_h(s,a)\le H-h+1$, then $\widehat Q_h(s,a)=(\overline Q_h(s,a))^+\ge \overline Q_h(s,a)$, and hence
\[
\iota_h(s,a)=(\mathbb{B}_h \widehat V_{h+1})(s,a)-\widehat Q_h(s,a)
\le (\mathbb{B}_h \widehat V_{h+1})(s,a)-\overline Q_h(s,a)
\le 2\Gamma_h^{\rm pool}(s,a)+(C_h+1)\Delta_h.
\]
If $\overline Q_h(s,a)>H-h+1$, then $\widehat Q_h(s,a)=H-h+1$, and using the bounded-reward assumption and $\widehat V_{h+1}\in[0,H-h]$ (thus $(\mathbb{B}_h\widehat V_{h+1})(s,a)\le H-h+1$), we obtain
\[
\iota_h(s,a)\le (\mathbb{B}_h \widehat V_{h+1})(s,a)-(H-h+1)\le 0
\le 2\Gamma_h^{\rm pool}(s,a)+(C_h+1)\Delta_h.
\]

\emph{Lower bound.}
If $\overline Q_h(s,a)<0$, then $\widehat Q_h(s,a)=0$ and thus
$\iota_h(s,a)=(\mathbb{B}_h\widehat V_{h+1})(s,a)\ge 0\ge -(C_h+1)\Delta_h$.
Otherwise $\overline Q_h(s,a)\ge 0$ and $\widehat Q_h(s,a)=\min\{\overline Q_h(s,a),H-h+1\}\le \overline Q_h(s,a)$, hence
\[
\iota_h(s,a)\ge (\mathbb{B}_h \widehat V_{h+1})(s,a)-\overline Q_h(s,a)\ge -(C_h+1)\Delta_h.
\]
\end{proof}

Let $C:=\sup_{h\in[H]} C_h$. Then $(C_h+1)\Delta_h\le (C+1)\Delta_h$ for every $h\in[H]$. Consequently, by Lemma~\ref{lemma:decomposition}, Lemma~\ref{le:bounded-evaluation-error-cluster}, and the greedy choice of $\widehat\pi$, we obtain
\begin{align*}
\operatorname{SubOpt}(\widehat\pi^{\mathrm{pool}};s) &\leq 2 \sum_{h=1}^H \mathbb{E}^{\pi^\star, P^{\pi^\star, \dagger}}\left[\Gamma_h^{\mathrm{pool}}\left(s_h, a_h\right) \mid s_1=s\right]+2(C+1)\sum_{h=1}^H \Delta_h\\
&=2 \beta^\mathrm{pool} \sum_{h=1}^H \sum_{i=1}^d \mathbb{E}^{\pi^{\star}, P^{\pi^{\star},\dagger}}\left[\max _m\left\|\phi_i\left(s_h, a_h\right) \mathbf{e}_i\right\|_{\left(\boldsymbol{\Lambda}_h^{\mathcal C_m}\right)^{-1}} \mid s_1=s\right]+2(C+1)\sum_{h=1}^H \Delta_h
\end{align*}
Finally, by Assumption~\ref{partial-coverage-cluster} and following the same steps as in the proof of Corollary~\ref{corollary-suboptimality}, we have
\[
\operatorname{SubOpt}(\widehat\pi^{\mathrm{pool}};s)
\leq
c^\natural d^2H^2
\sqrt{
\frac{
\log\left(2dMHN_{\max}^{\mathrm{pool}}/\xi\right)
}{
N_{\min}^{\mathrm{pool}}
}
}
+
C^\natural\sum_{h=1}^H\Delta_h,
\]
where $c^\natural>0$ is a constant depending only on $c$ and $c'$, and 
\[
C^\natural
=
2\left(C+1\right)
=
2\left(
1+
\sup_{h\in[H]}
\sup_{(\tilde s,a)\in\mathcal S\times\mathcal A}
\sum_{i=1}^d
\phi_i(\tilde s,a)
\sqrt{
\frac{1}{
1/N_{\max}^{\mathrm{pool}}+c'\sigma_{h,i}^2
}
}
\right).
\]
Thus, on an event of probability at least $1-\xi$, the stated bound holds whenever the realized dataset satisfies Assumption~\ref{partial-coverage-cluster} for the fixed initial state $s$, which proves the claimed implication.
\end{proof}

\section{Complete Cluster-Level GR-PEVI Algorithm}
\label{app:cluster-algorithm}

For completeness, Algorithm~\ref{alg:cluster-gr-pevi} presents the cluster-level implementation of GR-PEVI. It applies the original procedure to the aggregated datasets $\{\mathcal D_{\mathcal C_m}\}_{m=1}^M$, treating each cluster as a super-source. At each stage, $\widehat\pi_h^{\mathrm{pool}}$ is chosen to be a measurable deterministic maximizer.

\begin{algorithm}[htbp]
\caption{Cluster-Level Group-Robust Pessimistic Value Iteration}
\label{alg:cluster-gr-pevi}
\begin{algorithmic}[1]
\Require features $\boldsymbol{\phi}$,
horizon $H$, regularizer $\lambda>0$, pessimism parameter
$\beta^{\mathrm{pool}}>0$, cluster partition $\{\mathcal C_m\}_{m=1}^M$,
and aggregated datasets
$\mathcal D^{\mathrm{pool}}=\{\mathcal D_{\mathcal C_m}\}_{m=1}^M$.
\State $\widehat V_{H+1}^{\mathrm{pool}}(\cdot)\gets 0$
\For{$h=H,H-1,\ldots,1$}
  \State \textbf{(A) Cluster-wise ridge estimation and uncertainty calculation}
  \For{$m=1$ \textbf{to} $M$}
     \State $y_h^{(m,\tau)}
     \gets r_h^{(m,\tau)}
     +\widehat V_{h+1}^{\mathrm{pool}}(s_{h+1}^{(m,\tau)})$
     for $h<H$, and
     $y_H^{(m,\tau)}\gets r_H^{(m,\tau)}$,
     for all $\tau\in[N_m]$
     \State $\boldsymbol{\Lambda}_h^{\mathcal C_m}\gets\sum_{\tau=1}^{N_m}\boldsymbol{\phi}\bigl(s_h^{(m,\tau)},a_h^{(m,\tau)}\bigr)\boldsymbol{\phi}\bigl(s_h^{(m,\tau)},a_h^{(m,\tau)}\bigr)^\top+\lambda\mathbf I_d$
     \State $\widehat{\boldsymbol{\nu}}_h^{\mathcal C_m}\gets(\boldsymbol{\Lambda}_h^{\mathcal C_m})^{-1}\sum_{\tau=1}^{N_m}\boldsymbol{\phi}\bigl(s_h^{(m,\tau)},a_h^{(m,\tau)}\bigr)y_h^{(m,\tau)}$
     \State $\mathbf u_h^{\mathcal C_m}\gets\sqrt{\operatorname{diag}\bigl((\boldsymbol{\Lambda}_h^{\mathcal C_m})^{-1}\bigr)}$
    \Comment{$[\mathbf u_h^{\mathcal C_m}]_i=\|\mathbf e_i\|_{(\boldsymbol{\Lambda}_h^{\mathcal C_m})^{-1}}$}
  \EndFor
  \State \textbf{(B) Coordinate-wise min/max across clusters}
  \For{$i=1$ \textbf{to} $d$}
   \State $(\widehat{\boldsymbol{\eta}}_h)_i
   \gets
   \min_{m\in[M]}
   (\widehat{\boldsymbol{\nu}}_h^{\mathcal C_m})_i$
   \State $(\mathbf m_h^{\mathrm{pool}})_i
   \gets
   \max_{m\in[M]}
   (\mathbf u_h^{\mathcal C_m})_i$
\EndFor
  \State \textbf{(C) Compute $Q/\pi/V$ with data-dependent pessimism}
\Statex \hspace{\algorithmicindent} All updates below are defined pointwise.
\State $\overline Q_h^{\mathrm{pool}}(s,a)\gets
\boldsymbol{\phi}(s,a)^\top\widehat{\boldsymbol{\eta}}_h
-\beta^{\mathrm{pool}}\boldsymbol{\phi}(s,a)^\top
\mathbf m_h^{\mathrm{pool}}$
\State $\widehat Q_h^{\mathrm{pool}}(s,a)\gets
\min\{\overline Q_h^{\mathrm{pool}}(s,a),H-h+1\}^{+}$
\State $\widehat\pi_h^{\mathrm{pool}}(\cdot\mid s)\in
\underset{\nu\in\mathcal P(\mathcal A)}{\arg\max}\,
\mathbb E_{a\sim\nu}[\widehat Q_h^{\mathrm{pool}}(s,a)]$
\hfill {\small$\triangleright$ greedy}
\State $\widehat V_h^{\mathrm{pool}}(s)\gets
\mathbb E_{a\sim\widehat\pi_h^{\mathrm{pool}}(\cdot\mid s)}
[\widehat Q_h^{\mathrm{pool}}(s,a)]$
\EndFor
\State \textbf{Output:}
$\widehat\pi^{\mathrm{pool}}
=\{\widehat\pi_h^{\mathrm{pool}}\}_{h=1}^H$
\end{algorithmic}
\end{algorithm}

\section{Additional Simulation Studies}
\label{sec:supp-simulations}
\subsection{Implementation Details for Comparison Methods}
\label{app:baseline-details}
We provide additional implementation details for the four comparison methods. Pooled PEVI, Avg-PEVI, and Majority-PEVI use the PEVI procedure of \citet[Algorithm~2]{jin2021pessimism} as a building block. For a PEVI fit based on $N$ trajectories, we use ridge parameter $\lambda$ and pessimism coefficient $\beta_N=c dH\sqrt{\log\left(2dHN/\xi\right)}$, and clip the resulting pessimistic $Q$-values to $[0,H-h+1]$ at stage $h$. The tuning parameters used in each experiment are reported in the corresponding simulation or real-data section.

Pooled PEVI combines the trajectories from all sources into a single offline dataset and applies PEVI to the pooled data, with the pessimism coefficient computed using the pooled sample size $N_{\mathrm{pool}}=\sum_{k=1}^K N_k$.

Avg-PEVI and Majority-PEVI apply PEVI separately to each source $k\in[K]$, yielding source-specific policies $\{\widehat{\pi}^k\}_{k=1}^K$. At each state--stage pair $(s,h)$, Avg-PEVI selects uniformly among the $K$ source-specific action recommendations $\{\widehat{\pi}_h^k(s)\}_{k=1}^K$. For reproducibility, we fix the randomization within each replication using a seeded pseudo-random rule, so that the same source-specific recommendation is selected whenever a given state--stage pair is evaluated. Majority-PEVI uses the same source-specific policies but selects the most frequently recommended action,
\[
\widehat{\pi}_h^{\mathrm{Maj}}(s)
\in
\arg\max_{a\in\mathcal A}
\sum_{k=1}^K
\mathbf 1
\left\{
\widehat{\pi}_h^k(s)=a
\right\}.
\]
Ties are resolved using a seeded pseudo-random rule within each replication.

Operator-Avg PEVI aggregates source-specific pessimistic Bellman backups rather than source-specific policies. Starting from $\widehat V_{H+1}^{\mathrm{OA}}\equiv 0$, the procedure proceeds backward over $h=H,\ldots,1$, with all sources using the common continuation value $\widehat V_{h+1}^{\mathrm{OA}}$. For each source $k$, we compute
\[
\boldsymbol{\Lambda}_h^k
=
\sum_{\tau=1}^{N_k}
\boldsymbol{\phi}\left(s_h^{(k,\tau)},a_h^{(k,\tau)}\right)
\boldsymbol{\phi}\left(s_h^{(k,\tau)},a_h^{(k,\tau)}\right)^\top
+\lambda\mathbf I_d
\]
and
\[
\widehat{\mathbf w}_h^k
=
\left(\boldsymbol{\Lambda}_h^k\right)^{-1}
\sum_{\tau=1}^{N_k}
\boldsymbol{\phi}\left(s_h^{(k,\tau)},a_h^{(k,\tau)}\right)
\left[
r_h^{(k,\tau)}
+
\widehat V_{h+1}^{\mathrm{OA}}
\left(s_{h+1}^{(k,\tau)}\right)
\right].
\]
The corresponding source-specific pessimistic backup is
\[
\widetilde Q_h^k(s,a)
=
\min\left\{
\boldsymbol{\phi}(s,a)^\top\widehat{\mathbf w}_h^k
-
\beta_{N_k}
\left\|
\boldsymbol{\phi}(s,a)
\right\|_{(\boldsymbol{\Lambda}_h^k)^{-1}},
H-h+1
\right\}^+,
\]
where $\beta_{N_k}=c dH\sqrt{\log\left(2dHN_k/\xi\right)}$. These backups are averaged equally across sources:
\[
\widetilde Q_h^{\mathrm{OA}}(s,a)
=
\frac{1}{K}\sum_{k=1}^K\widetilde Q_h^k(s,a),
\qquad
\widehat V_h^{\mathrm{OA}}(s)
=
\max_{a\in\mathcal A}
\widetilde Q_h^{\mathrm{OA}}(s,a).
\]
The learned policy is greedy with respect to $\widetilde Q_h^{\mathrm{OA}}$, and $\widehat V_h^{\mathrm{OA}}$ is used as the shared continuation value at the preceding stage.

\subsection{Simulation Setup for the Main Multi-Source Experiment}
\label{sec:supp-main-simulation}
We provide additional details for the data-generating process used in Section~\ref{sec:benchmark-comparison}. Recall that the main experiment uses $K=3$ sources, state space $\mathcal S=[0,1]^p$ with $p=3$, $|\mathcal A|=10$, horizon $H=7$, and feature dimension $d=p|\mathcal A|=30$.

For a state $s\in[0,1]^p$, let $\mathbf 1_p$ denote the $p$-dimensional vector of ones, and define
\[
\overline{s}
=
\begin{cases}
s/\|s\|_1, & \|s\|_1>0,\\[2mm]
p^{-1}\mathbf 1_p, & \|s\|_1=0.
\end{cases}
\]
The feature vector $\boldsymbol\phi(s,a)\in\mathbb R^d$ is obtained by placing $\overline{s}$ in the block corresponding to action $a$ and setting all other blocks to zero. Thus, $\boldsymbol\phi(s,a)$ is nonnegative and has unit $\ell_1$ norm.

For each source $k$, stage $h$, and feature coordinate $i$, the reward coefficient is generated independently as $\theta_{h,i}^k \sim \operatorname{Unif}[0.1,0.9]$. The conditional mean reward is therefore $r_h^k(s,a)
=
\left\langle
\boldsymbol\phi(s,a),\boldsymbol\theta_h^k
\right\rangle$. Given $(s,a)$, the observed reward is generated from
\[
r_h^{(k,\tau)}
\mid s,a
\sim
\operatorname{Beta}
\left(
\kappa r_h^k(s,a),
\kappa\{1-r_h^k(s,a)\}
\right),
\qquad
\kappa=15,
\]
so that $\mathbb E[r_h^{(k,\tau)}\mid s,a]=r_h^k(s,a)$. Since $r_h^k(s,a)\in[0.1,0.9]$, both shape parameters are positive and the observed rewards take values in $(0,1)$.

We next specify the transition components. For each source $k$, feature coordinate $i\in[d]$, and state coordinate
$j\in[p]$, we independently generate source-level multiplicative factors
satisfying
\[
\log U_{i,j}^{k,a},\;
\log U_{i,j}^{k,b}
\overset{\mathrm{i.i.d.}}{\sim}
N(0,\sigma_{\mathrm{source}}^2),
\]
and define the baseline Beta shape parameters
\[
a_{i,j}^{k,0}
=
\max\{a_{\min},a_0U_{i,j}^{k,a}\},
\qquad
b_{i,j}^{k,0}
=
\max\{a_{\min},b_0U_{i,j}^{k,b}\}.
\]
At each stage $h$, independently generate
\[
\log V_{h,i,j}^{k,a},
\log V_{h,i,j}^{k,b}
\overset{\mathrm{i.i.d.}}{\sim}
N(0,\sigma_{\mathrm{stage}}^2),
\]
and set
\[
a_{h,i,j}^k
=
\max\{a_{\min},a_{i,j}^{k,0}V_{h,i,j}^{k,a}\},
\qquad
b_{h,i,j}^k
=
\max\{a_{\min},b_{i,j}^{k,0}V_{h,i,j}^{k,b}\}.
\]
The $i$th transition component is the product-Beta distribution
\[
\mu_{h,i}^k
=
\bigotimes_{j=1}^p
\operatorname{Beta}\!\left(a_{h,i,j}^k,b_{h,i,j}^k\right).
\]
The source-specific transition kernel is then defined as
\[
P_h^k(\cdot\mid s,a)
=
\sum_{i=1}^d
\phi_i(s,a)\mu_{h,i}^k(\cdot).
\]

For the main simulation, we use $a_0=b_0=2$, $a_{\min}=0.3$, $\sigma_{\mathrm{source}}=1.0$, and $\sigma_{\mathrm{stage}}=1.2$. The source-specific MDP parameters are generated once and held fixed across the $50$ replications. In each replication, new offline trajectories are generated under a uniform behavior policy, with initial states sampled uniformly from $[0,1]^p$. The source sample sizes are $(N_1,N_2,N_3)=(3000,2000,5000)$.

\subsection{Sensitivity to Transition Heterogeneity}
\label{sec:supp-mild-heterogeneity}

We repeat the simulation in Section~\ref{sec:benchmark-comparison} under a milder transition-heterogeneity regime, while keeping all other aspects of the data-generating process, sample sizes, learning procedures, and evaluation protocol unchanged. Specifically, we reduce the stage-level heterogeneity parameter from $\sigma_{\mathrm{stage}}=1.2$ to $0.4$ and the source-level heterogeneity parameter from $\sigma_{\mathrm{source}}=1.0$ to $0.3$.

Table~\ref{tab:mild-heterogeneity} reports the results under the milder transition-heterogeneity regime together with those from the main simulation setting for comparison. GR-PEVI continues to achieve the lowest mean robust suboptimality, and the competing methods generally improve relative to the main setting. In particular, the mean suboptimality of Operator-Avg PEVI decreases from $0.242$ to $0.196$, and that of Pooled PEVI decreases from $0.329$ to $0.274$, whereas GR-PEVI remains close to $0.02$. This pattern is consistent with a reduced advantage from guarding against cross-source transition variation when the transition mechanisms are more similar across sources.

\begin{table}[tb]
\centering
\caption{Robust suboptimality under the main and milder transition-heterogeneity settings. Entries are means over 50 Monte Carlo replications, with standard deviations in parentheses. Lower values indicate better performance.}
\label{tab:mild-heterogeneity}
\begin{tabular}{lcc}
\toprule
Method & Main setting & Milder transition heterogeneity \\
\midrule
GR-PEVI            & 0.017 (0.007) & 0.020 (0.006) \\
Pooled PEVI        & 0.329 (0.042) & 0.274 (0.034) \\
Avg-PEVI           & 0.951 (0.021) & 0.872 (0.020) \\
Majority-PEVI      & 0.895 (0.022) & 0.778 (0.025) \\
Operator-Avg PEVI  & 0.242 (0.021) & 0.196 (0.015) \\
\bottomrule
\end{tabular}
\end{table}

\subsection{Numerical Illustration of Cluster-Level Extension}
\label{app:cluster-simulation}

We provide an additional numerical experiment to illustrate the statistical--approximation trade-off underlying the cluster-level extension in Section~\ref{sec:cross-source}. The basic linear MDP construction follows the setup in Section~\ref{sec:supp-main-simulation}, including the action-specific simplex-valued feature map, the product-Beta transition components, and the generation of reward components.

We consider $K=4$ sources partitioned into two prespecified clusters, $\mathcal C_1=\{1,2\}$ and $\mathcal C_2=\{3,4\}$. For each cluster $m\in\{1,2\}$, we independently generate cluster-center reward coefficients $\overline{\boldsymbol\theta}_h^m$ and Beta transition shape parameters $\{\overline a_{h,i,j}^m,\overline b_{h,i,j}^m\}_{i\in[d],\,j\in[p]}$ for each stage $h\in[H]$. We also independently generate source-specific anchor reward coefficients $\widetilde{\boldsymbol\theta}_h^k$ and Beta transition shape parameters $\{\widetilde a_{h,i,j}^k,\widetilde b_{h,i,j}^k\}_{i\in[d],\,j\in[p]}$ for each $k\in[K]$ and $h\in[H]$, using the same parameter-generation mechanism.

For each source $k\in\mathcal C_m$, we define
\[
    \boldsymbol\theta_h^k(\delta)
    =
    (1-\delta)\overline{\boldsymbol\theta}_h^m
    +
    \delta\widetilde{\boldsymbol\theta}_h^k,
\]
and, for each $i\in[d]$ and $j\in[p]$,
\[
    a_{h,i,j}^k(\delta)
    =
    (1-\delta)\overline a_{h,i,j}^m
    +
    \delta\widetilde a_{h,i,j}^k,
    \qquad
    b_{h,i,j}^k(\delta)
    =
    (1-\delta)\overline b_{h,i,j}^m
    +
    \delta\widetilde b_{h,i,j}^k.
\]
The $i$th transition component is then defined as
\[
\mu_{h,i}^k(\delta)
=
\bigotimes_{j=1}^p
\operatorname{Beta}\!\left(
a_{h,i,j}^k(\delta),
b_{h,i,j}^k(\delta)
\right).
\]
The parameter $\delta\in[0,1]$ controls within-cluster heterogeneity. When $\delta=0$, the sources within each cluster share identical transition and reward components. As $\delta$ increases, each source moves toward its source-specific anchor, resulting in greater within-cluster variation. The two cluster centers are independently generated, thereby inducing between-cluster heterogeneity in the construction.

We use state space $\mathcal S=[0,1]^p$ with $p=3$, $|\mathcal A|=10$, and horizon $H=7$. The source sample sizes are $(N_1,N_2,N_3,N_4)=(300,3000,300,3000)$. Hence, the source-level minimum sample size is $N_{\min}=300$, and pooling within the two prespecified clusters yields cluster-level sample sizes of $3300$ each, so that $N_{\min}^{\mathrm{pool}}=3300$. We vary the within-cluster perturbation level over $\delta\in\{0,0.1,0.2,0.4,0.6\}$. Offline trajectories are generated under a uniformly random behavior policy, and all methods use $\lambda=1$, $\xi=0.05$, and $c=5\times10^{-4}$.

We compare source-level GR-PEVI, cluster-level GR-PEVI using the prespecified partition $\{\mathcal C_1,\mathcal C_2\}$, and Pooled PEVI. For cluster-level GR-PEVI, trajectories from sources within the same cluster are aggregated, and the two pooled datasets are treated as super-sources during policy learning.

All learned policies are evaluated under the original four-source robust objective. Thus, cluster aggregation only affects the estimation procedure but does not change the evaluation target. For each value of $\delta$, we run $R=50$ replications under fixed source MDPs, regenerating the offline dataset in each replication. Within each replication, expected robust suboptimality is estimated by averaging $V_1^{\star}(s)-V_1^{\widehat\pi}(s)$ over $200$ initial states sampled uniformly from $\mathcal S$.

\begin{figure}[t]
    \centering
    \includegraphics[width=0.9\textwidth]{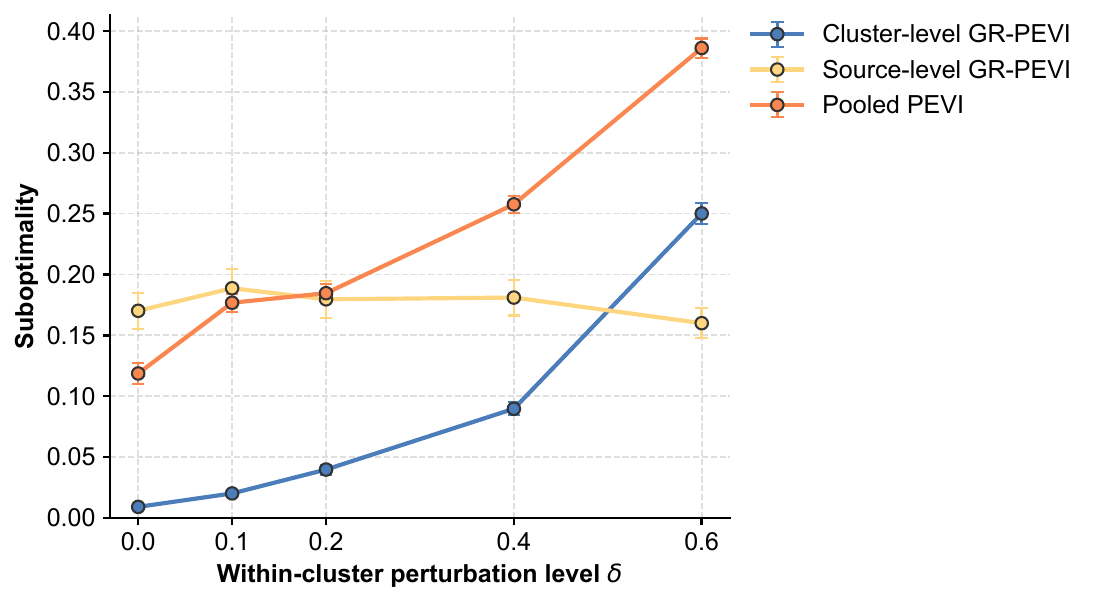}
    \caption{Comparison of robust suboptimality across methods under increasing within-cluster heterogeneity. Mean suboptimality with 95\% confidence intervals over $R=50$ replications.}
    \label{fig:cluster-tradeoff}
\end{figure}

Figure~\ref{fig:cluster-tradeoff} reports the mean and 95\% confidence interval across the $R$ replications. When $\delta$ is small, cluster-level GR-PEVI substantially improves over source-level GR-PEVI. In this regime, the sources within each cluster are sufficiently similar that pooling introduces little approximation error, and the larger cluster-level sample sizes reduce estimation uncertainty. As $\delta$ increases, the robust suboptimality of cluster-level GR-PEVI rises steadily, reflecting the growing discrepancy among the sources pooled within each cluster.

Source-level GR-PEVI remains comparatively stable across the values of $\delta$, since it estimates each source separately and does not rely on within-cluster similarity. Cluster-level GR-PEVI continues to outperform the source-level method for moderate heterogeneity, but becomes less favorable when $\delta$ is large. Pooled PEVI, by contrast, never achieves the lowest suboptimality across the entire range of $\delta$, consistent with its failure to account for cross-source heterogeneity. These results illustrate that cluster-level pooling can improve statistical efficiency when the proposed clusters are sufficiently homogeneous, while its advantage diminishes as the approximation error induced by within-cluster heterogeneity increases.

\section{Additional Details for the Real-Data Analysis}
\label{app:real-data-details}
\subsection{Disease-Modifying Therapy Classification}
\label{sec:app_DMT_tab}
Table~\ref{tab:ms_dmt} lists the disease-modifying therapies used to define the treatment-action categories in the real-data analysis. RxNorm-coded prescription records matching these medications were mapped to either the standard- or high-efficacy DMT class.
\begin{table}[htbp]
\centering
\caption{Disease-modifying therapies for multiple sclerosis grouped by efficacy category.}
\label{tab:ms_dmt}
\begin{tabular}{llr}
\toprule
\textbf{Generic name} & \textbf{Brand name(s)} & \textbf{RxNorm ID} \\
\midrule
\multicolumn{3}{l}{\textit{High efficacy}} \\
\midrule
Natalizumab           & Tysabri            &  354770 \\
Ocrelizumab           & Ocrevus            & 1876366 \\
Ofatumumab            & Arzerra; Kesimpta  &  712566 \\
Mitoxantrone          & Novantrone         &    7005 \\
Alemtuzumab           & Campath; Lemtrada  &  117055 \\
Cladribine            & Mavenclad          &   44157 \\
Rituximab             & Rituxan            &  121191 \\
Cyclophosphamide      & Cytoxan            &    3002 \\
\midrule
\multicolumn{3}{l}{\textit{Standard efficacy}} \\
\midrule
Dimethyl fumarate     & Tecfidera          & 1373478 \\
Diroximel fumarate    & Vumerity           & 2261783 \\
Monomethyl fumarate   & Bafiertam          & 1546433 \\
Glatiramer acetate    & Copaxone; Glatopa  &  214582 \\
Daclizumab            & Zenapax            &  190353 \\
Interferon beta-1a    & Avonex; Rebif      &   75917 \\
Interferon beta-1b    & Betaseron; Extavia &   72257 \\
Peginterferon beta-1a & Plegridy           & 1546168 \\
Teriflunomide         & Aubagio            & 1310520 \\
Fingolimod            & Gilenya; Tascenso  & 1012892 \\
Ozanimod              & Zeposia            & 2288236 \\
Siponimod             & Mayzent            & 2121085 \\
\bottomrule
\end{tabular}
\end{table}

\subsection{PDDS Imputation}
\label{app:pdds-imputation}
Because PDDS was not recorded in every month, missing monthly PDDS values were imputed within each train--test split. We fitted a weighted XGBoost model \citep{chen2016xgboost} to training patient-month records with observed PDDS, pooling MGB and UPMC records with weights that balanced the two healthcare systems, and applied the fitted model to missing values in both the training and test sets. Imputed values were truncated to $[0,8]$ without rounding, and observed PDDS values were left unchanged. The imputation was repeated for each of the 50 random splits.

\subsection{Fitted Q-Evaluation and WAIPW Estimation}
\label{appendix:ope}
Let $\mathcal{D}_{k,b}^{\mathrm{tr}}$ and $\mathcal{D}_{k,b}^{\mathrm{te}}$ denote the training and held-out trajectory sets from source $k$, with corresponding sample sizes $n_{k,b}^{\mathrm{tr}}$ and $n_{k,b}^{\mathrm{te}}$. For notational simplicity, we suppress the split index $b$ throughout this section and write these quantities as $\mathcal{D}_{k}^{\mathrm{tr}}$, $\mathcal{D}_{k}^{\mathrm{te}}$, $n_k^{\mathrm{tr}}$, and $n_k^{\mathrm{te}}$.

\subsubsection{Fitted Q-evaluation}
For a fixed deterministic target policy $\pi=\{\pi_h\}_{h=1}^{H}$, source $k$, and stage $h\in[H]$, define the (non-robust) action-value and value functions of $\pi$ under the source-$k$ MDP as
\[
Q_{k,h}^{\pi}(s,a)
=
\mathbb{E}_{P^k,\pi}
\left[
\sum_{t=h}^{H} r_t^k(s_t,a_t)
\,\middle|\,
s_h=s,\ a_h=a
\right],
\qquad
V_{k,h}^{\pi}(s)
=
Q_{k,h}^{\pi}\bigl(s,\pi_h(s)\bigr),
\]
with $V_{k,H+1}^{\pi}\equiv 0$. We estimate $Q_{k,h}^{\pi}$ using a linear model in the shared state--action representation, clipped to the feasible return range:
\[
\widehat{Q}_{k,h}^{\pi}(s,a)
=
\min\left\{
\boldsymbol{\phi}(s,a)^\top
\widehat{\mathbf{w}}_{k,h}^{\pi},
\,H-h+1
\right\}^{+},
\qquad
\widehat{V}_{k,h}^{\pi}(s)
=
\widehat{Q}_{k,h}^{\pi}\bigl(s,\pi_h(s)\bigr).
\]

The coefficients $\{\widehat{\mathbf{w}}_{k,h}^{\pi}\}_{h=1}^{H}$ are estimated backward in time using the source-$k$ training data $\mathcal{D}_{k}^{\mathrm{tr}}$. Starting from $\widehat{V}_{k,H+1}^{\pi}\equiv0$, for $h=H,H-1,\ldots,1$, the regression target for training trajectory $\tau$ is
\[
y_h^{(k,\tau)}
=
r_h^{(k,\tau)}
+
\widehat{V}_{k,h+1}^{\pi}
\left(s_{h+1}^{(k,\tau)}\right),
\]
which reduces to $y_H^{(k,\tau)}=r_H^{(k,\tau)}$ at the last stage, and the coefficient vector is estimated by ridge regression:
\[
\widehat{\mathbf{w}}_{k,h}^{\pi}
=
\operatorname*{arg\,min}_{\mathbf{w}\in\mathbb{R}^{d}}
\left\{
\sum_{\tau=1}^{n_k^{\mathrm{tr}}}
\left[
y_h^{(k,\tau)}
-
\boldsymbol{\phi}
\left(s_h^{(k,\tau)},a_h^{(k,\tau)}\right)^\top
\mathbf{w}
\right]^2
+
\lambda_{\mathrm{FQE}}\|\mathbf{w}\|_2^2
\right\},
\]
with $\lambda_{\mathrm{FQE}} = 1$. The clipped estimates $\widehat{Q}_{k,h}^{\pi}$ and $\widehat{V}_{k,h}^{\pi}$ are used both in the backward recursion and in the WAIPW estimator below.

\subsubsection{WAIPW estimation}
\label{subsec:waipw}
For held-out trajectory $\tau$ and stage $h\in[H]$, define the temporal-difference residual at stage $h$ as 
\[
\widehat{\delta}_{k,h}^{\pi,(\tau)}
=
r_h^{(k,\tau)}
+
\widehat{V}_{k,h+1}^{\pi}
\left(s_{h+1}^{(k,\tau)}\right)
-
\widehat{Q}_{k,h}^{\pi}
\left(s_h^{(k,\tau)},a_h^{(k,\tau)}\right).
\]

Let $\widehat{b}_{k,h}(a\mid s)$ denote the behavior propensity at stage $h$, estimated separately for each source and stage using the training data. We fitted $\ell_2$-regularized multinomial logistic regression models on standardized state covariates with inverse regularization strength $C=0.1$. When only one action was observed, we used a constant propensity model. 
 
For the deterministic target policy $\pi$ and stage $h\in[H]$, define the stage-specific importance ratio as  
\[
\omega_{k,h}^{\pi,(\tau)}
=
\frac{
\mathbf{1}
\left\{
a_h^{(k,\tau)}
=
\pi_h\left(s_h^{(k,\tau)}\right)
\right\}
}{
\max
\left\{
\widehat{b}_{k,h}
\left(a_h^{(k,\tau)}\mid s_h^{(k,\tau)}\right),
b_{\min}
\right\}
},
\]
where $b_{\min}=0.1$ is the propensity floor. The truncated cumulative importance weights are defined recursively by  
\[
\widetilde{\rho}_{k,1:h}^{\pi,(\tau)}
=
\min
\left\{
\widetilde{\rho}_{k,1:h-1}^{\pi,(\tau)}
\omega_{k,h}^{\pi,(\tau)},
\rho_{\max}
\right\},
\qquad
\widetilde{\rho}_{k,1:0}^{\pi,(\tau)}=1,
\]
with truncation threshold $\rho_{\max}=100$.

The source-specific WAIPW value estimator is 
\[
\widehat{V}_{k}^{\mathrm{WAIPW}}(\pi)
=
\frac{1}{n_k^{\mathrm{te}}}
\sum_{\tau=1}^{n_k^{\mathrm{te}}}
\widehat{V}_{k,1}^{\pi}
\left(s_1^{(k,\tau)}\right)
+
\sum_{h=1}^{H}
\frac{
\displaystyle
\sum_{\tau=1}^{n_k^{\mathrm{te}}}
\widetilde{\rho}_{k,1:h}^{\pi,(\tau)}
\widehat{\delta}_{k,h}^{\pi,(\tau)}
}{
\displaystyle
\sum_{\tau=1}^{n_k^{\mathrm{te}}}
\widetilde{\rho}_{k,1:h}^{\pi,(\tau)}
}.
\]
If the denominator of the residual-correction term is zero at a particular stage, the corresponding term is set to zero.

For a learned policy $\widehat{\pi}$, the empirical worst-source performance measure used in the main analysis is 
\[ 
\widehat{V}_{\mathrm{worst}}(\widehat{\pi}) = \min_{k\in[K]} \widehat{V}_{k}^{\mathrm{WAIPW}}(\widehat{\pi}), 
\] 
where $K=3$, and $k=1,2,3$ correspond to MGB-young, MGB-old, and UPMC, respectively. 

\section{Additional Real-Data Analysis: Sepsis}
\label{appendix:sepsis}
As an additional real-data analysis, we considered fluid-management decisions for patients with sepsis using the MIMIC-IV EHR database \citep{raghu2017continuous,johnson2020mimiciv}. The sepsis cohort was extracted using publicly available SQL code\footnote{\url{https://github.com/yugangjia/Team-Sepsis/blob/main/Sepsis_cohort.sql}}, with ICU care units treated as distinct data sources. We focused on six ICU care units with relatively large numbers of trajectories: Medical/Surgical ICU (MICU/SICU), Trauma SICU (TSICU), Surgical ICU (SICU), Medical ICU (MICU), Coronary Care Unit (CCU), and Cardiac Vascular ICU (CVICU). The resulting dataset contained 24,719 trajectories, with source-specific sample sizes of 4,828, 2,898, 3,509, 5,800, 2,282, and 5,402, respectively.

We constructed finite-horizon trajectories with $H=5$ stages, each corresponding to a four-hour interval. The clinical state consisted of five variables: hemoglobin, potassium, body weight, temperature, and systolic blood pressure. For each data split, missing state variables were imputed using medians estimated from the pooled training data. The resulting state variables were then winsorized using the 1st and 99th percentiles of the pooled training data and min--max scaled to $[0,1]$ using the same training-derived bounds. Letting $z_h=(z_{h1},\ldots,z_{h5})^\top$ denote the five preprocessed clinical variables, we represented the state as $S_h=(1,z_h^\top)^\top\in\mathbb R^6$. The action space $\mathcal A=\{0,1,2\}$ represented three levels of intravenous fluid administration within each four-hour stage: no fluid, low positive fluid, and high positive fluid. For each split, positive fluid inputs were divided at the median of the positive intravenous fluid inputs in the pooled training data. Action $1$ corresponded to positive inputs below the cutoff and action $2$ to inputs at or above the cutoff, while nonpositive or missing fluid inputs were assigned to action $0$. The stage-wise reward was defined as
\[
R_h=1-\operatorname{clip}\left(\frac{L_h}{32},0,1\right),
\]
where $L_h$ denotes the lactate level observed at stage $h$. Thus, larger rewards correspond to lower lactate levels. At each stage, observations with recorded lactate were used for the Bellman regressions in policy learning and FQE. 

For WAIPW evaluation, let $M_h^{(k,\tau)}$ indicate whether the reward of trajectory $\tau$ from source $k$ was observed at stage $h$, with $M_h^{(k,\tau)}=1$ if the reward is observed and $0$ otherwise. The WAIPW estimator in Section~\ref{subsec:waipw} was adapted to
\[
\widehat V_k^{\mathrm{WAIPW}}(\pi)
=
\frac{1}{n_k^{\mathrm{te}}}
\sum_{\tau=1}^{n_k^{\mathrm{te}}}
\widehat V_{k,1}^{\pi}\bigl(s_1^{(k,\tau)}\bigr)
+
\sum_{h=1}^{H}
\frac{
\sum_{\tau=1}^{n_k^{\mathrm{te}}}
M_h^{(k,\tau)}
\widetilde{\rho}_{k,1:h}^{\pi,(\tau)}
\widehat{\delta}_{k,h}^{\pi,(\tau)}
}{
\sum_{\tau=1}^{n_k^{\mathrm{te}}}
M_h^{(k,\tau)}
\widetilde{\rho}_{k,1:h}^{\pi,(\tau)}
}.
\]
The residual correction at each stage was computed using only observations with recorded lactate, with importance weights normalized over the corresponding subset. Because lactate was not observed at every stage, the resulting WAIPW estimates may be affected by the lactate observation mechanism and associated selection bias.

To construct the feature representation used by the group-linear model, we defined
\[
g(S_h)
=
\bigl(
1,z_{h1},\ldots,z_{h5},
z_{h1}^2,\ldots,z_{h5}^2
\bigr)^\top
\in\mathbb R^{11},
\qquad
\psi(S_h)
=
\frac{g(S_h)}{\|g(S_h)\|_1}.
\]
We then used the action-block construction, assigning the normalized state features to the block corresponding to action $a$, with all other action blocks set to zero. This yielded a nonnegative simplex-valued feature map of dimension $d=33$. Following the same analysis framework as in the MS study, we compared GR-PEVI with Pooled PEVI, Avg-PEVI, Majority-PEVI, and Operator-Avg PEVI. For the sepsis analysis, each source was independently split into training and test sets in an approximately 50:50 ratio. The policy-learning and WAIPW tuning parameters were otherwise the same as those used in the MS analysis.  For each split, the worst-source value was defined as the minimum of the six source-specific WAIPW estimates on the held-out data. We repeated the procedure over 50 random splits. Table~\ref{tab:sepsis_results} summarizes the resulting performance. GR-PEVI achieved the highest mean and median worst-source values among the five methods and attained the best worst-source performance in 38 of the 50 splits. Compared with Operator-Avg PEVI, the strongest baseline in the worst-source comparison, GR-PEVI improved the worst-source WAIPW value by 0.0284 on average across splits.

\begin{table}[t]
\centering
\caption{Worst-source WAIPW over 50 train--test splits. SD is computed across splits; win count is the number of splits with the highest value; GR-PEVI win rate is the proportion of paired splits in which GR-PEVI outperformed each baseline.}
\label{tab:sepsis_results}
\begin{tabular}{lccccc}
\toprule
Method & Mean & SD & Median & Win count & GR-PEVI win rate \\
\midrule
GR-PEVI
& \textbf{4.6629} & 0.0247 & \textbf{4.6576} & \textbf{38/50} & -- \\
Operator-Avg PEVI
& 4.6345 & 0.0128 & 4.6369 & 6/50 & 86\% \\
Majority-PEVI
& 4.6269 & 0.0237 & 4.6277 & 4/50 & 84\% \\
Avg-PEVI
& 4.6095 & 0.0348 & 4.6187 & 2/50 & 92\% \\
Pooled PEVI
& 4.3767 & 0.0577 & 4.3824 & 0/50 & 100\% \\
\bottomrule
\end{tabular}
\end{table}

To examine the source-specific patterns underlying the worst-source results, Table~\ref{tab:sepsis_sourcewise} reports the relative improvements of each learned policy's WAIPW estimate over the empirical cumulative reward under the observed behavior policy, expressed as a percentage. Because lactate was not observed at every stage, the empirical cumulative reward under the observed behavior policy was computed by first averaging the observed rewards within each stage and source on the held-out data and then summing these stage-specific averages over the horizon. GR-PEVI achieved positive mean relative improvements in all six sources, with the largest improvements among the methods in TSICU, SICU, and CVICU. It outperformed Avg-PEVI, Majority-PEVI, and Operator-Avg PEVI in terms of mean relative improvement in five of the six sources, with CCU being the exception. Pooled PEVI attained the largest mean improvements in MICU/SICU and MICU but showed a negative mean improvement in TSICU, whereas Operator-Avg PEVI performed best in CCU.

\begin{table}[t]
\centering
\caption{Source-specific relative improvements (\%) over the observed behavior policy across 50 train--test splits. Entries report the mean (SD) across splits; bold entries indicate the largest mean in each source.}
\label{tab:sepsis_sourcewise}
\small
\setlength{\tabcolsep}{7pt}
\renewcommand{\arraystretch}{1.08}
\resizebox{\textwidth}{!}{%
\begin{tabular}{@{}lcccccc@{}}
\toprule
Method & MICU/SICU & TSICU & SICU & MICU & CCU & CVICU \\
\midrule
GR-PEVI
& $+1.61$ (0.65)
& $\mathbf{+2.07}$ (0.89)
& $\mathbf{+2.18}$ (0.71)
& $+1.75$ (0.55)
& $+1.32$ (0.59)
& $\mathbf{+1.90}$ (0.62) \\

Operator-Avg PEVI
& $+0.70$ (0.25)
& $+0.92$ (0.45)
& $+0.64$ (0.39)
& $+1.22$ (0.24)
& $\mathbf{+2.71}$ (0.79)
& $+0.53$ (0.30) \\

Avg-PEVI
& $+1.30$ (0.34)
& $+0.32$ (0.96)
& $+0.84$ (0.73)
& $+1.59$ (0.45)
& $+1.98$ (1.15)
& $+0.68$ (0.44) \\

Majority-PEVI
& $+1.56$ (0.28)
& $+0.65$ (0.68)
& $+0.80$ (0.51)
& $+1.69$ (0.31)
& $+2.35$ (0.83)
& $+0.72$ (0.31) \\

Pooled PEVI
& $\mathbf{+1.86}$ (0.43)
& $-5.05$ (1.24)
& $+0.45$ (0.86)
& $\mathbf{+2.26}$ (0.46)
& $+2.55$ (0.82)
& $+1.53$ (0.54) \\
\bottomrule
\end{tabular}%
}
\end{table}

\end{document}